\documentclass[%
 reprint,
 superscriptaddress,
 amsmath,amssymb,
 aps,
prb,
floatfix,
longbibliography,
]{revtex4-1}

\usepackage{graphicx}
\usepackage{dcolumn}
\usepackage{bm}
\usepackage[normalem]{ulem}
\usepackage{hyperref}
\hypersetup{colorlinks = true, citecolor = blue, breaklinks = true}

\usepackage{textcomp}
\usepackage{multirow}
\usepackage{xfrac}

\usepackage[version=4]{mhchem}

\usepackage[resetlabels]{multibib}
\newcites{SI}{Supporting Information}

\begin{document}

\title{Interplay between polarization, strain and defect-pairs in Fe-doped \texorpdfstring{\ce{SrMnO_{3-\delta}}}{SrMnO$_{3-\delta}$}}

\affiliation{%
Department of Chemistry and Biochemistry, University of Bern, Freiestrasse 3, CH-3012 Bern, Switzerland 
}%
\affiliation{%
National Centre for Computational Design and Discovery of Novel Materials (MARVEL), Switzerland
}%

\author{Chiara Ricca}
\affiliation{%
Department of Chemistry and Biochemistry, University of Bern, Freiestrasse 3, CH-3012 Bern, Switzerland 
}%
\affiliation{%
National Centre for Computational Design and Discovery of Novel Materials (MARVEL), Switzerland
}%

\author{Ulrich Aschauer}
\email{ulrich.aschauer@dcb.unibe.ch}
\affiliation{%
Department of Chemistry and Biochemistry, University of Bern, Freiestrasse 3, CH-3012 Bern, Switzerland 
}%
\affiliation{%
National Centre for Computational Design and Discovery of Novel Materials (MARVEL), Switzerland
}%
\date{\today}

\begin{abstract}
Defect chemistry, strain, and structural, magnetic and electronic degrees of freedom constitute a rich space for the design of functional properties in transition metal oxides. Here, we show that it is possible to engineer polarity and ferroelectricity in non-polar perovskite oxides via polar defect pairs formed by anion vacancies coupled to substitutional cations. We use a self-consistent site-dependent DFT+$U$ approach that accounts for local structural and chemical changes upon defect creation and which is crucial to reconcile predictions with the available experimental data. Our results for Fe-doped oxygen-deficient \ce{SrMnO3} show that substitutional Fe and oxygen vacancies can promote polarity due to an off-center displacement of the defect charge resulting in a net electric dipole moment, which polarizes the lattice in the defect neighborhood. The formation of these defects and the resulting polarization can be tuned by epitaxial strain, resulting in enhanced polarization also for strain values lower than the ones necessary to induce a polar phase transition in undoped \ce{SrMnO3}. For high enough defect concentrations, these defect dipoles couple in a parallel fashion, thus enabling defect- and strain-based engineering of ferroelectricity in \ce{SrMnO3}. 
\end{abstract}

\maketitle

\section{Introduction\label{sec:intro}}

The interplay between electric polarization, magnetism, strain and the defect chemistry constitutes a rich phase diagram for the design and control of novel functional properties in transition-metal perovskites~\cite{fuchigami2009, tuller2011, kalinin2012, kalinin2013functional, chandrasekaran2013, bivskup2014, bhattacharya2014magnetic, becher2015strain, marthinsen2016coupling,griffin2017defect, rojac2017domain}. In particular, strain imposed, for example, by lattice matching with the substrate during coherent epitaxial growth of thin films is an established route to engineer polarity and ferroelectricity in non-polar complex oxides~\cite{lee2010epitaxial, chandrasekaran2013, bhattacharya2014magnetic, becher2015strain, marthinsen2016coupling, griffin2017defect}. Defect engineering can tailor the ferroelectric response by introducing polar defect pairs. In particular, substitutional defects coupled to oxygen vacancies (\ce{V_O}), such as \ce{Fe_{Ti}-V_{O}} defect pairs, were shown to align with the direction of the lattice polarization in ferroelectric \ce{PbTiO3}~\cite{chandrasekaran2013} or to promote ferroelectricity in paraelectric \ce{SrTiO3}~\cite{Wang2017nano}. Polar distortions, strain and stoichiometry can couple or compete in determining the material properties as shown for oxygen-deficient \ce{SrMnO3} (SMO) thin films~\citep{marthinsen2016coupling}, the material we also use a model system in the present study.

\begin{figure}
	\centering
	\includegraphics[width=\columnwidth]{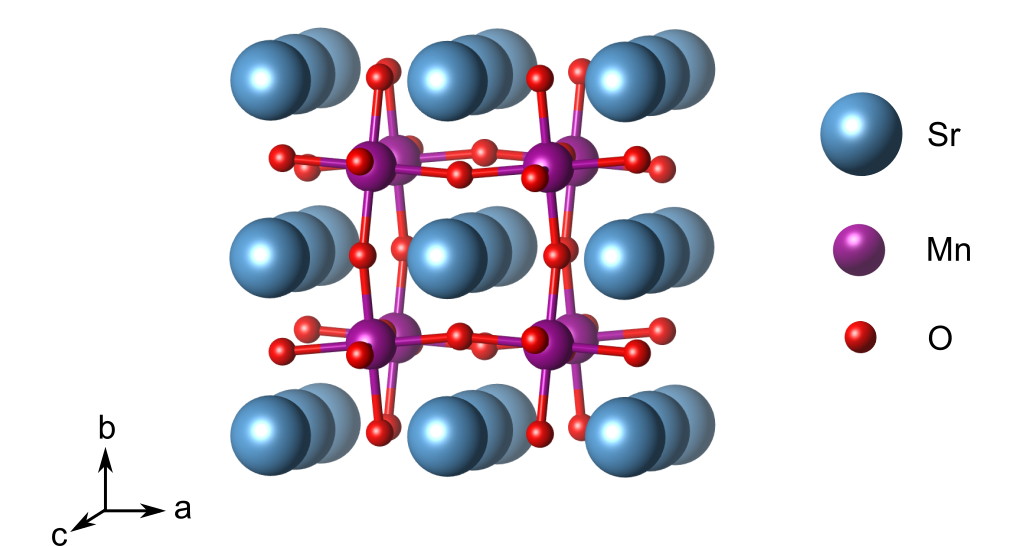}
	\caption{($2\times2\times2$) \textit{Pnma} supercell of stoichiometric \ce{SrMnO3}.}
	\label{fig:SMO_structure_bulk}
\end{figure}

Bulk SMO occurs in a hexagonal structure~\cite{Syono1969Structure}, but the perovskite phase of SMO (space group \textit{Pnma}, see Fig. \ref{fig:SMO_structure_bulk}) with G-type antiferromagnetic (AFM) order~\cite{chmaissem2001relationship} can be stabilized at low temperature in thin films~\cite{Kobayashi:2010by}. It was predicted from theory that biaxial epitaxial strain induces a polar distortion in SMO, mainly associated with Mn ions displacing from their high-symmetry positions, the magnitude of the distortion and hence the ferroelectric polarization increasing with increasing biaxial strain~\cite{lee2010epitaxial, becher2015strain, marthinsen2016coupling}. This is caused by softening of the in-plane polar modes that become unstable for tensile strains larger than about~2\%. Compressive strains larger than~4\% are, instead, necessary to induce ferroelectricity in the direction perpendicular to the strain plane. The strain response is different in the competing FM phase, requiring smaller compressive but larger tensile strains to trigger polar instabilities~\cite{ricca2021ferroelectricity}. At the same time, tensile strain favors the formation of oxygen-vacancies, the presence of which was, however, found to be detrimental to ferroelectricity~\cite{marthinsen2016coupling}. 

Fe-doping of oxygen deficient (\ce{SrMnO_{3-\delta}}) thin-films could be a strategy to reverse this behavior, since ferroelectricity is expected to be enhanced both by the decrease in crystal symmetry due to the aliovalent Fe ion and especially by the formation of a defect dipole due to the possible association of the substitutional Fe with an oxygen vacancy~\cite{Wang2017nano}. \ce{SrMn_{1-x}Fe_xO_{3-\delta}} was synthesized in the past in an attempt to obtain manganites with a mixture of $d^3$ (\ce{Mn^{+4}}) and $d^4$ (\ce{Fe^{+4}}) cations~\cite{FAWCETT2000821}, as found in other manganites showing colossal magnetoresistance, \textit{e.g.} the family of doped \ce{Ln_{1-x}(Ca/Sr)_xMnO3}~\cite{Jin413, Millis1998Nature}. At room temperature \ce{SrMn_{1-x}Fe_xO_{3-\delta}} phases were found to adopt a cubic perovskite structure with a disordered arrangement of Mn and Fe transition-metal cations in the octahedral sites due to their similar ionic radii. Oxygen vacancies observed during synthesis in air are generally created to maintain charge balance after the aliovalent substitution and introduce \ce{Fe^{+3}} cations in the structure as suggested by iodometric and M{\"o}ssbauer measurements~\cite{Battle1988, FAWCETT2000821, BattleJM9960601187}. M{\"o}ssbauer data suggest as well that oxygen vacancies are predominantly found in the vicinity of the \ce{Fe^{+3}} ions~\cite{Battle1988, FAWCETT2000821, BattleJM9960601187}. \ce{SrMn_{1-x}Fe_xO_{3-\delta}} samples show AFM behavior for both low ($x\leq0.3$) and high ($x\geq 0.9$) Fe-doping, while a spin glass state caused by the interaction between \ce{Fe^{+3/+4}} and \ce{Mn^{+3/+4}} was observed for intermediate dopant concentrations~\cite{Battle1988, FAWCETT2000821}.

With the aim of understanding the mechanism underlying emerging polarization and the interaction between polar defect pairs, epitaxial strain and the electronic, structural and magnetic degrees of freedom, we used DFT+$U$ calculations to investigate the properties of Fe-doped oxygen deficient SMO thin films. Our results suggest that defect engineering, through controlling the concentration and distribution of polar defect pairs formed by anion vacancies coupled to substitutional cations, constitutes a parameter to design multiferroic materials. Defect couples can, indeed, promote polarity and ferroelectricity in non-polar perovskites due to an off-center displacement of the defect: the spatially separated substitutional \ce{Fe_{Mn}} (negatively charged) and \ce{V_O} (positively charged) offset the charge center from the geometry center of the lattice, resulting in an electric dipole moment along the direction from \ce{Fe_{Mn}} to \ce{V_O} already in unstrained and hence non-polar SMO. Other effects related to the appearance of reduced \ce{Mn^{+3}}, negatively charged with respect to the Mn lattice sites, should also be taken into account, since \ce{Mn^{+3}-V_O} pairs can result in additional dipoles that affect the overall ferroelectric response. Finally, the defect-pair dipole can couple with applied epitaxial strain favoring the transition to a polar phase, even for strains lower than the ones necessary to stabilize the polar structure in stoichiometric SMO.

\section{Methods\label{sec:compdetails}}

All DFT calculations were performed with the {\sc{Quantum ESPRESSO}} distribution~\cite{giannozzi2009quantum,Giannozzi2017}. PBEsol~\cite{perdew2008pbesol} was used as exchange-correlation functional together with ultrasoft pseudopotentials \cite{vanderbilt1990soft} with Sr($4s$, $4p$, $5s$), Mn($3p$, $4s$, $3d$), and O($2s$, $2p$) valence states \footnote{Ultrasoft pseudopotentials from the PSLibrary were taken from \url{www.materialscloud.org}: Sr.pbesol-spn-rrkjus\textunderscore psl.1.0.0.UPF, Mn.pbesol-spn-rrkjus\textunderscore psl.0.3.1.UPF, and O.pbesol-n-rrkjus\textunderscore psl.1.0.0.UPF}. A kinetic-energy cutoff of 70~Ry for wave functions and 840~Ry for spin-charge density and potentials were applied. A Gaussian smearing with a broadening parameter of 0.01~Ry was used in all cases.

SMO was simulated as a 40-atom $2\times2\times2$ supercell of the 5-atom primitive cubic cell. A shifted $6\times6\times6$ Monkhorst-Pack \cite{monkhorst1976special} \textbf{k}-point grid was used to sample the Brillouin zone. Both bulk and thin film geometries of the G-type AFM and ferromagnetic (FM) phases of SMO were considered. For stoichiometric bulk calculations, both lattice parameters and atomic positions were relaxed, while thin-film geometries with biaxial epitaxial strain in the \textit{ac}-plane imposed by a cubic substrate were computed following the procedure described in Ref.~\onlinecite{Rondinelli:2011jk}. Prior to defect creation, the atoms were displaced along the eigenvectors of the polar phonon modes computed for  unstrained stoichiometric SMO. Defects were then created by removing one oxygen atom (\ce{V_O}, concentration 4.2\%) and at the same time substituting one Mn with a Fe ion (\ce{Fe_{Mn}}, concentration 12.5\%). Different possible relative arrangements of the substitutional Fe with respect to the \ce{V_O} were taken into account (see Fig.~\ref{fig:SMO_structure_defect}). M{\"o}ssbauer experiments have shown that iron is present as \ce{Fe^{+3}} when associated with the doubly positively charged \ce{V_O} and the substitutional defect is thus negatively charged. The calculations were hence performed considering the positive charge state of this defect pair (\ce{Fe^'_{Mn}-V_O^{..}} in Kr{\"o}ger-Vink notation~\cite{KROGER1956307}, where the prime and dot symbols indicate, respectively, a charge of -1 and +1 relative to the respective lattice site). This was obtained by adjusting the number of electrons and by applying a background charge to ensure neutrality of the unit cell, as required by calculations under periodic-boundary conditions to avoid divergences in the electrostatic potential. For simplicity, we will refer to the defect pairs in this charge state simply as \ce{Fe_{Mn}-V_O}. Finally, for defective cells, atomic positions were optimized while keeping the lattice vectors fixed at optimized values of the non-defective system. In all calculations, atomic forces were converged to within $5\times10^{-2}$~eV/\AA, while energies were converged to within $1.4\times10^{-5}$~eV. Auxiliary calculations using $2\sqrt2\times2\sqrt2\times2$ , $3\times3\times3$, and $4\times4\times4$ supercells with 80, 135, and 320 atoms (and $3\times3\times4$, $3\times3\times 3$ \textbf{k}-meshes and $\Gamma$ point sampling of the Brillouin zone, respectively) were performed to investigate the interaction of two defect pairs and the influence of the defect concentration on the predicted polarization.

A Hubbard $U$ correction~\cite{anisimov1991band, anisimov1997first, Dudarev1998} was applied in all calculations. For stoichiometric bulk systems, where all Mn sites are crystallographically and chemically equivalent, we used global self-consistent $U$ parameters ($U_\mathrm{SC}$) computed for the G-AFM and FM phases of SMO in Ref.~\cite{Ricca2019} using density-functional perturbation theory (DFPT)~\cite{Timrov2018}, as implemented in hp.x of {\sc{Quantum ESPRESSO}}~\cite{giannozzi2009quantum,Giannozzi2017}. Self-consistent site-dependent $U$ parameters ($U_\mathrm{SC-SD}$) were instead computed for defective systems by perturbing the inequivalent sites resulting from defect formation (atoms were selected to be perturbed if their unperturbed atomic occupations differed by more than $10^{-3}$)~\cite{Ricca2019}. DFPT calculations were performed with a $\Gamma$-point sampling of the \textbf{q}-space~\cite{Timrov2018} in the 40-atom cell. A convergence threshold of 0.01 eV was applied for the self-consistence of $U$ values. In all cases, atomic orbitals were used to construct occupation matrices and projectors in the DFT+$U$ scheme. For simplicity, $U_\mathrm{SC-SD}$ values (see SI Sec. \ref{sec:Uscsd}) have only been computed for the unstrained stoichiometric and defective geometries, since even 4\% tensile strain changes $U_\mathrm{SC}$ by only 0.01 eV compared to zero strain~\cite{Ricca2019}.

The strain-dependent formation energy ($E_\textrm{f}$) of a \ce{Fe_{Mn}-V_O} defect-pair in the $q=+1$ charge state was calculated as described in Ref.~\onlinecite{freysoldt2014first}:
\begin{multline}
E_\textrm{f}(\epsilon, \mu_i, q) = E_\textrm{tot,def}(\epsilon, q) - E_\textrm{tot,stoic}(\epsilon) - \sum_i n_i \mu_i \\
+ q \, E_\textrm{Fermi}(\epsilon) + E_\textrm{corr}(\epsilon) \,.
\label{eq:formenerg}
\end{multline}
Here $E_\textrm{tot,def}$ and $E_\textrm{tot,stoic}$ are the DFT total energies of the defective system with a charge $q$ and of the stoichiometric cell, respectively. $\epsilon$ is the applied strain, $E_\textrm{Fermi}$ is the Fermi energy relative to the valence band maximum of the defect-free system, which can take values between zero and the band-gap of the material, while $n_i$ indicates the number of atoms of a certain specie $i$ that is added ($n_i > 0$) or removed ($n_i < 0$) from the supercell to form the defect, while $\mu_i$ is its chemical potential. Finally, $E_\textrm{corr}$ is a corrective term necessary to align the electrostatic potential of the defective cell with the one of the neutral stoichiometric system obtained by calculating the difference in electrostatic potential between the neutral defect-free cell and the charged defective one averaged in spheres around atomic sites located far from the defect~\cite{lany2008}. No further finite-size corrections were applied since the defect concentrations we simulate are realistic for this material. Different synthesis conditions can be accommodated by adjusting the set of chemical potentials $\mu_i = \mu_i^0 + \Delta \mu_i$ for each element by assuming equilibrium with a physical reservoir such as a gas or a bulk phase. We expressed $\mu_{\ce{Fe}}$ and $\mu_{\ce{Mn}}$ as a function of $\mu_{\ce{O}}$. For this latter, we used \ce{O2} as a reference, $\mu_{\ce{O}} = \frac{1}{2}E(\ce{O2})+ \Delta\mu_{\ce{O}}$, while bounds on $\Delta\mu_{\ce{O}}$ were derived imposing the stability of SMO ($\Delta\mu_{\ce{Sr}}+ \Delta\mu_{\ce{Mn}} + 3\Delta\mu_{\ce{O}} \le \Delta H_f(\ce{SMO})$) against decomposition into elemental Sr/Mn species ($\Delta\mu_{\ce{Sr/Mn}}\le 0$) and against the formation of competing phases like \ce{SrO} ($\Delta\mu_{\ce{Sr}} + \Delta\mu_{\ce{O}} \le \Delta H_f(\ce{SrO})$), and \ce{MnO} ($\Delta\mu_{\ce{Mn}} + \Delta\mu_{\ce{O}} \le \Delta H_f(\ce{MnO})$). For the Fe impurity, stability against solubility-limiting phases, such as \ce{FeO}, were instead considered to relate $\mu_{\ce{Fe}}$ to $\mu_{\ce{O}}$~\cite{vanderwalle2004sol}. The computed heat of formation ($\Delta H_f$) of the transition-metal oxides were corrected according to Ref.~\onlinecite{ceder20011} to account for the mixing of DFT and DFT+$U$ total energies in the derivation of the formation enthalpy. We will show results in the oxygen-poor limit with $\Delta\mu_{\ce{Mn}}=$ -1.77~eV and $\Delta\mu_{\ce{Fe}}=$ -1.43~eV and for a Fermi energy equal to the band gap of unstrained SMO.

The polarization $\vec{P}$ was computed using a point-charge model:
\begin{equation}
	\vec{P}=\sum_i \vec{r}_i q_i\,,
	\label{eq:polarization1}
\end{equation}
where $\vec{r}_i$ is the position of atom $i$ and $q_i$ is its nominal charge: +2 for Sr, -2 for O, and +4 or +3 for stoichiometric or reduced Mn and Fe sites. The charge applied on each Mn and Fe ion was defined on the base of its oxidation state computed through the method introduced by Sit \textit{et al.}~\cite{Sit2011}. The polarization, being a multivalued quantity, has been corrected by an integer number of polarization quanta $\vec{Q}$, computed as:
\begin{equation}
	\vec{Q}=\frac{e}{V} \begin{bmatrix} a\\b\\c \end{bmatrix}\,,
	\label{eq:polarization2}
\end{equation}
with $a$, $b$, and $c$ being the lattice parameters, $V$ the volume of the unit cell, and $e$ the elementary charge. Results obtained with this method include polarization contributions of the lattice and the defect dipole, but neglect electronic redistribution effects compared to other approaches such as the Berry Phase formalism~\cite{berryphase1,berryphase2}. However, the metallic nature of some defective SMO cells did not allow the application of the Berry phase approach.

\section{Results and Discussion\label{sec:results}}

%
\begin{figure}
	\centering
	\includegraphics[width=\columnwidth]{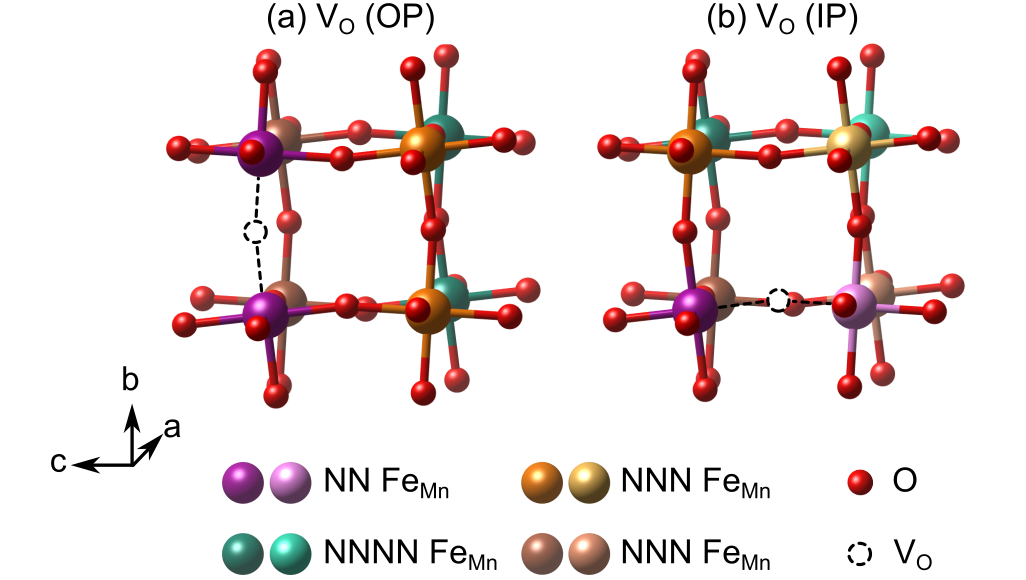}
	\caption{Schematic representation of the possible relative arrangements of \ce{Fe_{Mn}-V_O} defect pairs in the 40-atom SMO cell in the case of a (a) out-of-plane (OP) or (b) in-plane (IP) oxygen vacancy. Purple, orange and brown, and green refer to configurations in which \ce{Fe_{Mn}} is in next-neighbors (NN), next-nearest neighbors (NNN), and next next-nearest neighbors position (NNNN) to the \ce{V_O}. As opposed to (a), the two Mn sites at the same distance from the \ce{V_O} in (b) are not symmetry equivalent as indicated by the slightly different shade for each color.}
	\label{fig:SMO_structure_defect}
\end{figure}
\ce{SrMn_{1-x}Fe_xO_{3-\delta}} is modeled using all possible symmetry inequivalent \ce{Fe_{Mn}-V_O} configurations within a $2\times2\times2$ SMO supercell and (see Fig.~\ref{fig:SMO_structure_defect}). In particular, there are two symmetry-distinct oxygen atoms in this structure: an out-of-plane \ce{V_O} with the broken \ce{Mn-O-Mn} bond perpendicular to the biaxial strain ($ac$) plane (OP, see Fig.~\ref{fig:SMO_structure_defect}a) and an in-plane O position with the broken \ce{Mn-O-Mn} bond in the $ac$ plane (IP, see Fig.~\ref{fig:SMO_structure_defect}b). For \ce{V_O^{OP}}, the Mn sites lying at nearest neighbors positions to the defect (NN, in violet in Fig.~\ref{fig:SMO_structure_defect}a) or far away from it (NNNN, in green in Fig.~\ref{fig:SMO_structure_defect}a) are equivalent by symmetry, while two different groups of Mn ions can be distinguished for substitution sites in next-nearest neighbor positions to the vacancy (NNN, in orange and brown in Fig.~\ref{fig:SMO_structure_defect}a) for a total of 4 symmetry-inequivalent \ce{Fe_{Mn}-V_O^{OP}} defect pairs. Instead, for a \ce{V_O^{IP}}, the two Mn atoms lying at NN, NNN, or NNNN positions correspond each to two symmetry-distinct substitution sites for a total of 8 \ce{Fe_{Mn}-V_O^{IP}} symmetry-inequivalent configurations (see Fig.~\ref{fig:SMO_structure_defect}b).

In agreement with M\"ossbauer studies~\cite{Battle1988}, our DFT+$U_\mathrm{SC-SD}$ calculations for the insulating AFM phase show a partial reduction of Mn adjacent to the oxygen vacancy. This implies that for a \ce{Fe_{Mn}-V_O} defect pair two dipoles exist in the structure, one pointing from the \ce{Fe^'_{Mn}} to the \ce{V_O} and one pointing from the \ce{Mn^'_{Mn}} to the \ce{V_O^{..}} as discussed in more detail in SI Sec. \ref{sec:Uscsd}. This will have implications on the magnetism and polarization as discussed in Secs.~\ref{sec:magneticorder} and \ref{sec:unstrained_pol}. In the FM phase a partial reduction of one or two Mn sites is observed, which we associated with its metallic nature.

\subsection{Relative Stability and Formation Energy\label{sec:stability}}

%
\begin{figure}
	\centering
	\includegraphics[width=\columnwidth]{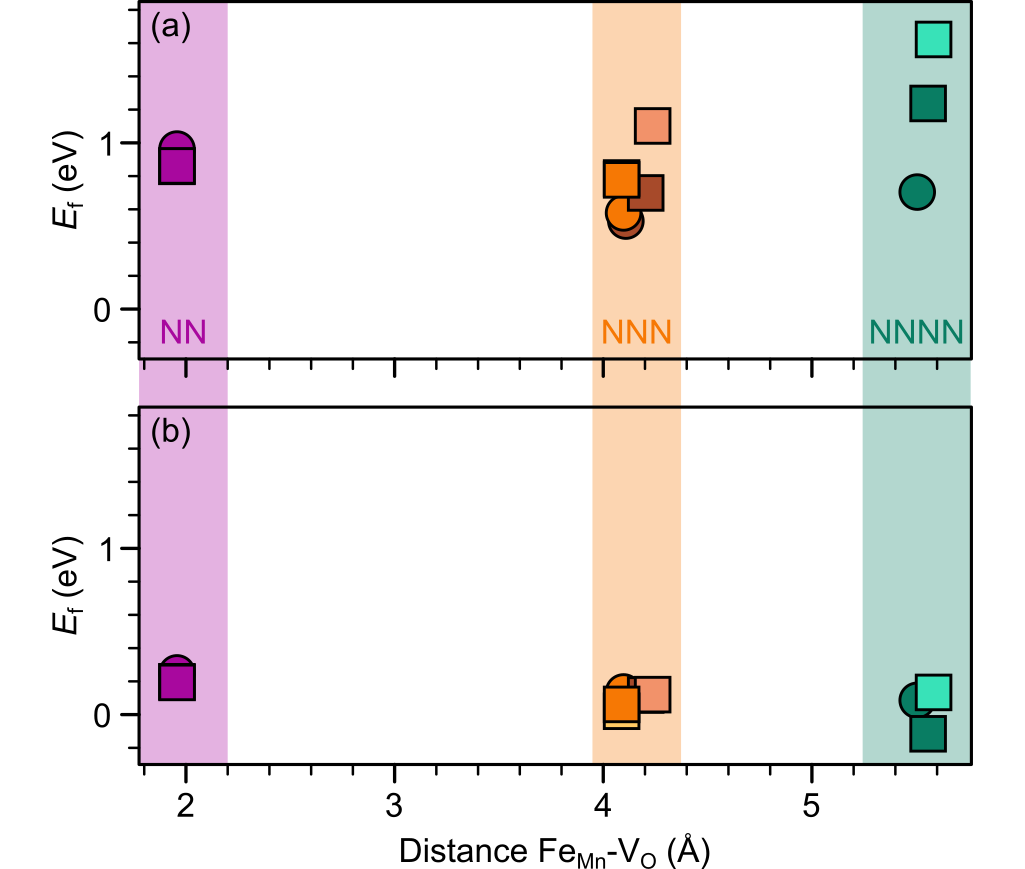}
	\caption{Formation energy ($E_\textrm{f}$) computed for \ce{Fe_{Mn}-V_O} configurations in unstrained (a) AFM and (b) FM SMO as a function of the \ce{Fe_{Mn}-V_O} distance. Circle and square symbols refer to configurations with \ce{V_O^{OP}} and \ce{V_O^{IP}}, respectively. See color code in Fig.~\ref{fig:SMO_structure_defect}.}
	\label{fig:ef_nostrain}
\end{figure}
We begin by investigating the relative stability of the \ce{Fe_{Mn}-V_O} configurations in bulk AFM and FM SMO. We note that we are mainly interested in relative formation-energy differences for the different configurations and in strain-induced changes, rather than absolute defect pair formation energies, which have been derived in O-poor conditions and thus correspond to a lower limit for $E_\textrm{f}$. In the AFM phase, the most stable configurations are the NNN \ce{Fe_{Mn}-V_O^{OP}}, as can be seen from Fig.~\ref{fig:ef_nostrain}a, where we report the formation energy ($E_\textrm{f}$) computed for the different \ce{Fe_{Mn}-V_O} arrangements as a function of the distance between the \ce{Fe_{Mn}} and the \ce{V_O}. NN \ce{Fe_{Mn}-V_O}, the NNNN \ce{Fe_{Mn}-V_O^{OP}}, and the majority of the NNN \ce{Fe_{Mn}-V_O^{IP}} have formation energies higher by about 0.1-0.4~eV, while NNNN \ce{Fe_{Mn}-V_O^{IP}} configurations are about 0.7-1.0~eV less stable. These results indicate that the \ce{V_O} are preferentially in the neighborhood of the substitutional iron, as suggested by spectroscopic results~\cite{BattleJM9960601187}, even though not necessarily in its first coordination shell. More importantly, our data suggest that some disorder is expected, as also indicated by experiments~\cite{FAWCETT2000821}. This is particularly interesting because different configurations correspond to different orientations of the electric dipole associate to the \ce{Fe_{Mn}-V_O} pair, which - as we will discuss in Sec.~\ref{sec:unstrained_pol} - is responsible for the polarization in unstrained \ce{SrMn_{1-x}Fe_xO_{3-\delta}}. Hence, the presence of different energetically similar configurations could potentially lead to a switchable polarization and defect-induced ferroelectricity if the defect dipoles couple in a parallel fashion, which we will explore in Sec.~\ref{sec:concentration}.

A different behavior is observed, instead, in the FM phase (see Fig.~\ref{fig:ef_nostrain}b), where not only are \ce{Fe_{Mn}-V_O^{IP}} found to be generally more stable regardless of the distance between the substitutional iron and the vacancy, but where the most stable configuration is a NNNN \ce{Fe_{Mn}-V_O^{IP}} defect pair, likely because these configurations lead to \ce{Mn^{3+}/Mn^{4+}} interactions that stabilize the FM phase, as we will discuss in more detail in Sec.~\ref{sec:magneticorder}. Furthermore, the average difference in $E_\textrm{f}$ between the defect configurations in FM SMO is only 0.2~eV and $E_\textrm{f}$ in the FM phase are on average 0.8~eV lower than in the AFM phase. This can be explained by the smaller energetic cost to accommodate the two excess electrons associated with the \ce{V_O} on delocalized Mn/Fe states in the metallic FM phase.

\begin{figure}
	\centering
	\includegraphics[width=\columnwidth]{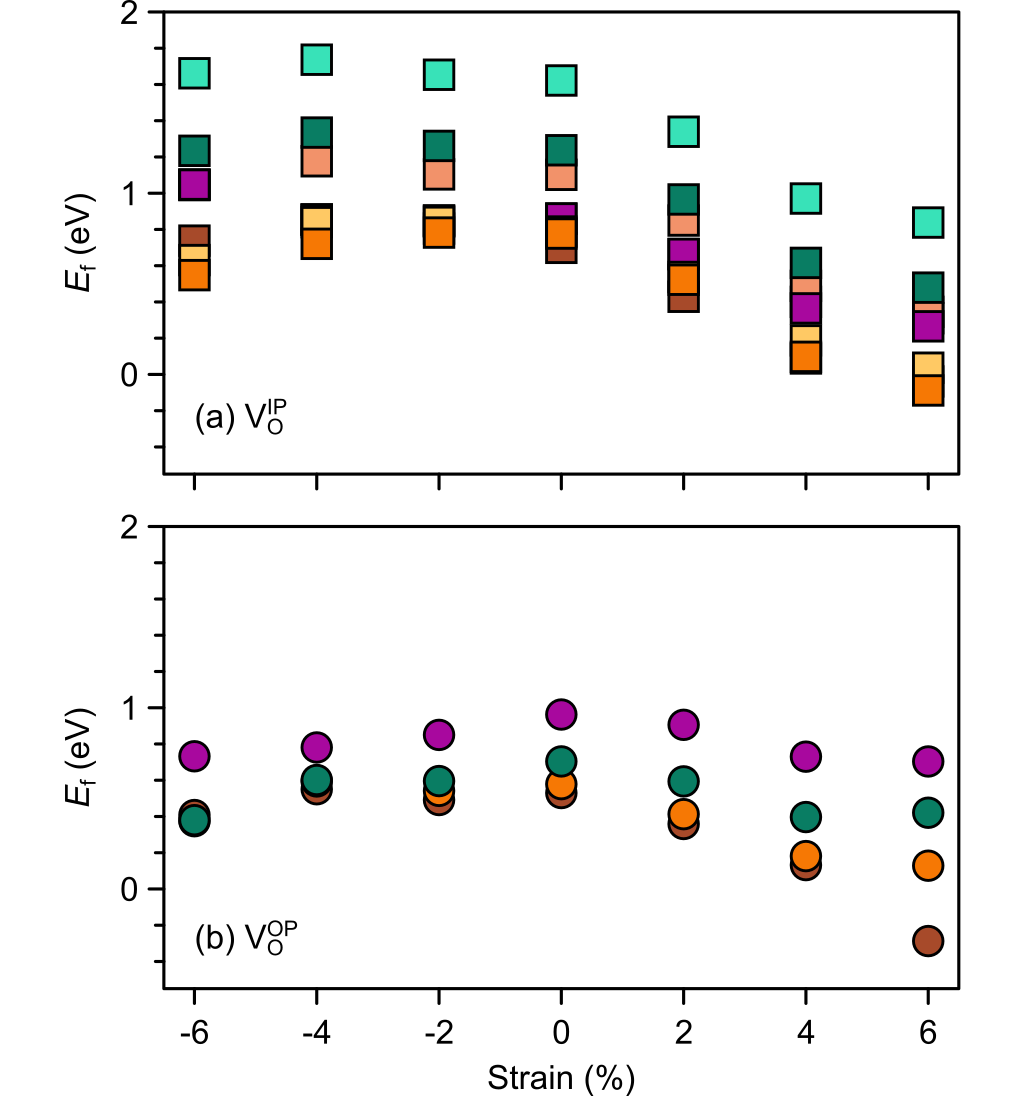}
	\caption{Strain dependent formation energy ($E_\textrm{f}$) of (a) \ce{Fe_{Mn}-V_O^{IP}} and (b) \ce{Fe_{Mn}-V_O^{OP}} defect pairs in AFM SMO. See color code in Fig.~\ref{fig:SMO_structure_defect}.}
	\label{fig:ef_strain}
\end{figure}
In SMO thin films, the defect formation energy and consequently the defect concentration depend on volume changes induced by biaxial strain~\cite{aschauer2013strain, aschauer2016interplay, marthinsen2016coupling}. Biaxial strain also breaks the symmetry~\cite{Rondinelli:2011jk} and could thus allow strain-controlled ordering of defects on inequivalent sites~ \cite{marthinsen2016coupling, aschauer2013strain, aschauer2016interplay}. Hence, we now consider the interplay between \ce{Fe_{Mn}-V_O} defects, strain, and magnetism in SMO. Fig.~\ref{fig:ef_strain} shows the changes in $E_\textrm{f}$ for defect pairs as a function of the applied strain. In the AFM phase, tensile strain results in a reduction of the formation energy of \ce{Fe_{Mn}-V_O}, consistent with the chemical expansion~\cite{adler2004} due to reduced transition metal sites. Defect pairs with \ce{V_O^{IP}} are found to be more sensitive to tensile strain, 4\% strain resulting in a reduction of $E_\textrm{f}$ by about 0.4-0.7~eV for \ce{Fe_{Mn}-V_O^{IP}} compared to only 0.2-0.4~eV for \ce{Fe_{Mn}-V_O^{OP}}. Consequently, the energy difference between the NNN \ce{Fe_{Mn}-V_O^{OP}} and \ce{Fe_{Mn}-V_O^{IP}} configurations is reduced, favoring disorder. Under compressive strain, instead, the formation energy of \ce{Fe_{Mn}-V_O} defect pairs remains fairly constant (average changes of about 0.05~eV) as already observed for \ce{V_O} in SMO because of crystal field effects~\cite{aschauer2013strain} and \ce{Fe_{Mn}-V_O^{OP}} are slightly stabilized with respect to \ce{Fe_{Mn}-V_O^{IP}}.

Unsurprisingly, as shown in SI Fig. \ref{fig:ef_strainFM}, the FM phase exhibits a reduced sensitivity of the formation energy to strain, which is rationalized by its metallicity. We also note that for the FM phase, under compressive strain, the formation energy of \ce{Fe_{Mn}-V_O^{IP}} defects increases as expected from volume arguments, which can be explained in terms of a reduced sensitivity of the metallic FM phase to crystal field effects, allowing volume effects to dominate~\cite{aschauer2013strain}.

\subsection{Magnetic Order\label{sec:magneticorder}}

%
\begin{figure}
	\centering
	\includegraphics[width=\columnwidth]{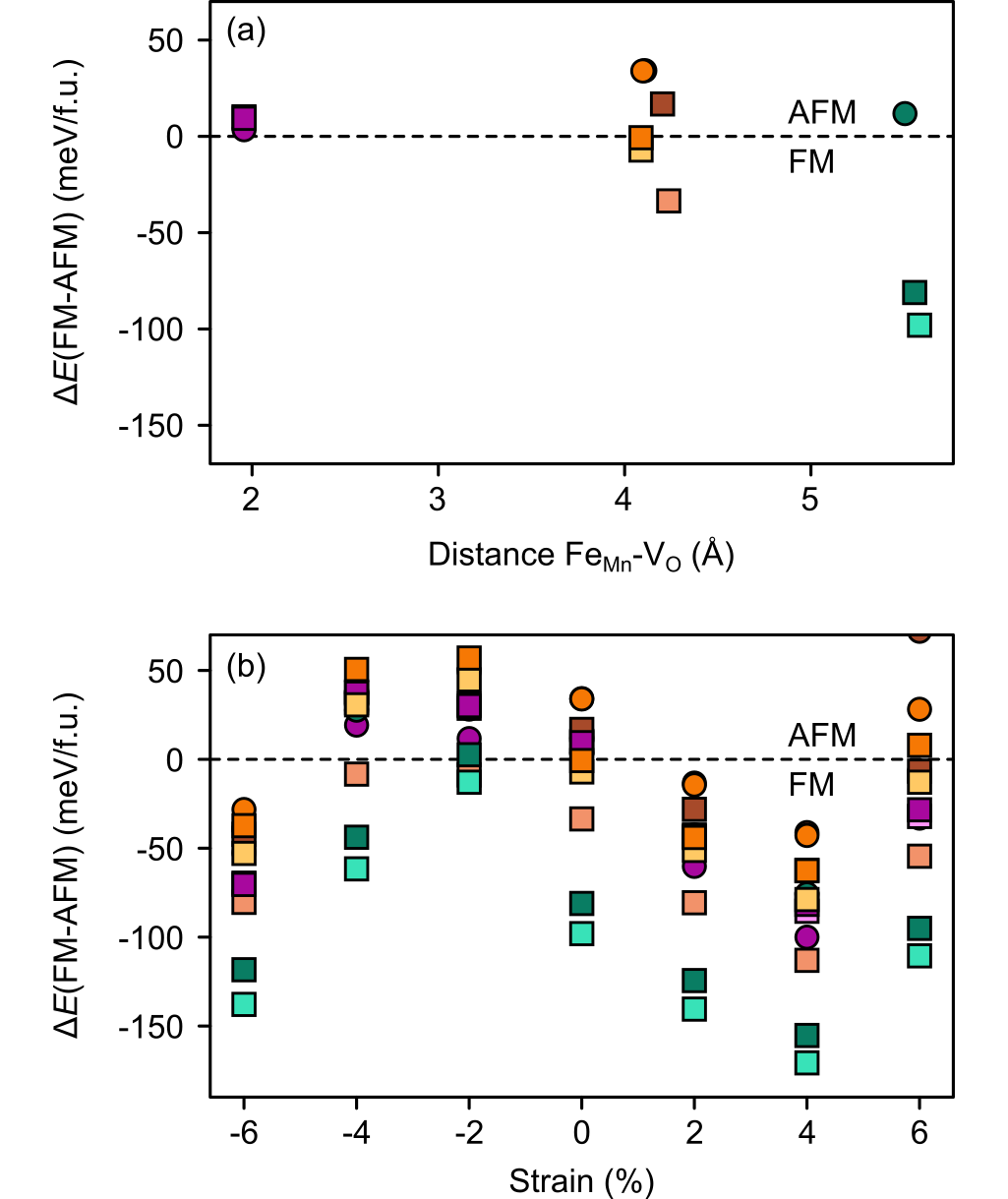}
	\caption{Total energy differences ($\Delta E(\textrm{FM-AFM})$) per formula unit between the defective cells with FM an AFM order. AFM is more stable for positive and FM for negative differences. (a) $\Delta E(\textrm{FM-AFM})$ reported with respect to the the \ce{Fe_{Mn}-V_O} distance in each defective configuration in the unstrained SMO structure. (b) Changes in $\Delta E(\textrm{FM-AFM})$ as a function of strain for all the considered configurations. Circle and square symbols refer to data obtained for \ce{V_O^{OP}} and \ce{V_O^{IP}}, respectively. See color code in Fig.~\ref{fig:SMO_structure_defect}.}
	\label{fig:magnetic_order}
\end{figure}
Bulk stoichiometric SMO has a G-type AFM ground state. DFT+$U$ calculations have shown a 4.2\% oxygen vacancy concentration to induce a magnetic phase transition from AFM to FM~\cite{marthinsen2016coupling, Ricca2019}, which is explained by the vacancy-induced \ce{Mn^{4+}-Mn^{3+}} double exchange coupling. The properties of Fe-doped oxygen-deficient SMO are more complex due to the presence of Fe transition-metal atoms. Indeed, for the \ce{Mn^{4+}}, \ce{Mn^{3+}}, and \ce{Fe^{3+}} ions present in the simulated cells, the interactions between neighboring \ce{Mn^{4+}-Mn^{4+}} and \ce{Mn^{3+}-Mn^{3+}} are AFM, between \ce{Mn^{4+}-Mn^{3+}} are FM, while those between \ce{Mn^{4+}-Fe^{3+}} are AFM through the $\pi$ orbitals and FM through the $\sigma$ orbitals~\cite{FAWCETT2000821}. Fig.~\ref{fig:magnetic_order}a shows the total-energy difference between the unstrained AFM and FM phases with different \ce{Fe_{Mn}-V_O} defect pairs as a function of the distance between the two defects. The most stable configurations (see Fig.~\ref{fig:ef_nostrain}) favour the AFM order. Indeed, all the configurations with a \ce{V_O^{OP}}, and the NN and the majority of the NNN \ce{Fe_{Mn}-V_O^{IP}} configurations prefer this magnetic phase. Only the NNNN \ce{Fe_{Mn}-V_O^{IP}} defects strongly favor the FM phase since the \ce{Mn^{4+}-Mn^{3+}} interactions are promoted due to the larger distance between the \ce{Fe^{3+}} ion and the reduced \ce{Mn^{3+}} site. This result is in line with the experimental data reporting \ce{SrMn_{1-x}Fe_xO_{3-\delta}} with Fe concentration close to the one of our study ($x=0.125$) to show AFM behavior. Interestingly, the DFT+$U_\textrm{SC-SD}$ approach including local chemical changes on the transition-metal atoms around the defect is fundamental to predict defect-induced magnetic properties, since DFT+$U_\textrm{SC}$ with global $U_\textrm{SC}$ of stoichiometric SMO predicts a preferential FM order for all configurations (see SI Fig.~\ref{fig:magnetic_order_USC}). This result can be explained by increasing $U$ favoring the FM order, which conversely implies that the decreased $U_\textrm{SC-SD}$ values on the reduced \ce{Mn^{3+}} will locally destabilize the FM order~\cite{Ricca2019}.

Biaxial strain beyond a critical value of 2\% is known to induce a AFM to FM transition in stoichiometric SMO~\cite{marthinsen2016coupling, Ricca2019}. We now consider the interplay between this magnetic phase transition and the \ce{Fe_{Mn}-V_O} defect pair (see Fig.~\ref{fig:magnetic_order}b). Unsurprisingly, tensile strain stabilizes the FM phase even in presence of the defect pair, affecting all configurations in almost the same way and in an approximately linear fashion up to 4\% strain. Larger tensile strain result, instead, in a stabilization of the AFM phase for the most stable configurations and in a reduction of the FM stabilization for the others. This observation can be explained considering the stronger sensitivity of the AFM phase to volume changes, which results, as previously discussed, in a larger reduction of the formation energy and consequently in a stabilization of the AFM phase for large tensile strains. Compressive strain up to -2\% favors the AFM order: at -2\% the majority of the configurations show an AFM ground state and for the remaining cases the two magnetic orders are very close in energy. The preference for the AFM, instead, decreases for larger compressive strain due to the increased stability of the FM phase in compressively strained stoichiometric SMO films~\cite{Ricca2019}.

\subsection{Polarization in unstrained \ce{SrMn_{1-x}Fe_xO_{3-\delta}}}\label{sec:unstrained_pol}

We now consider the polarization in Fe-doped oxygen deficient SMO by starting from the unstrained geometries. Polarization can arise due to symmetry reduction by the defect pair but also due to the formation of a defect dipole ($\vec{D}$): the spatially separated substitutional \ce{Fe^'_{Mn}} (negatively charged) and \ce{V_O^{..}} (positively charged) result in the charge center being offset from the geometry center of the lattice and hence an electric dipole moment along the direction from \ce{Fe_{Mn}} to \ce{V_O}~\cite{Wang2017nano}. As detailed in SI Sec.~\ref{sec:Uscsd}, the situation is further complicated in the AFM phase by the reduction of one Mn ion (\ce{Mn^'_{Mn}}) adjacent to \ce{V_O}, which induces an additional defect dipole ($\vec{D'}$) from the negatively charged reduced \ce{Mn^'_{Mn}} to the positively charged \ce{V_O^{..}}. For simplicity, we will concentrate the following discussion mainly on results obtained for the AFM order. Indeed, the metallic nature of the FM phase and the consequent partial reduction of more than one Mn (cf. SI Sec.~\ref{sec:Uscsd}) result in a more complex behavior, which may not be properly described within the simple approach we use to estimate the polarization based on nominal charges (see SI Sec.~\ref{sec:SI_pol}). Results for the FM phase are reported in SI Sec.~\ref{sec:polFM}

\begin{figure}
	\centering
	\includegraphics[width=\columnwidth]{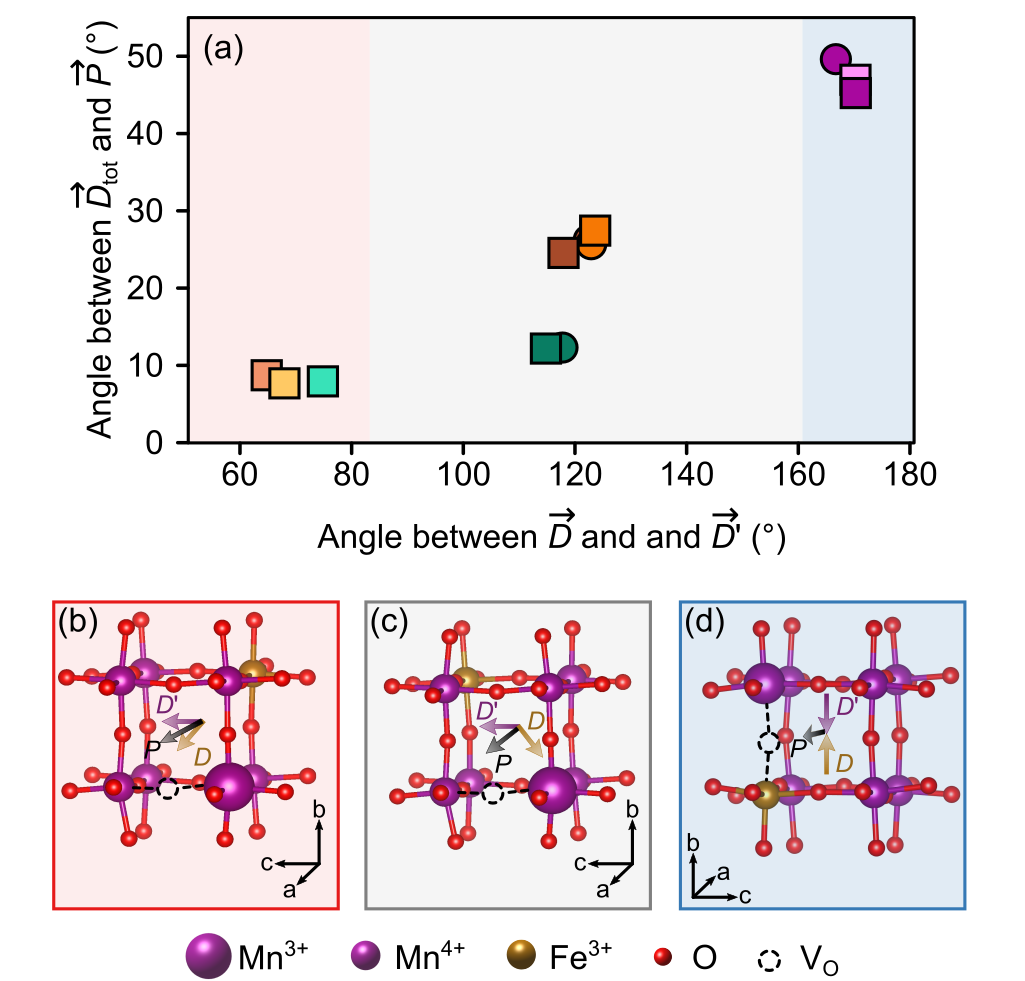}
	\caption{(a) Angle between the total dipole $\vec{D}_\textrm{tot}$ and the computed polarization $\vec{P}$ with respect to the angle between the \ce{Fe^'_{Mn}-V_O^{..}} defect dipole $\vec{D}$ and the \ce{Mn^'_{Mn}-V_O^{..}} dipole $\vec{D'}$ for the different defect-pair configurations. Circle and square symbols refer to data obtained for \ce{V_O^{OP}} and \ce{V_O^{IP}}, respectively. See color code in Fig.~\ref{fig:SMO_structure_defect}. Schematic representation of the $\vec{D}$, $\vec{D'}$, and $\vec{P}$ vectors for the cases in which the angle between $\vec{D}$ and $\vec{D'}$ is about (b)~60$^{\circ}$, (c)~120$^{\circ}$ or (d)~180$^{\circ}$.}
	\label{fig:Angle_Dip_Pol}
\end{figure}
\begin{figure}
	\centering
	\includegraphics[width=\columnwidth]{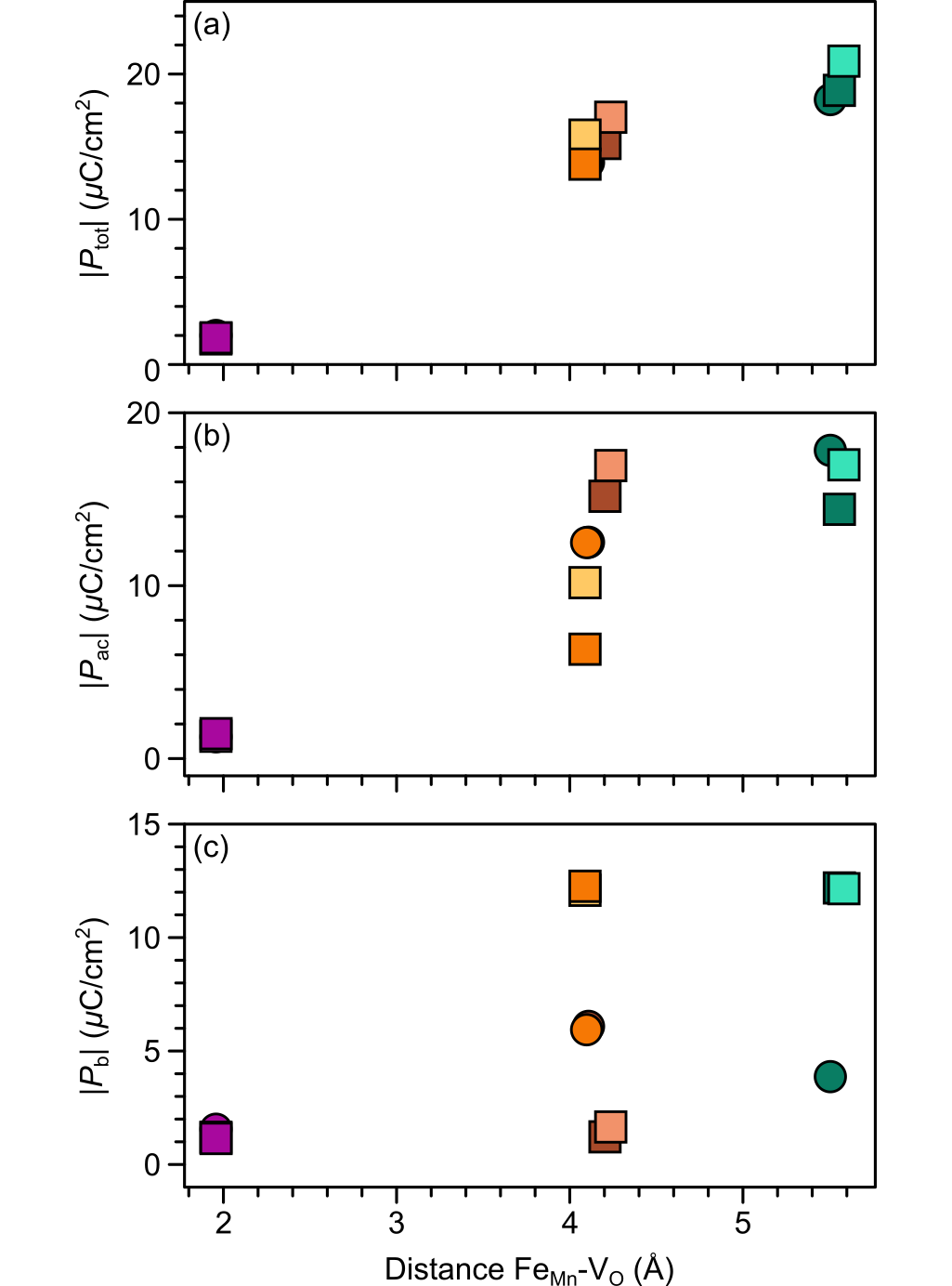}
	\caption{Magnitude of (a) the total polarization vector, as well as (b) its in-plane ($|P_{ac}|$), and (c) out-of-plane ($|P_{b}|$) components for the different defect configurations in the AFM phase. Circle and square symbols refer to data obtained for \ce{V_O^{OP}} and \ce{V_O^{IP}}, respectively. See color code in Fig.~\ref{fig:SMO_structure_defect}.}
	\label{fig:Pol_AFM_unstrained}
\end{figure}
The computed polarization $\vec{P}$ is roughly aligned with the vector sum $\vec{D}+\vec{D'}= \vec{D}_\textrm{tot}$, forming with $\vec{D}_\textrm{tot}$ an angle ranging from about 6$^{\circ}$ to 50$^{\circ}$ for the different configurations as shown in Fig.~\ref{fig:Angle_Dip_Pol}a. The alignment between $\vec{D}_\textrm{tot}$ and $\vec{P}$ depends on the relative geometric arrangement of \ce{Fe^'_{Mn}}, \ce{Mn^'_{Mn}}, and \ce{V_O}. In particular, due to geometric constraints, the angle between $\vec{D}$ and $\vec{D'}$ can either be around 60$^{\circ}$ (see Fig.~\ref{fig:Angle_Dip_Pol}b), 120$^{\circ}$ (see Fig.~\ref{fig:Angle_Dip_Pol}c), or 180$^{\circ}$ (see Fig.~\ref{fig:Angle_Dip_Pol}d). The smaller this angle, the stronger the combination of the two dipoles and consequently the larger the alignment between $\vec{P}$ and $\vec{D}_\textrm{tot}$ (see Fig.~\ref{fig:Angle_Dip_Pol}a and also SI Sec.~\ref{sec:SI_pol}). For example, for NN \ce{Fe_{Mn}-V_O} configurations, the \ce{Fe^'_{Mn}} and \ce{Mn^'_{Mn}} ions are located at the two sites adjacent to \ce{V_O}, resulting in antiparallel $\vec{D}$ and $\vec{D'}$ dipoles (see Fig.~\ref{fig:Angle_Dip_Pol}d). As a result of this peculiar arrangement of the defect dipoles and of the smaller distance between the substitutional iron and the vacancy, NN \ce{Fe_{Mn}-V_O} configurations are associated with the smallest total polarization ($P_\textrm{tot}$) of about 2~$\mu\textrm{C/cm}^2$ (see Fig.~\ref{fig:Pol_AFM_unstrained} a). Unsurprisingly, $P_\textrm{tot}$ slightly increases with increasing \ce{Fe_{Mn}-V_O} distance up to 15-20~$\mu\textrm{C/cm}^2$ (see Fig.~\ref{fig:Pol_AFM_unstrained}a). These polarizations are of similar magnitude than the ones in conventional ferroelectrics such as \ce{BaTiO3}~\cite{Ederer2005Effect}. The in-plane component of the polarization ($P_{ac}$, see Fig.~\ref{fig:Pol_AFM_unstrained}b) behaves similarly and is generally larger than the out-of-plane component $P_b$ (see Fig.~\ref{fig:Pol_AFM_unstrained}c). $P_b$ reflects, instead, the relative arrangement of the \ce{Fe_{Mn}} and \ce{V_O} defects along the $b$-axis: for NNN \ce{Fe_{Mn}-V_O^{IP}} configurations, the defects can belong to the same atomic layer (see configurations in brown in Fig.~\ref{fig:SMO_structure_defect}b) showing $P_b$ values lower than 1~$\mu\textrm{C/cm}^2$ or they can belong to atomic layers separated by about 3.8~\AA\ along $b$ (see configurations in orange and green in Fig.~\ref{fig:SMO_structure_defect}b) with $P_b$ values of about 12~$\mu\textrm{C/cm}^2$. NNN and NNNN \ce{Fe_{Mn}-V_O^{OP}} have instead intermediate $P_b$ values of about 8-10~$\mu\textrm{C/cm}^2$ since \ce{Fe_{Mn}} and \ce{V_O} defects always belong to atomic planes at a distance of about 1.9~\AA.

These results suggest that the mechanism underlying the observed polarization in unstrained SMO is the off-centering of the charge due to the separation of the \ce{Fe^'_{Mn}/Mn^'_{Mn}} and the \ce{V_O^{..}} defects.

\subsection{Polarization and defect concentration}\label{sec:concentration}

%
\begin{figure}
 \centering
 \includegraphics[width=\columnwidth]{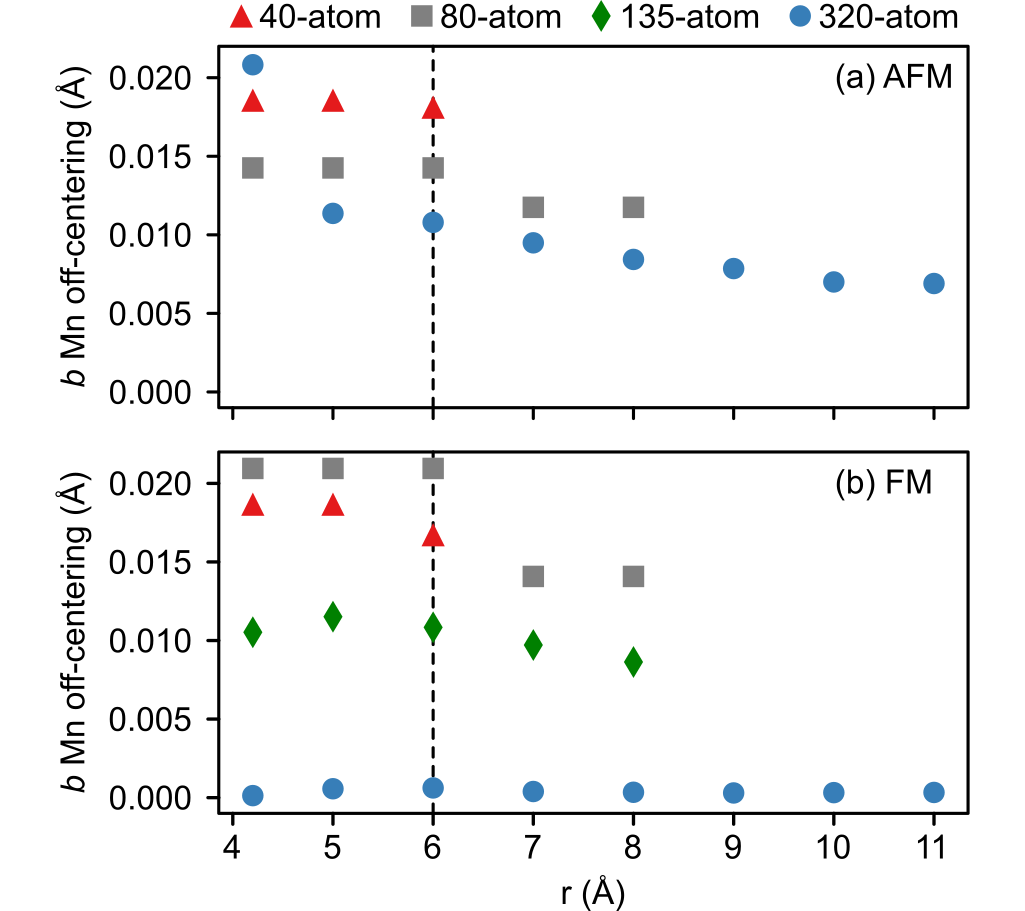}
 \caption{Total off-centering along the $b$-axis for Mn atoms lying within a sphere of radius $r$ and centered on the center of mass of the defect pair. Results for different cell sizes in both unstrained (a) AFM and (b) FM SMO.}
\label{fig:Mnoffs_cell_size}
\end{figure}
In order to further investigate the mechanism underlying the polarization induced by the defect pair, we now examine how the defect concentration impacts the polarization of Fe-doped oxygen-deficient SMO. SI Fig.~\ref{fig:Pol_cell_size} shows that the polarization decreases with increasing cell size (\textit{i.e.} with decreasing the defect concentration). This suggests the polarization to originate from a local change in symmetry around the defect pair due to the defect dipoles. Indeed, \ce{Fe_{Mn}-V_O} defects induce small displacements of the atoms in the vicinity of the defect pair from their high-symmetry positions. For example, Fig.~\ref{fig:Mnoffs_cell_size} shows the Mn displacements along the $b$-axis (with the largest polarization component) for the defect configuration considered in SI Fig.~\ref{fig:Pol_cell_size} and computed in supercells of different size. These off-centerings have been computed excluding the two Mn adjacent to the oxygen vacancy to avoid artifacts due to the relaxation of these undercoordinated sites. In general, the larger the polarization in SI Fig.~\ref{fig:Pol_cell_size}, the larger the Mn off-centerings. More importantly, the Mn displacements are generally larger and constant for sites lying within 6~\AA\ from the defect pair and decrease afterwards, pointing to a spatially limited effect of the defect pair.

\begin{figure}
 \centering
 \includegraphics[width=\columnwidth]{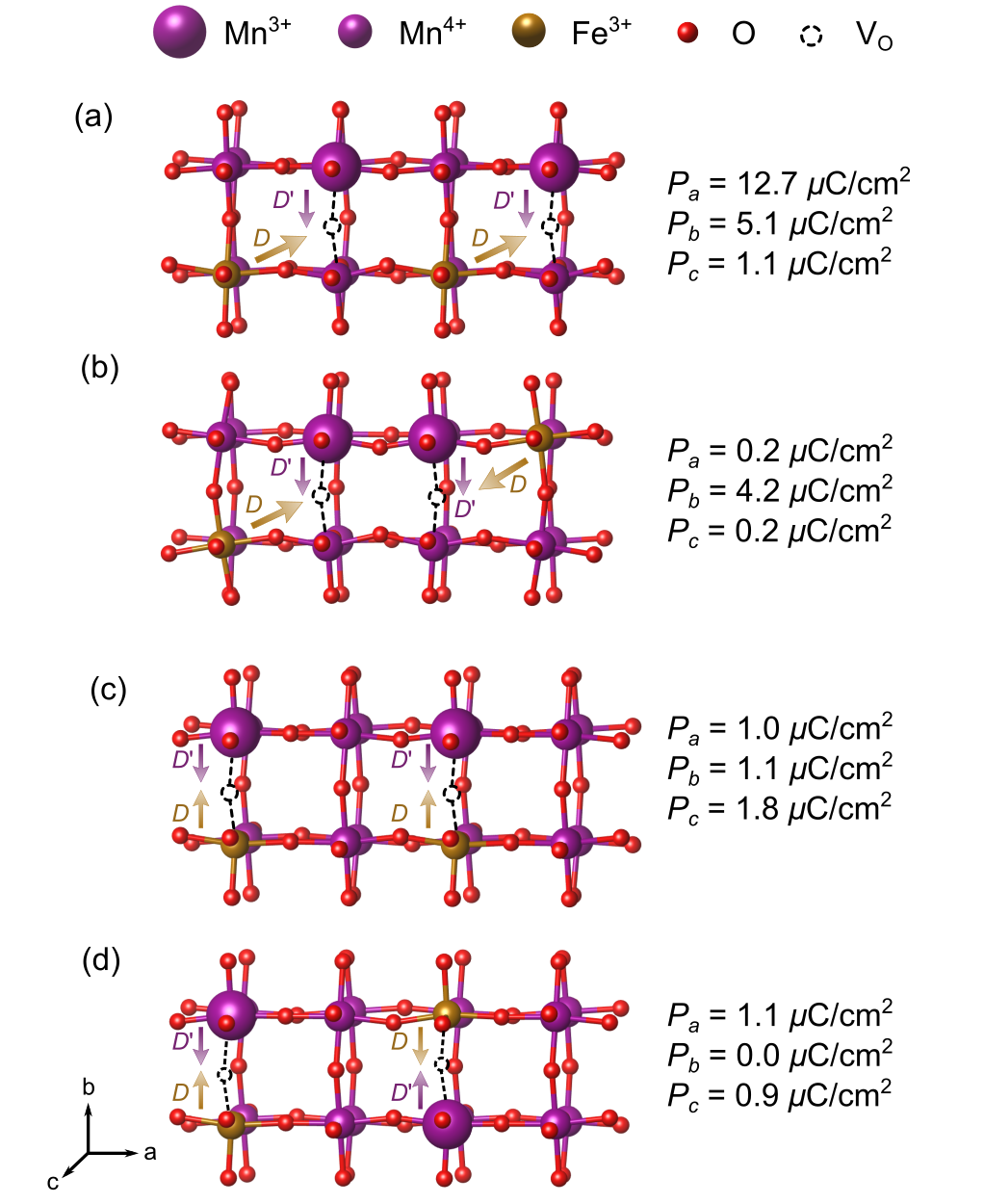}
 \caption{$4\times2\times2$ supercells with two aligned (a, c) and anti-aligned (b, d) \ce{Fe_{Mn}-V_O} defect pairs. (a, b) for NNN and (c, d) for NN \ce{Fe_{Mn}-V_O^{OP}}. Gold and purple arrows indicate the direction of the \ce{Fe^'_{Mn}-V_O^{..}} and \ce{Mn^'_{Mn}-V_O^{..}} defect dipoles, respectively.}
\label{fig:double_defect}
\end{figure}

High defect concentrations may thus promote macroscopic polarization but the possibility of different orientations of neighboring defect dipoles should be taken into account. For this reason, we performed additional calculations in a $4\times2\times2$ supercell containing two \ce{Fe_{Mn}-V_O} defect pairs and investigated the cases in which the \ce{Fe_{Mn}-V_O} defect dipoles lie parallel or anti-parallel to one another. Due to the importance of elastic effects for defects in close proximity~\cite{chandrasekaran2013, varvenne2013}, both atomic positions and lattice parameters were allowed to relax in these calculations.

Fig.~\ref{fig:double_defect} schematically illustrates the structures and orientation of the defect dipoles for the most stable NNN \ce{Fe_{Mn}-V_O^{OP}} configuration in which the defect dipoles lie mainly in the $ac$-plane (Figs.~\ref{fig:double_defect}a and b) or along the $b$-axis (Figs.~\ref{fig:double_defect}c and d), for the NN \ce{Fe_{Mn}-V_O^{OP}} configuration. In order to explain the obtained polarization in presence of the two parallel/anti-parallel dipoles, the orientation of both the \ce{Fe^'_{Mn}-V_O^{..}} and \ce{Mn^'_{Mn}-V_O^{..}} defect dipoles has to be considered. As discussed in Sec.~\ref{sec:unstrained_pol}, for configurations like NN \ce{Fe_{Mn}-V_O^{OP}}, in which the substitutional Fe is adjacent to \ce{V_O}, the resulting polarization lies almost along the direction given by the combination of the \ce{Fe^'_{Mn}-V_O^{..}} and \ce{Mn^'_{Mn}-V_O^{..}} and has nearly equal components along the $a$, $b$, and $c$-directions (see Fig. \ref{fig:double_defect}c). Introducing two anti-parallel NN \ce{Fe_{Mn}-V_O^{OP}} results in two opposite \ce{Fe^'_{Mn}-V_O^{..}} and \ce{Mn^'_{Mn}-V_O^{..}} dipoles along the $b$-axis, and consequently in quenching the polarization component along this axis (see Fig.~\ref{fig:double_defect}d). Instead, for the NNN \ce{Fe_{Mn}-V_O^{OP}} configuration, the polarization is larger along $a$ due to the longer \ce{Fe^'_{Mn}-V_O^{..}} dipole with a smaller component along $b$ also due to the \ce{Mn^'_{Mn}-V_O^{..}} dipole (see Fig.~\ref{fig:double_defect}a). Two anti-parallel NNN \ce{Fe_{Mn}-V_O^{OP}} defect pairs result in quenching the large component of the polarization along $a$, but in only a small decrease along $b$, due to the fact that the \ce{Mn^'_{Mn}-V_O^{..}} dipoles are still aligned along this axis (see Fig.~\ref{fig:double_defect}b). In both cases, the parallel alignment of the defect dipoles is energetically favored by 0.02~eV for the NN \ce{Fe_{Mn}-V_O^{OP}} and even more (by 0.26~eV) for NNN \ce{Fe_{Mn}-V_O^{OP}}, suggesting a parallel coupling of the defect dipoles even at room temperature.

\subsection{Interplay between polarization and strain\label{sec:strained_pol}}

%
\begin{figure}
 \centering
 \includegraphics[width=\columnwidth]{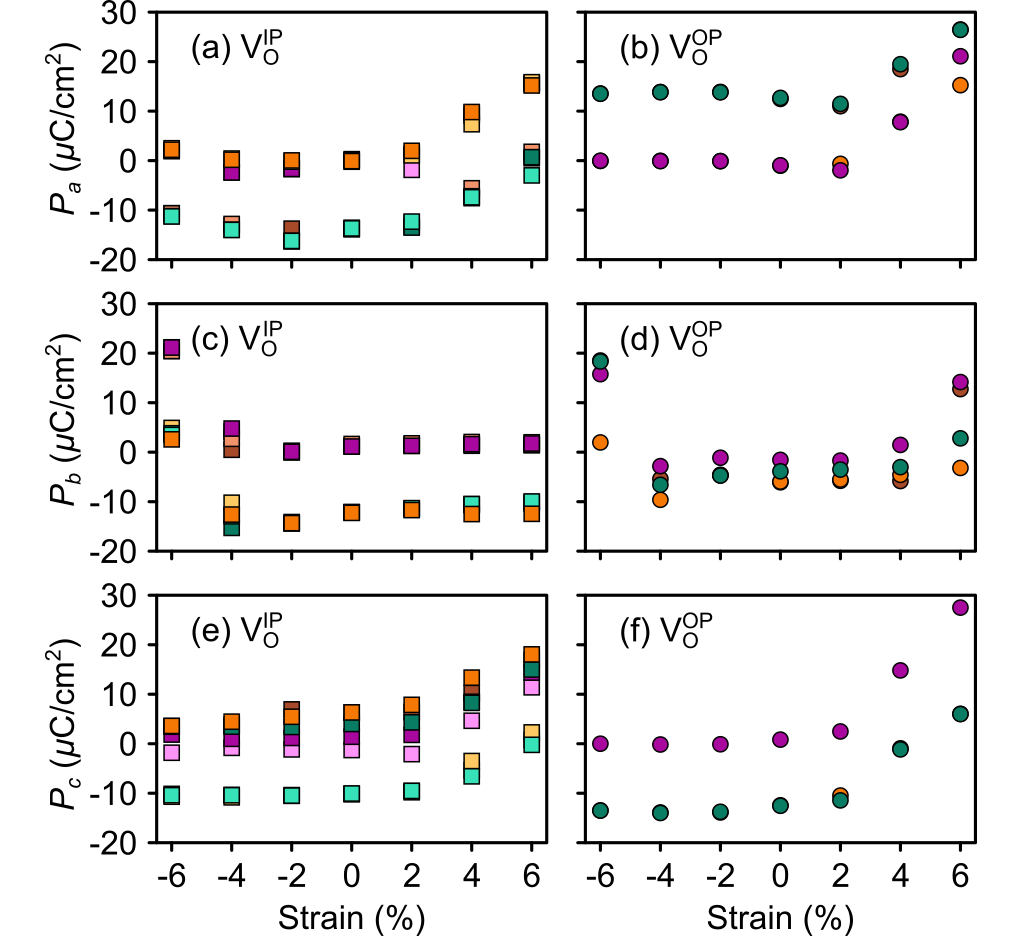}
\caption{Strain dependence of the $a$- (a-b) ,$b$- (c-d), and $c$- (e-f) components of the polarization for the different defect-pair configurations in AFM SMO. (a), (c), and (e) for \ce{Fe_{Mn}-V_O^{OP}} and (b), (d), and (f) for \ce{Fe_{Mn}-V_O^{IP}} defect pairs. Circle and square symbols refer to data obtained for \ce{V_O^{OP}} and \ce{V_O^{IP}}, respectively. See color code in Fig.~\ref{fig:SMO_structure_defect}.}
\label{fig:Pol_AFM}
\end{figure}

It is established that epitaxial strain imposed for example by lattice matching with a substrate during coherent epitaxial growth of thin films breaks the symmetry and affects competing energy contributions. Hence, it constitutes a viable strategy to induce ferroelectric properties in non-polar oxides~\cite{lee2010epitaxial, becher2015strain}. In this section, we discuss how the \ce{Fe_{Mn}-V_O} defect chemistry interacts with strain and with the magnetic properties in determining the polar properties of SMO thin films.

\begin{figure}
 \centering
 \includegraphics[width=\columnwidth]{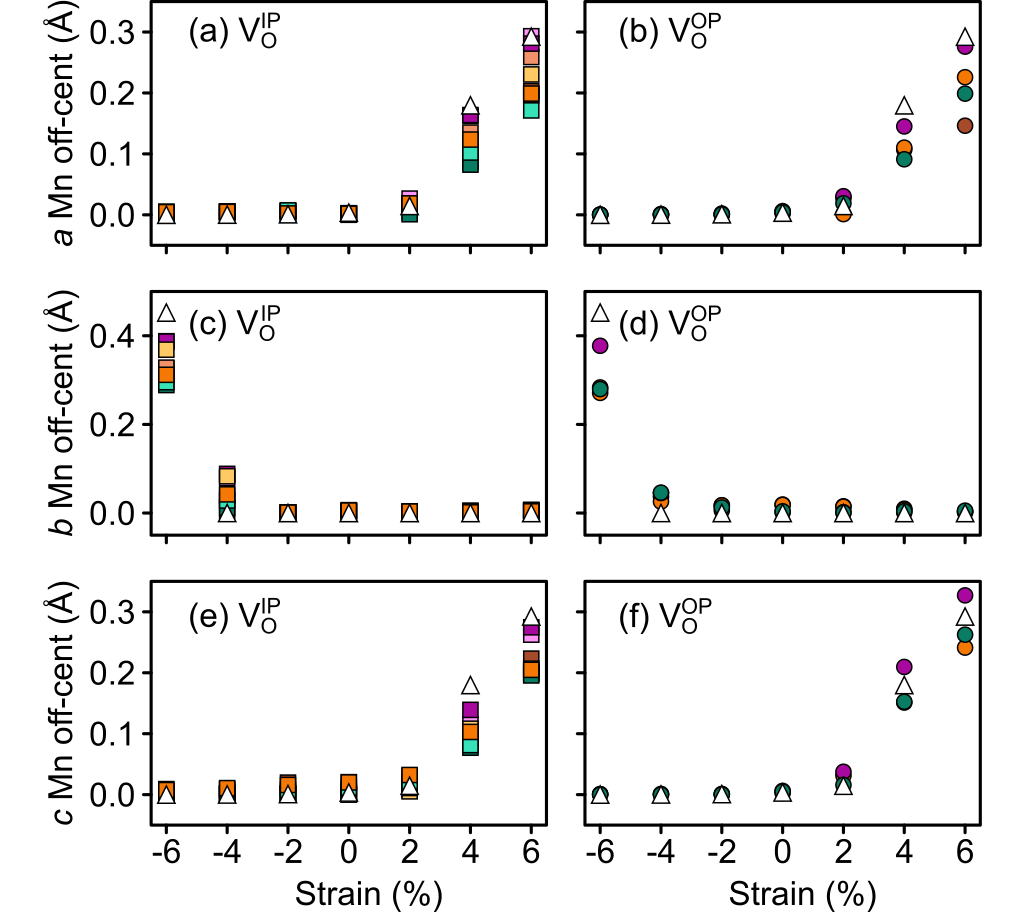}
\caption{Strain dependence of the average Mn off-centering long the $a$- (a-b), $b$- (c-d), and $c$-axis (e-f) for the different defect-pair configurations in the AFM phase of SMO. (a), (c), and (e) plots for \ce{Fe_{Mn}-V_O^{OP}} and (b), (d), and (f) plots for \ce{Fe_{Mn}-V_O^{IP}} defects. Circle and square symbols refer to data obtained for \ce{V_O^{OP}} and \ce{V_O^{IP}}, respectively. See color code in Fig.~\ref{fig:SMO_structure_defect}. The white triangles correspond, instead, to the Mn off-centerings in the stoichiometric SMO cells.}
\label{fig:Mnoff_AFM_strain}
\end{figure}

For tensile strained AFM Fe-doped oxygen-deficient SMO, we observe a general increase of the in-plane components of the polarization ($P_a$ and $P_c$, see Fig.~\ref{fig:Pol_AFM}): at 4\% strain by about 7~$\mu\textrm{C/cm}^2$ for $P_a$ and 12/7~$\mu\textrm{C/cm}^2$ for $P_c$ for \ce{Fe_{Mn}-V_O^{OP}}/\ce{Fe_{Mn}-V_O^{IP}} defects. This increase in polarization is accompanied by an average increase of the Mn-off-centering by about 0.07~and 0.15~\AA\ at +4\% strain along the $a$- and $c$-axis, respectively (see Fig.~\ref{fig:Mnoff_AFM_strain}a, b and e, f). Conversely, compressive strain is associated with a increase (up to 3~$\mu\textrm{C/cm}^2$, see Fig.~\ref{fig:Pol_AFM}c and d) of the out-of-plane component of the polarization associated with Mn off-centerings of about 0.03-0.1\AA\ along the $b$-axis (see Fig.~\ref{fig:Mnoff_AFM_strain}c and d) already for -4\% strain. Interestingly, this strain is smaller than the -6\% predicted necessary to destabilize the polar out-of-plane phonon mode in AFM SMO~\cite{ricca2021ferroelectricity}. Indeed, the larger Mn displacements computed in presence of defect pairs compared to the stoichiometric case (see Fig.~\ref{fig:Mnoff_AFM_strain}) in a strain range between -4\% and 2\% strain further highlight the ability of defect pairs to favor the polar phase transition. 

Interestingly, in the FM phase, polarization was found to be roughly constant as a function of the applied strain (cf. SI Fig.~\ref{fig:Pol_FM}), between -2 and 4 \% strain with small Mn off-centerings of the same magnitude as in the unstrained structure (cf. SI Fig.~\ref{fig:Mnoff_FM_strain}). A smaller increase of $P_b$/$P_{ac}$ and of the Mn off-centerings along $b$-axis/$ac$-plane starting for strain of about -4\%/6\% are observed also in the FM phase. The different behavior of the FM phase can be explained by the larger electronic screening of the defect dipole in the metallic FM phase and the strain dependence of the polar modes in stoichiometric SMO, where the IP modes soften only for large tensile strains beyond 6\% and the OP mode becomes unstable at 2\% compressive strain~\cite{ricca2021ferroelectricity}.

In summary, not only do these results suggest that doping SMO thin films with Fe can reverse the suppression of the ferroelectricity by oxygen vacancies generally present in the samples~\cite{marthinsen2016coupling}, but also that \ce{Fe_{Mn}-V_O} defect pairs can, depending on the magnetic order, couple with strain to favor the polar phase transition.

\section{Conclusions\label{sec:concls}}

In the present work we used DFT+$U_\textrm{SC-SD}$ calculations to investigate the potential of inducing ferroelectricity in \ce{SrMnO3} (SMO) through polar defect pairs, formed by a substitutional Fe atom and an oxygen vacancy. We further studied the interplay of these defect pairs with epitaxial strain and the magnetic phase. DFT+$U_\textrm{SC-SD}$ is fundamental to describe electronic-structure changes upon defect formation and to reconcile predicted magnetic properties with the available experimental data.

Our results suggest that defect engineering via polar defect pairs constitutes a parameter to design multiferroic materials. Ferroelectricity in nominally non-polar SMO can arise due to an off-center displacement of the defect charge resulting in a net electric dipole moment along the direction from the negatively  charged substitutional \ce{Fe^'_{Mn}} and \ce{Mn^'_{Mn}} sites to the positively charged \ce{V_O}. Furthermore, the defect pairs lead to a small off-centering of the Mn atoms in the defect neighborhood from their high-symmetry positions. 

Epitaxial strain strain can couple with \ce{Fe_{Mn}-V_O} not only reducing the defect formation energy (and hence increasing the defect concentration) under tensile strain, but more importantly inducing polarity either in-plane or out-of-plane for tensile and compressive strain respectively, already for strains smaller than those required to induce a polar phase transition in the defect-free material.

These results and the fact that local defect-induced dipoles couple in a parallel fashion, establishes polar defect pairs as a promising route to engineer ferroelectricity in nominally non-polar transition metal oxides. 

\section*{Acknowledgments\label{sec:acknow}}

This research was supported by the NCCR MARVEL, funded by the Swiss National Science Foundation. Computational resources were provided by the University of Bern (on the HPC cluster UBELIX, http://www.id.unibe.ch/hpc), by the Swiss National Supercomputing Center (CSCS) under project ID mr26 and by SuperMUC at GCS@LRZ, Germany, for which we acknowledge PRACE for awarding us access.

\bibliography{references}

\begin{thebibliography}{5}%
\makeatletter
\providecommand \@ifxundefined [1]{%
 \@ifx{#1\undefined}
}%
\providecommand \@ifnum [1]{%
 \ifnum #1\expandafter \@firstoftwo
 \else \expandafter \@secondoftwo
 \fi
}%
\providecommand \@ifx [1]{%
 \ifx #1\expandafter \@firstoftwo
 \else \expandafter \@secondoftwo
 \fi
}%
\providecommand \natexlab [1]{#1}%
\providecommand \enquote  [1]{``#1''}%
\providecommand \bibnamefont  [1]{#1}%
\providecommand \bibfnamefont [1]{#1}%
\providecommand \citenamefont [1]{#1}%
\providecommand \href@noop [0]{\@secondoftwo}%
\providecommand \href [0]{\begingroup \@sanitize@url \@href}%
\providecommand \@href[1]{\@@startlink{#1}\@@href}%
\providecommand \@@href[1]{\endgroup#1\@@endlink}%
\providecommand \@sanitize@url [0]{\catcode `\\12\catcode `\$12\catcode
  `\&12\catcode `\#12\catcode `\^12\catcode `\_12\catcode `\%12\relax}%
\providecommand \@@startlink[1]{}%
\providecommand \@@endlink[0]{}%
\providecommand \url  [0]{\begingroup\@sanitize@url \@url }%
\providecommand \@url [1]{\endgroup\@href {#1}{\urlprefix }}%
\providecommand \urlprefix  [0]{URL }%
\providecommand \Eprint [0]{\href }%
\providecommand \doibase [0]{http://dx.doi.org/}%
\providecommand \selectlanguage [0]{\@gobble}%
\providecommand \bibinfo  [0]{\@secondoftwo}%
\providecommand \bibfield  [0]{\@secondoftwo}%
\providecommand \translation [1]{[#1]}%
\providecommand \BibitemOpen [0]{}%
\providecommand \bibitemStop [0]{}%
\providecommand \bibitemNoStop [0]{.\EOS\space}%
\providecommand \EOS [0]{\spacefactor3000\relax}%
\providecommand \BibitemShut  [1]{\csname bibitem#1\endcsname}%
\let\auto@bib@innerbib\@empty
\bibitem [{\citenamefont {Ricca}\ \emph {et~al.}(2019)\citenamefont {Ricca},
  \citenamefont {Timrov}, \citenamefont {Cococcioni}, \citenamefont {Marzari},\
  and\ \citenamefont {Aschauer}}]{Ricca2019SI}%
  \BibitemOpen
  \bibfield  {author} {\bibinfo {author} {\bibfnamefont {C.}~\bibnamefont
  {Ricca}}, \bibinfo {author} {\bibfnamefont {I.}~\bibnamefont {Timrov}},
  \bibinfo {author} {\bibfnamefont {M.}~\bibnamefont {Cococcioni}}, \bibinfo
  {author} {\bibfnamefont {N.}~\bibnamefont {Marzari}}, \ and\ \bibinfo
  {author} {\bibfnamefont {U.}~\bibnamefont {Aschauer}},\ }\href {\doibase
  10.1103/PhysRevB.99.094102} {\bibfield  {journal} {\bibinfo  {journal} {Phys.
  Rev. B}\ }\textbf {\bibinfo {volume} {99}},\ \bibinfo {pages} {094102}
  (\bibinfo {year} {2019})}\BibitemShut {NoStop}%
\bibitem [{\citenamefont {Sit}\ \emph {et~al.}(2011)\citenamefont {Sit},
  \citenamefont {Car}, \citenamefont {Cohen},\ and\ \citenamefont
  {Selloni}}]{Sit2011SI}%
  \BibitemOpen
  \bibfield  {author} {\bibinfo {author} {\bibfnamefont {P.~H.~L.}\
  \bibnamefont {Sit}}, \bibinfo {author} {\bibfnamefont {R.}~\bibnamefont
  {Car}}, \bibinfo {author} {\bibfnamefont {M.~H.}\ \bibnamefont {Cohen}}, \
  and\ \bibinfo {author} {\bibfnamefont {A.}~\bibnamefont {Selloni}},\ }\href
  {\doibase 10.1021/ic2013107} {\bibfield  {journal} {\bibinfo  {journal}
  {Inorganic Chemistry}\ }\textbf {\bibinfo {volume} {50}},\ \bibinfo {pages}
  {10259} (\bibinfo {year} {2011})}\BibitemShut {NoStop}%
\bibitem [{\citenamefont {Battle}\ \emph {et~al.}(1996)\citenamefont {Battle},
  \citenamefont {Davison}, \citenamefont {Gibb},\ and\ \citenamefont
  {Vente}}]{BattleJM9960601187SI}%
  \BibitemOpen
  \bibfield  {author} {\bibinfo {author} {\bibfnamefont {P.~D.}\ \bibnamefont
  {Battle}}, \bibinfo {author} {\bibfnamefont {C.~M.}\ \bibnamefont {Davison}},
  \bibinfo {author} {\bibfnamefont {T.~C.}\ \bibnamefont {Gibb}}, \ and\
  \bibinfo {author} {\bibfnamefont {J.~F.}\ \bibnamefont {Vente}},\ }\href
  {\doibase 10.1039/JM9960601187} {\bibfield  {journal} {\bibinfo  {journal}
  {J. Mater. Chem.}\ }\textbf {\bibinfo {volume} {6}},\ \bibinfo {pages} {1187}
  (\bibinfo {year} {1996})}\BibitemShut {NoStop}%
\bibitem [{\citenamefont {Aschauer}\ \emph {et~al.}(2013)\citenamefont
  {Aschauer}, \citenamefont {Pfenninger}, \citenamefont {Selbach},
  \citenamefont {Grande},\ and\ \citenamefont
  {Spaldin}}]{aschauer2013strainSI}%
  \BibitemOpen
  \bibfield  {author} {\bibinfo {author} {\bibfnamefont {U.}~\bibnamefont
  {Aschauer}}, \bibinfo {author} {\bibfnamefont {R.}~\bibnamefont
  {Pfenninger}}, \bibinfo {author} {\bibfnamefont {S.~M.}\ \bibnamefont
  {Selbach}}, \bibinfo {author} {\bibfnamefont {T.}~\bibnamefont {Grande}}, \
  and\ \bibinfo {author} {\bibfnamefont {N.~A.}\ \bibnamefont {Spaldin}},\
  }\href {\doibase 10.1103/PhysRevB.88.054111} {\bibfield  {journal} {\bibinfo
  {journal} {Phys. Rev. B}\ }\textbf {\bibinfo {volume} {88}},\ \bibinfo
  {pages} {054111} (\bibinfo {year} {2013})}\BibitemShut {NoStop}%
\bibitem [{\citenamefont {Ricca}\ \emph {et~al.}(2021)\citenamefont {Ricca},
  \citenamefont {Berkowitz},\ and\ \citenamefont
  {Aschauer}}]{ricca2021ferroelectricitySI}%
  \BibitemOpen
  \bibfield  {author} {\bibinfo {author} {\bibfnamefont {C.}~\bibnamefont
  {Ricca}}, \bibinfo {author} {\bibfnamefont {D.}~\bibnamefont {Berkowitz}}, \
  and\ \bibinfo {author} {\bibfnamefont {U.}~\bibnamefont {Aschauer}},\
  }\href@noop {} {\enquote {\bibinfo {title} {{F}erroelectricity promoted by
  cation/anion divacancies in \ce{SrMnO3}},}\ } (\bibinfo {year} {2021}),\
  \Eprint {http://arxiv.org/abs/2105.09360} {arXiv:2105.09360
  [cond-mat.mtrl-sci]} \BibitemShut {NoStop}%
\end{thebibliography}%


\begin{thebibliography}{47}%
\makeatletter
\providecommand \@ifxundefined [1]{%
 \@ifx{#1\undefined}
}%
\providecommand \@ifnum [1]{%
 \ifnum #1\expandafter \@firstoftwo
 \else \expandafter \@secondoftwo
 \fi
}%
\providecommand \@ifx [1]{%
 \ifx #1\expandafter \@firstoftwo
 \else \expandafter \@secondoftwo
 \fi
}%
\providecommand \natexlab [1]{#1}%
\providecommand \enquote  [1]{``#1''}%
\providecommand \bibnamefont  [1]{#1}%
\providecommand \bibfnamefont [1]{#1}%
\providecommand \citenamefont [1]{#1}%
\providecommand \href@noop [0]{\@secondoftwo}%
\providecommand \href [0]{\begingroup \@sanitize@url \@href}%
\providecommand \@href[1]{\@@startlink{#1}\@@href}%
\providecommand \@@href[1]{\endgroup#1\@@endlink}%
\providecommand \@sanitize@url [0]{\catcode `\\12\catcode `\$12\catcode
  `\&12\catcode `\#12\catcode `\^12\catcode `\_12\catcode `\%12\relax}%
\providecommand \@@startlink[1]{}%
\providecommand \@@endlink[0]{}%
\providecommand \url  [0]{\begingroup\@sanitize@url \@url }%
\providecommand \@url [1]{\endgroup\@href {#1}{\urlprefix }}%
\providecommand \urlprefix  [0]{URL }%
\providecommand \Eprint [0]{\href }%
\providecommand \doibase [0]{http://dx.doi.org/}%
\providecommand \selectlanguage [0]{\@gobble}%
\providecommand \bibinfo  [0]{\@secondoftwo}%
\providecommand \bibfield  [0]{\@secondoftwo}%
\providecommand \translation [1]{[#1]}%
\providecommand \BibitemOpen [0]{}%
\providecommand \bibitemStop [0]{}%
\providecommand \bibitemNoStop [0]{.\EOS\space}%
\providecommand \EOS [0]{\spacefactor3000\relax}%
\providecommand \BibitemShut  [1]{\csname bibitem#1\endcsname}%
\let\auto@bib@innerbib\@empty
\bibitem [{\citenamefont {Fuchigami}\ \emph {et~al.}(2009)\citenamefont
  {Fuchigami}, \citenamefont {Gai}, \citenamefont {Ward}, \citenamefont {Yin},
  \citenamefont {Snijders}, \citenamefont {Plummer},\ and\ \citenamefont
  {Shen}}]{fuchigami2009}%
  \BibitemOpen
  \bibfield  {author} {\bibinfo {author} {\bibfnamefont {K.}~\bibnamefont
  {Fuchigami}}, \bibinfo {author} {\bibfnamefont {Z.}~\bibnamefont {Gai}},
  \bibinfo {author} {\bibfnamefont {T.~Z.}\ \bibnamefont {Ward}}, \bibinfo
  {author} {\bibfnamefont {L.~F.}\ \bibnamefont {Yin}}, \bibinfo {author}
  {\bibfnamefont {P.~C.}\ \bibnamefont {Snijders}}, \bibinfo {author}
  {\bibfnamefont {E.~W.}\ \bibnamefont {Plummer}}, \ and\ \bibinfo {author}
  {\bibfnamefont {J.}~\bibnamefont {Shen}},\ }\bibfield  {title} {\enquote
  {\bibinfo {title} {{T}unable metallicity of the \ce{La_{5/8}Ca_{3/8}MnO3}
  (001) surface by an oxygen overlayer},}\ }\href {\doibase
  10.1103/PhysRevLett.102.066104} {\bibfield  {journal} {\bibinfo  {journal}
  {Phys. Rev. Lett.}\ }\textbf {\bibinfo {volume} {102}},\ \bibinfo {pages}
  {066104} (\bibinfo {year} {2009})}\BibitemShut {NoStop}%
\bibitem [{\citenamefont {Tuller}\ and\ \citenamefont
  {Bishop}(2011)}]{tuller2011}%
  \BibitemOpen
  \bibfield  {author} {\bibinfo {author} {\bibfnamefont {H.~L.}\ \bibnamefont
  {Tuller}}\ and\ \bibinfo {author} {\bibfnamefont {S.~R.}\ \bibnamefont
  {Bishop}},\ }\bibfield  {title} {\enquote {\bibinfo {title} {{P}oint defects
  in oxides: tailoring materials through defect engineering},}\ }\href
  {\doibase 10.1146/annurev-matsci-062910-100442} {\bibfield  {journal}
  {\bibinfo  {journal} {Annu. Rev. Mater. Res.}\ }\textbf {\bibinfo {volume}
  {41}},\ \bibinfo {pages} {369--398} (\bibinfo {year} {2011})}\BibitemShut
  {NoStop}%
\bibitem [{\citenamefont {Kalinin}\ \emph {et~al.}(2012)\citenamefont
  {Kalinin}, \citenamefont {Borisevich},\ and\ \citenamefont
  {Fong}}]{kalinin2012}%
  \BibitemOpen
  \bibfield  {author} {\bibinfo {author} {\bibfnamefont {S.~V.}\ \bibnamefont
  {Kalinin}}, \bibinfo {author} {\bibfnamefont {A.}~\bibnamefont {Borisevich}},
  \ and\ \bibinfo {author} {\bibfnamefont {D.}~\bibnamefont {Fong}},\
  }\bibfield  {title} {\enquote {\bibinfo {title} {{B}eyond condensed matter
  physics on the nanoscale: the role of ionic and electrochemical phenomena in
  the physical functionalities of oxide materials},}\ }\href {\doibase
  10.1021/nn304930x} {\bibfield  {journal} {\bibinfo  {journal} {ACS Nano}\
  }\textbf {\bibinfo {volume} {6}},\ \bibinfo {pages} {10423--10437} (\bibinfo
  {year} {2012})}\BibitemShut {NoStop}%
\bibitem [{\citenamefont {Kalinin}\ and\ \citenamefont
  {Spaldin}(2013)}]{kalinin2013functional}%
  \BibitemOpen
  \bibfield  {author} {\bibinfo {author} {\bibfnamefont {S.~V.}\ \bibnamefont
  {Kalinin}}\ and\ \bibinfo {author} {\bibfnamefont {N.~A.}\ \bibnamefont
  {Spaldin}},\ }\bibfield  {title} {\enquote {\bibinfo {title} {{F}unctional
  ion defects in transition metal oxides},}\ }\href {\doibase
  10.1126/science.1243098} {\bibfield  {journal} {\bibinfo  {journal}
  {Science}\ }\textbf {\bibinfo {volume} {341}},\ \bibinfo {pages} {858--859}
  (\bibinfo {year} {2013})}\BibitemShut {NoStop}%
\bibitem [{\citenamefont {Chandrasekaran}\ \emph {et~al.}(2013)\citenamefont
  {Chandrasekaran}, \citenamefont {Damjanovic}, \citenamefont {Setter},\ and\
  \citenamefont {Marzari}}]{chandrasekaran2013}%
  \BibitemOpen
  \bibfield  {author} {\bibinfo {author} {\bibfnamefont {A.}~\bibnamefont
  {Chandrasekaran}}, \bibinfo {author} {\bibfnamefont {D.}~\bibnamefont
  {Damjanovic}}, \bibinfo {author} {\bibfnamefont {N.}~\bibnamefont {Setter}},
  \ and\ \bibinfo {author} {\bibfnamefont {N.}~\bibnamefont {Marzari}},\
  }\bibfield  {title} {\enquote {\bibinfo {title} {{D}efect ordering and
  defect--domain-wall interactions in \ce{PbTiO3}: {A} first-principles
  study},}\ }\href {\doibase 10.1103/PhysRevB.88.214116} {\bibfield  {journal}
  {\bibinfo  {journal} {Phys. Rev. B}\ }\textbf {\bibinfo {volume} {88}},\
  \bibinfo {pages} {214116} (\bibinfo {year} {2013})}\BibitemShut {NoStop}%
\bibitem [{\citenamefont {Bi{\v s}kup}\ \emph {et~al.}(2014)\citenamefont
  {Bi{\v s}kup}, \citenamefont {Salafranca}, \citenamefont {Mehta},
  \citenamefont {Oxley}, \citenamefont {Suzuki}, \citenamefont {Pennycook},
  \citenamefont {Pantelides},\ and\ \citenamefont {Varela}}]{bivskup2014}%
  \BibitemOpen
  \bibfield  {author} {\bibinfo {author} {\bibfnamefont {N.}~\bibnamefont
  {Bi{\v s}kup}}, \bibinfo {author} {\bibfnamefont {J.}~\bibnamefont
  {Salafranca}}, \bibinfo {author} {\bibfnamefont {V.}~\bibnamefont {Mehta}},
  \bibinfo {author} {\bibfnamefont {M.~P.}\ \bibnamefont {Oxley}}, \bibinfo
  {author} {\bibfnamefont {Y.}~\bibnamefont {Suzuki}}, \bibinfo {author}
  {\bibfnamefont {S.~J.}\ \bibnamefont {Pennycook}}, \bibinfo {author}
  {\bibfnamefont {S.~T.}\ \bibnamefont {Pantelides}}, \ and\ \bibinfo {author}
  {\bibfnamefont {M.}~\bibnamefont {Varela}},\ }\bibfield  {title} {\enquote
  {\bibinfo {title} {{I}nsulating ferromagnetic \ce{LaCoO_{3-\delta}} films: a
  phase induced by ordering of oxygen vacancies},}\ }\href {\doibase
  10.1103/PhysRevLett.112.087202} {\bibfield  {journal} {\bibinfo  {journal}
  {Phys. Rev. Lett.}\ }\textbf {\bibinfo {volume} {112}},\ \bibinfo {pages}
  {087202} (\bibinfo {year} {2014})}\BibitemShut {NoStop}%
\bibitem [{\citenamefont {Bhattacharya}\ and\ \citenamefont
  {May}(2014)}]{bhattacharya2014magnetic}%
  \BibitemOpen
  \bibfield  {author} {\bibinfo {author} {\bibfnamefont {A.}~\bibnamefont
  {Bhattacharya}}\ and\ \bibinfo {author} {\bibfnamefont {S.~J.}\ \bibnamefont
  {May}},\ }\bibfield  {title} {\enquote {\bibinfo {title} {{M}agnetic oxide
  heterostructures},}\ }\href {\doibase 10.1146/annurev-matsci-070813-113447}
  {\bibfield  {journal} {\bibinfo  {journal} {Annu. Rev. Mater. Res.}\ }\textbf
  {\bibinfo {volume} {44}},\ \bibinfo {pages} {65--90} (\bibinfo {year}
  {2014})}\BibitemShut {NoStop}%
\bibitem [{\citenamefont {Becher}\ \emph {et~al.}(2015)\citenamefont {Becher},
  \citenamefont {Maurel}, \citenamefont {Aschauer}, \citenamefont {Lilienblum},
  \citenamefont {Mag{\'e}n}, \citenamefont {Meier}, \citenamefont {Langenberg},
  \citenamefont {Trassin}, \citenamefont {Blasco}, \citenamefont {Krug},
  \citenamefont {Algarabel}, \citenamefont {Spaldin}, \citenamefont {Pardo},\
  and\ \citenamefont {Fiebig}}]{becher2015strain}%
  \BibitemOpen
  \bibfield  {author} {\bibinfo {author} {\bibfnamefont {C.}~\bibnamefont
  {Becher}}, \bibinfo {author} {\bibfnamefont {L.}~\bibnamefont {Maurel}},
  \bibinfo {author} {\bibfnamefont {U.}~\bibnamefont {Aschauer}}, \bibinfo
  {author} {\bibfnamefont {M.}~\bibnamefont {Lilienblum}}, \bibinfo {author}
  {\bibfnamefont {C.}~\bibnamefont {Mag{\'e}n}}, \bibinfo {author}
  {\bibfnamefont {D.}~\bibnamefont {Meier}}, \bibinfo {author} {\bibfnamefont
  {E.}~\bibnamefont {Langenberg}}, \bibinfo {author} {\bibfnamefont
  {M.}~\bibnamefont {Trassin}}, \bibinfo {author} {\bibfnamefont
  {J.}~\bibnamefont {Blasco}}, \bibinfo {author} {\bibfnamefont {I.~P.}\
  \bibnamefont {Krug}}, \bibinfo {author} {\bibfnamefont {P.~A.}\ \bibnamefont
  {Algarabel}}, \bibinfo {author} {\bibfnamefont {N.~A.}\ \bibnamefont
  {Spaldin}}, \bibinfo {author} {\bibfnamefont {J.~A.}\ \bibnamefont {Pardo}},
  \ and\ \bibinfo {author} {\bibfnamefont {M.}~\bibnamefont {Fiebig}},\
  }\bibfield  {title} {\enquote {\bibinfo {title} {{S}train-induced coupling of
  electrical polarization and structural defects in \ce{SrMnO3} films},}\
  }\href {\doibase 10.1038/nnano.2015.108} {\bibfield  {journal} {\bibinfo
  {journal} {Nat. Nanotechnol.}\ }\textbf {\bibinfo {volume} {10}},\ \bibinfo
  {pages} {661} (\bibinfo {year} {2015})}\BibitemShut {NoStop}%
\bibitem [{\citenamefont {Marthinsen}\ \emph {et~al.}(2016)\citenamefont
  {Marthinsen}, \citenamefont {Faber}, \citenamefont {Aschauer}, \citenamefont
  {Spaldin},\ and\ \citenamefont {Selbach}}]{marthinsen2016coupling}%
  \BibitemOpen
  \bibfield  {author} {\bibinfo {author} {\bibfnamefont {A.}~\bibnamefont
  {Marthinsen}}, \bibinfo {author} {\bibfnamefont {C.}~\bibnamefont {Faber}},
  \bibinfo {author} {\bibfnamefont {U.}~\bibnamefont {Aschauer}}, \bibinfo
  {author} {\bibfnamefont {N.~A.}\ \bibnamefont {Spaldin}}, \ and\ \bibinfo
  {author} {\bibfnamefont {S.~M.}\ \bibnamefont {Selbach}},\ }\bibfield
  {title} {\enquote {\bibinfo {title} {{C}oupling and competition between
  ferroelectricity, magnetism, strain, and oxygen vacancies in \ce{AMnO3}
  perovskites},}\ }\href {\doibase 10.1557/mrc.2016.30} {\bibfield  {journal}
  {\bibinfo  {journal} {MRS Commun.}\ }\textbf {\bibinfo {volume} {6}},\
  \bibinfo {pages} {182--191} (\bibinfo {year} {2016})}\BibitemShut {NoStop}%
\bibitem [{\citenamefont {Griffin}\ \emph {et~al.}(2017)\citenamefont
  {Griffin}, \citenamefont {Reidulff}, \citenamefont {Selbach},\ and\
  \citenamefont {Spaldin}}]{griffin2017defect}%
  \BibitemOpen
  \bibfield  {author} {\bibinfo {author} {\bibfnamefont {S.~M.}\ \bibnamefont
  {Griffin}}, \bibinfo {author} {\bibfnamefont {M.}~\bibnamefont {Reidulff}},
  \bibinfo {author} {\bibfnamefont {S.~M.}\ \bibnamefont {Selbach}}, \ and\
  \bibinfo {author} {\bibfnamefont {N.~A.}\ \bibnamefont {Spaldin}},\
  }\bibfield  {title} {\enquote {\bibinfo {title} {{D}efect chemistry as a
  crystal structure design parameter: {I}ntrinsic point defects and \ce{Ga}
  substitution in \ce{InMnO3}},}\ }\href {\doibase
  10.1021/acs.chemmater.6b04207} {\bibfield  {journal} {\bibinfo  {journal}
  {Chem. Mater.}\ }\textbf {\bibinfo {volume} {29}},\ \bibinfo {pages}
  {2425--2434} (\bibinfo {year} {2017})}\BibitemShut {NoStop}%
\bibitem [{\citenamefont {Rojac}\ and\ \citenamefont
  {Damjanovic}(2017)}]{rojac2017domain}%
  \BibitemOpen
  \bibfield  {author} {\bibinfo {author} {\bibfnamefont {T.}~\bibnamefont
  {Rojac}}\ and\ \bibinfo {author} {\bibfnamefont {D.}~\bibnamefont
  {Damjanovic}},\ }\bibfield  {title} {\enquote {\bibinfo {title} {{D}omain
  walls and defects in ferroelectric materials},}\ }\href {\doibase
  10.7567/JJAP.56.10PA01} {\bibfield  {journal} {\bibinfo  {journal} {Jpn. J.
  Appl. Phys.}\ }\textbf {\bibinfo {volume} {56}},\ \bibinfo {pages} {10PA01}
  (\bibinfo {year} {2017})}\BibitemShut {NoStop}%
\bibitem [{\citenamefont {Lee}\ and\ \citenamefont
  {Rabe}(2010)}]{lee2010epitaxial}%
  \BibitemOpen
  \bibfield  {author} {\bibinfo {author} {\bibfnamefont {J.~H.}\ \bibnamefont
  {Lee}}\ and\ \bibinfo {author} {\bibfnamefont {K.~M.}\ \bibnamefont {Rabe}},\
  }\bibfield  {title} {\enquote {\bibinfo {title} {{E}pitaxial-strain-induced
  multiferroicity in \ce{SrMnO3} from first principles},}\ }\href {\doibase
  10.1103/PhysRevLett.104.207204} {\bibfield  {journal} {\bibinfo  {journal}
  {Phys. Rev. Lett.}\ }\textbf {\bibinfo {volume} {104}},\ \bibinfo {pages}
  {207204} (\bibinfo {year} {2010})}\BibitemShut {NoStop}%
\bibitem [{\citenamefont {Wang}\ \emph {et~al.}(2017)\citenamefont {Wang},
  \citenamefont {Tang}, \citenamefont {Liu}, \citenamefont {Jiang},\ and\
  \citenamefont {Jiang}}]{Wang2017nano}%
  \BibitemOpen
  \bibfield  {author} {\bibinfo {author} {\bibfnamefont {Y.-G.}\ \bibnamefont
  {Wang}}, \bibinfo {author} {\bibfnamefont {X.-G.}\ \bibnamefont {Tang}},
  \bibinfo {author} {\bibfnamefont {Q.-X.}\ \bibnamefont {Liu}}, \bibinfo
  {author} {\bibfnamefont {Y.-P.}\ \bibnamefont {Jiang}}, \ and\ \bibinfo
  {author} {\bibfnamefont {L.-L.}\ \bibnamefont {Jiang}},\ }\bibfield  {title}
  {\enquote {\bibinfo {title} {{R}oom temperature tunable multiferroic
  properties in sol-gel-derived nanocrystalline
  \ce{Sr(Ti_{1-x}Fe_x)O_{3-\delta}} thin films},}\ }\href {\doibase
  10.3390/nano7090264} {\bibfield  {journal} {\bibinfo  {journal}
  {Nanomaterials}\ }\textbf {\bibinfo {volume} {7}},\ \bibinfo {pages} {264}
  (\bibinfo {year} {2017})}\BibitemShut {NoStop}%
\bibitem [{\citenamefont {Syono}\ \emph {et~al.}(1969)\citenamefont {Syono},
  \citenamefont {Akimoto},\ and\ \citenamefont {Kohn}}]{Syono1969Structure}%
  \BibitemOpen
  \bibfield  {author} {\bibinfo {author} {\bibfnamefont {Y.}~\bibnamefont
  {Syono}}, \bibinfo {author} {\bibfnamefont {Y.}~\bibnamefont {Akimoto}}, \
  and\ \bibinfo {author} {\bibfnamefont {K.}~\bibnamefont {Kohn}},\ }\bibfield
  {title} {\enquote {\bibinfo {title} {{S}tructure relations of hexagonal
  perovskite-like compounds \ce{ABX3} at high pressure},}\ }\href {\doibase
  10.1143/jpsj.26.993} {\bibfield  {journal} {\bibinfo  {journal} {Journal of
  the Physical Society of Japan}\ }\textbf {\bibinfo {volume} {26}},\ \bibinfo
  {pages} {993--999} (\bibinfo {year} {1969})}\BibitemShut {NoStop}%
\bibitem [{\citenamefont {Chmaissem}\ \emph {et~al.}(2001)\citenamefont
  {Chmaissem}, \citenamefont {Dabrowski}, \citenamefont {Kolesnik},
  \citenamefont {Mais}, \citenamefont {Brown}, \citenamefont {Kruk},
  \citenamefont {Prior}, \citenamefont {Pyles},\ and\ \citenamefont
  {Jorgensen}}]{chmaissem2001relationship}%
  \BibitemOpen
  \bibfield  {author} {\bibinfo {author} {\bibfnamefont {O.}~\bibnamefont
  {Chmaissem}}, \bibinfo {author} {\bibfnamefont {B.}~\bibnamefont
  {Dabrowski}}, \bibinfo {author} {\bibfnamefont {S.}~\bibnamefont {Kolesnik}},
  \bibinfo {author} {\bibfnamefont {J.}~\bibnamefont {Mais}}, \bibinfo {author}
  {\bibfnamefont {D.~E.}\ \bibnamefont {Brown}}, \bibinfo {author}
  {\bibfnamefont {R.}~\bibnamefont {Kruk}}, \bibinfo {author} {\bibfnamefont
  {P.}~\bibnamefont {Prior}}, \bibinfo {author} {\bibfnamefont
  {B.}~\bibnamefont {Pyles}}, \ and\ \bibinfo {author} {\bibfnamefont {J.~D.}\
  \bibnamefont {Jorgensen}},\ }\bibfield  {title} {\enquote {\bibinfo {title}
  {{R}elationship between structural parameters and the {N}{\'e}el temperature
  in \ce{Sr_{1-x}Ca_xMnO3} $(0\leq x \leq 1)$ and \ce{Sr_{1-y}Ba_yMnO3} $(y
  \leq 0.2)$},}\ }\href {\doibase 10.1103/PhysRevB.64.134412} {\bibfield
  {journal} {\bibinfo  {journal} {Phys. Rev. B}\ }\textbf {\bibinfo {volume}
  {64}},\ \bibinfo {pages} {134412} (\bibinfo {year} {2001})}\BibitemShut
  {NoStop}%
\bibitem [{\citenamefont {Kobayashi}\ \emph {et~al.}(2010)\citenamefont
  {Kobayashi}, \citenamefont {Tokuda}, \citenamefont {Ohnishi}, \citenamefont
  {Mizoguchi}, \citenamefont {Shibata}, \citenamefont {Sato}, \citenamefont
  {Ikuhara},\ and\ \citenamefont {Yamamoto}}]{Kobayashi:2010by}%
  \BibitemOpen
  \bibfield  {author} {\bibinfo {author} {\bibfnamefont {S.}~\bibnamefont
  {Kobayashi}}, \bibinfo {author} {\bibfnamefont {Y.}~\bibnamefont {Tokuda}},
  \bibinfo {author} {\bibfnamefont {T.}~\bibnamefont {Ohnishi}}, \bibinfo
  {author} {\bibfnamefont {T.}~\bibnamefont {Mizoguchi}}, \bibinfo {author}
  {\bibfnamefont {N.}~\bibnamefont {Shibata}}, \bibinfo {author} {\bibfnamefont
  {Y.}~\bibnamefont {Sato}}, \bibinfo {author} {\bibfnamefont {Y.}~\bibnamefont
  {Ikuhara}}, \ and\ \bibinfo {author} {\bibfnamefont {T.}~\bibnamefont
  {Yamamoto}},\ }\bibfield  {title} {\enquote {\bibinfo {title} {{C}ation
  off-stoichiometric \ce{SrMnO_{3-\delta}} thin film grown by pulsed laser
  deposition},}\ }\href {\doibase 10.1007/s10853-010-5103-2} {\bibfield
  {journal} {\bibinfo  {journal} {J. Mater. Sci.}\ }\textbf {\bibinfo {volume}
  {46}},\ \bibinfo {pages} {4354--4360} (\bibinfo {year} {2010})}\BibitemShut
  {NoStop}%
\bibitem [{\citenamefont {Ricca}\ \emph {et~al.}(2021)\citenamefont {Ricca},
  \citenamefont {Berkowitz},\ and\ \citenamefont
  {Aschauer}}]{ricca2021ferroelectricity}%
  \BibitemOpen
  \bibfield  {author} {\bibinfo {author} {\bibfnamefont {C.}~\bibnamefont
  {Ricca}}, \bibinfo {author} {\bibfnamefont {D.}~\bibnamefont {Berkowitz}}, \
  and\ \bibinfo {author} {\bibfnamefont {U.}~\bibnamefont {Aschauer}},\
  }\href@noop {} {\enquote {\bibinfo {title} {{F}erroelectricity promoted by
  cation/anion divacancies in \ce{SrMnO3}},}\ } (\bibinfo {year} {2021}),\
  \Eprint {http://arxiv.org/abs/2105.09360} {arXiv:2105.09360
  [cond-mat.mtrl-sci]} \BibitemShut {NoStop}%
\bibitem [{\citenamefont {Fawcett}\ \emph {et~al.}(2000)\citenamefont
  {Fawcett}, \citenamefont {Veith}, \citenamefont {Greenblatt}, \citenamefont
  {Croft},\ and\ \citenamefont {Nowik}}]{FAWCETT2000821}%
  \BibitemOpen
  \bibfield  {author} {\bibinfo {author} {\bibfnamefont {I.~D.}\ \bibnamefont
  {Fawcett}}, \bibinfo {author} {\bibfnamefont {G.~M.}\ \bibnamefont {Veith}},
  \bibinfo {author} {\bibfnamefont {M.}~\bibnamefont {Greenblatt}}, \bibinfo
  {author} {\bibfnamefont {M.}~\bibnamefont {Croft}}, \ and\ \bibinfo {author}
  {\bibfnamefont {I.}~\bibnamefont {Nowik}},\ }\bibfield  {title} {\enquote
  {\bibinfo {title} {{P}roperties of the perovskites,
  \ce{SrMn_{1-x}Fe_xO_{3-\delta}} (x=1/3, 1/2, 2/3)},}\ }\href {\doibase
  10.1016/S1293-2558(00)01097-9} {\bibfield  {journal} {\bibinfo  {journal}
  {Solid State Sciences}\ }\textbf {\bibinfo {volume} {2}},\ \bibinfo {pages}
  {821--831} (\bibinfo {year} {2000})}\BibitemShut {NoStop}%
\bibitem [{\citenamefont {Jin}\ \emph {et~al.}(1994)\citenamefont {Jin},
  \citenamefont {Tiefel}, \citenamefont {McCormack}, \citenamefont {Fastnacht},
  \citenamefont {Ramesh},\ and\ \citenamefont {Chen}}]{Jin413}%
  \BibitemOpen
  \bibfield  {author} {\bibinfo {author} {\bibfnamefont {S.}~\bibnamefont
  {Jin}}, \bibinfo {author} {\bibfnamefont {T.~H.}\ \bibnamefont {Tiefel}},
  \bibinfo {author} {\bibfnamefont {M.}~\bibnamefont {McCormack}}, \bibinfo
  {author} {\bibfnamefont {R.~A.}\ \bibnamefont {Fastnacht}}, \bibinfo {author}
  {\bibfnamefont {R.}~\bibnamefont {Ramesh}}, \ and\ \bibinfo {author}
  {\bibfnamefont {L.~H.}\ \bibnamefont {Chen}},\ }\bibfield  {title} {\enquote
  {\bibinfo {title} {{T}housandfold change in resistivity in magnetoresistive
  \ce{La-Ca-Mn-O} films},}\ }\href {\doibase 10.1126/science.264.5157.413}
  {\bibfield  {journal} {\bibinfo  {journal} {Science}\ }\textbf {\bibinfo
  {volume} {264}},\ \bibinfo {pages} {413--415} (\bibinfo {year}
  {1994})}\BibitemShut {NoStop}%
\bibitem [{\citenamefont {Millis}(1998)}]{Millis1998Nature}%
  \BibitemOpen
  \bibfield  {author} {\bibinfo {author} {\bibfnamefont {A.~J.}\ \bibnamefont
  {Millis}},\ }\bibfield  {title} {\enquote {\bibinfo {title} {{L}attice
  effects in magnetoresistive manganese perovskites},}\ }\href {\doibase
  10.1038/32348} {\bibfield  {journal} {\bibinfo  {journal} {Nature}\ }\textbf
  {\bibinfo {volume} {392}},\ \bibinfo {pages} {147--150} (\bibinfo {year}
  {1998})}\BibitemShut {NoStop}%
\bibitem [{\citenamefont {Battle}\ \emph {et~al.}(1988)\citenamefont {Battle},
  \citenamefont {Gibb},\ and\ \citenamefont {Jones}}]{Battle1988}%
  \BibitemOpen
  \bibfield  {author} {\bibinfo {author} {\bibfnamefont {P.~D.}\ \bibnamefont
  {Battle}}, \bibinfo {author} {\bibfnamefont {T.~C.}\ \bibnamefont {Gibb}}, \
  and\ \bibinfo {author} {\bibfnamefont {C.~W.}\ \bibnamefont {Jones}},\
  }\bibfield  {title} {\enquote {\bibinfo {title} {{T}he structural and
  magnetic properties of \ce{SrMnO3}: {A} reinvestigation},}\ }\href {\doibase
  10.1016/0022-4596(88)90331-3} {\bibfield  {journal} {\bibinfo  {journal}
  {Journal of Solid State Chemistry France}\ }\textbf {\bibinfo {volume}
  {74}},\ \bibinfo {pages} {60--66} (\bibinfo {year} {1988})}\BibitemShut
  {NoStop}%
\bibitem [{\citenamefont {Battle}\ \emph {et~al.}(1996)\citenamefont {Battle},
  \citenamefont {Davison}, \citenamefont {Gibb},\ and\ \citenamefont
  {Vente}}]{BattleJM9960601187}%
  \BibitemOpen
  \bibfield  {author} {\bibinfo {author} {\bibfnamefont {P.~D.}\ \bibnamefont
  {Battle}}, \bibinfo {author} {\bibfnamefont {C.~M.}\ \bibnamefont {Davison}},
  \bibinfo {author} {\bibfnamefont {T.~C.}\ \bibnamefont {Gibb}}, \ and\
  \bibinfo {author} {\bibfnamefont {J.s~F.}\ \bibnamefont {Vente}},\ }\bibfield
   {title} {\enquote {\bibinfo {title} {{S}tructural chemistry of
  \ce{SrMn_{1-x}Fe_xO_{3-\delta}}, $x\approx 0.3$},}\ }\href {\doibase
  10.1039/JM9960601187} {\bibfield  {journal} {\bibinfo  {journal} {J. Mater.
  Chem.}\ }\textbf {\bibinfo {volume} {6}},\ \bibinfo {pages} {1187--1190}
  (\bibinfo {year} {1996})}\BibitemShut {NoStop}%
\bibitem [{\citenamefont {Giannozzi}\ \emph {et~al.}(2009)\citenamefont
  {Giannozzi}, \citenamefont {Baroni}, \citenamefont {Bonini}, \citenamefont
  {Calandra}, \citenamefont {Car}, \citenamefont {Cavazzoni}, \citenamefont
  {Ceresoli}, \citenamefont {Chiarotti}, \citenamefont {Cococcioni},
  \citenamefont {Dabo}, \citenamefont {Corso}, \citenamefont {de~Gironcoli},
  \citenamefont {Fabris}, \citenamefont {Fratesi}, \citenamefont {Gebauer},
  \citenamefont {Gerstmann}, \citenamefont {Gougoussis}, \citenamefont
  {Kokalj}, \citenamefont {Lazzeri}, \citenamefont {Martin-Samos},
  \citenamefont {Marzari}, \citenamefont {Mauri}, \citenamefont {Mazzarello},
  \citenamefont {Paolini}, \citenamefont {Pasquarello}, \citenamefont
  {Paulatto}, \citenamefont {Sbraccia}, \citenamefont {Scandolo}, \citenamefont
  {Sclauzero}, \citenamefont {Seitsonen}, \citenamefont {Smogunov},
  \citenamefont {Umari},\ and\ \citenamefont
  {Wentzcovitch}}]{giannozzi2009quantum}%
  \BibitemOpen
  \bibfield  {author} {\bibinfo {author} {\bibfnamefont {P.}~\bibnamefont
  {Giannozzi}}, \bibinfo {author} {\bibfnamefont {S.}~\bibnamefont {Baroni}},
  \bibinfo {author} {\bibfnamefont {N.}~\bibnamefont {Bonini}}, \bibinfo
  {author} {\bibfnamefont {M.}~\bibnamefont {Calandra}}, \bibinfo {author}
  {\bibfnamefont {R.}~\bibnamefont {Car}}, \bibinfo {author} {\bibfnamefont
  {C.}~\bibnamefont {Cavazzoni}}, \bibinfo {author} {\bibfnamefont
  {D.}~\bibnamefont {Ceresoli}}, \bibinfo {author} {\bibfnamefont {G.~L.}\
  \bibnamefont {Chiarotti}}, \bibinfo {author} {\bibfnamefont {M.}~\bibnamefont
  {Cococcioni}}, \bibinfo {author} {\bibfnamefont {I.}~\bibnamefont {Dabo}},
  \bibinfo {author} {\bibfnamefont {A.~Dal}\ \bibnamefont {Corso}}, \bibinfo
  {author} {\bibfnamefont {S.}~\bibnamefont {de~Gironcoli}}, \bibinfo {author}
  {\bibfnamefont {S.}~\bibnamefont {Fabris}}, \bibinfo {author} {\bibfnamefont
  {G.}~\bibnamefont {Fratesi}}, \bibinfo {author} {\bibfnamefont
  {R.}~\bibnamefont {Gebauer}}, \bibinfo {author} {\bibfnamefont
  {U.}~\bibnamefont {Gerstmann}}, \bibinfo {author} {\bibfnamefont
  {C.}~\bibnamefont {Gougoussis}}, \bibinfo {author} {\bibfnamefont
  {A.}~\bibnamefont {Kokalj}}, \bibinfo {author} {\bibfnamefont
  {M.}~\bibnamefont {Lazzeri}}, \bibinfo {author} {\bibfnamefont
  {L.}~\bibnamefont {Martin-Samos}}, \bibinfo {author} {\bibfnamefont
  {N.}~\bibnamefont {Marzari}}, \bibinfo {author} {\bibfnamefont
  {F.}~\bibnamefont {Mauri}}, \bibinfo {author} {\bibfnamefont
  {R.}~\bibnamefont {Mazzarello}}, \bibinfo {author} {\bibfnamefont
  {S.}~\bibnamefont {Paolini}}, \bibinfo {author} {\bibfnamefont
  {A.}~\bibnamefont {Pasquarello}}, \bibinfo {author} {\bibfnamefont
  {L.}~\bibnamefont {Paulatto}}, \bibinfo {author} {\bibfnamefont
  {C.}~\bibnamefont {Sbraccia}}, \bibinfo {author} {\bibfnamefont
  {S.}~\bibnamefont {Scandolo}}, \bibinfo {author} {\bibfnamefont
  {G.}~\bibnamefont {Sclauzero}}, \bibinfo {author} {\bibfnamefont {A.~P.}\
  \bibnamefont {Seitsonen}}, \bibinfo {author} {\bibfnamefont {A.}~\bibnamefont
  {Smogunov}}, \bibinfo {author} {\bibfnamefont {P.}~\bibnamefont {Umari}}, \
  and\ \bibinfo {author} {\bibfnamefont {R.~M.}\ \bibnamefont {Wentzcovitch}},\
  }\bibfield  {title} {\enquote {\bibinfo {title} {\textsc{QUANTUM} {ESPRESSO}:
  a modular and open-source software project for quantum simulations of
  materials},}\ }\href {\doibase 10.1088/0953-8984/21/39/395502} {\bibfield
  {journal} {\bibinfo  {journal} {J. Phys.: Condens. Matter.}\ }\textbf
  {\bibinfo {volume} {21}},\ \bibinfo {pages} {395502} (\bibinfo {year}
  {2009})}\BibitemShut {NoStop}%
\bibitem [{\citenamefont {Giannozzi}\ \emph {et~al.}(2017)\citenamefont
  {Giannozzi}, \citenamefont {Andreussi}, \citenamefont {Brumme}, \citenamefont
  {Bunau}, \citenamefont {Buongiorno~Nardelli}, \citenamefont {Calandra},
  \citenamefont {Car}, \citenamefont {Cavazzoni}, \citenamefont {Ceresoli},
  \citenamefont {Cococcioni}, \citenamefont {Colonna}, \citenamefont
  {Carnimeo}, \citenamefont {Dal~Corso}, \citenamefont {De~Gironcoli},
  \citenamefont {Delugas}, \citenamefont {Distasio}, \citenamefont {Ferretti},
  \citenamefont {Floris}, \citenamefont {Fratesi}, \citenamefont {Fugallo},
  \citenamefont {Gebauer}, \citenamefont {Gerstmann}, \citenamefont {Giustino},
  \citenamefont {Gorni}, \citenamefont {Jia}, \citenamefont {Kawamura},
  \citenamefont {Ko}, \citenamefont {Kokalj}, \citenamefont
  {K{\"u}c{\"u}kbenli}, \citenamefont {Lazzeri}, \citenamefont {Marsili},
  \citenamefont {Marzari}, \citenamefont {Mauri}, \citenamefont {Nguyen},
  \citenamefont {Nguyen}, \citenamefont {Otero-De-La-Roza}, \citenamefont
  {Paulatto}, \citenamefont {Ponc{\'e}}, \citenamefont {Rocca}, \citenamefont
  {Sabatini}, \citenamefont {Santra}, \citenamefont {Schlipf}, \citenamefont
  {Seitsonen}, \citenamefont {Smogunov}, \citenamefont {Timrov}, \citenamefont
  {Thonhauser}, \citenamefont {Umari}, \citenamefont {Vast}, \citenamefont
  {Wu},\ and\ \citenamefont {Baroni}}]{Giannozzi2017}%
  \BibitemOpen
  \bibfield  {author} {\bibinfo {author} {\bibfnamefont {P.}~\bibnamefont
  {Giannozzi}}, \bibinfo {author} {\bibfnamefont {O.}~\bibnamefont
  {Andreussi}}, \bibinfo {author} {\bibfnamefont {T.}~\bibnamefont {Brumme}},
  \bibinfo {author} {\bibfnamefont {O.}~\bibnamefont {Bunau}}, \bibinfo
  {author} {\bibfnamefont {M.}~\bibnamefont {Buongiorno~Nardelli}}, \bibinfo
  {author} {\bibfnamefont {M.}~\bibnamefont {Calandra}}, \bibinfo {author}
  {\bibfnamefont {R.}~\bibnamefont {Car}}, \bibinfo {author} {\bibfnamefont
  {C.}~\bibnamefont {Cavazzoni}}, \bibinfo {author} {\bibfnamefont
  {D.}~\bibnamefont {Ceresoli}}, \bibinfo {author} {\bibfnamefont
  {M.}~\bibnamefont {Cococcioni}}, \bibinfo {author} {\bibfnamefont
  {N.}~\bibnamefont {Colonna}}, \bibinfo {author} {\bibfnamefont
  {I.}~\bibnamefont {Carnimeo}}, \bibinfo {author} {\bibfnamefont
  {A.}~\bibnamefont {Dal~Corso}}, \bibinfo {author} {\bibfnamefont
  {S.}~\bibnamefont {De~Gironcoli}}, \bibinfo {author} {\bibfnamefont
  {P.}~\bibnamefont {Delugas}}, \bibinfo {author} {\bibfnamefont {R.~A.}\
  \bibnamefont {Distasio}}, \bibinfo {author} {\bibfnamefont {A.}~\bibnamefont
  {Ferretti}}, \bibinfo {author} {\bibfnamefont {A.}~\bibnamefont {Floris}},
  \bibinfo {author} {\bibfnamefont {G.}~\bibnamefont {Fratesi}}, \bibinfo
  {author} {\bibfnamefont {G.}~\bibnamefont {Fugallo}}, \bibinfo {author}
  {\bibfnamefont {R.}~\bibnamefont {Gebauer}}, \bibinfo {author} {\bibfnamefont
  {U.}~\bibnamefont {Gerstmann}}, \bibinfo {author} {\bibfnamefont
  {F.}~\bibnamefont {Giustino}}, \bibinfo {author} {\bibfnamefont
  {T.}~\bibnamefont {Gorni}}, \bibinfo {author} {\bibfnamefont
  {J.}~\bibnamefont {Jia}}, \bibinfo {author} {\bibfnamefont {M.}~\bibnamefont
  {Kawamura}}, \bibinfo {author} {\bibfnamefont {H.~Y.}\ \bibnamefont {Ko}},
  \bibinfo {author} {\bibfnamefont {A.}~\bibnamefont {Kokalj}}, \bibinfo
  {author} {\bibfnamefont {E.}~\bibnamefont {K{\"u}c{\"u}kbenli}}, \bibinfo
  {author} {\bibfnamefont {M.}~\bibnamefont {Lazzeri}}, \bibinfo {author}
  {\bibfnamefont {M.}~\bibnamefont {Marsili}}, \bibinfo {author} {\bibfnamefont
  {N.}~\bibnamefont {Marzari}}, \bibinfo {author} {\bibfnamefont
  {F.}~\bibnamefont {Mauri}}, \bibinfo {author} {\bibfnamefont {N.~L.}\
  \bibnamefont {Nguyen}}, \bibinfo {author} {\bibfnamefont {H.~V.}\
  \bibnamefont {Nguyen}}, \bibinfo {author} {\bibfnamefont {A.}~\bibnamefont
  {Otero-De-La-Roza}}, \bibinfo {author} {\bibfnamefont {L.}~\bibnamefont
  {Paulatto}}, \bibinfo {author} {\bibfnamefont {S.}~\bibnamefont {Ponc{\'e}}},
  \bibinfo {author} {\bibfnamefont {D.}~\bibnamefont {Rocca}}, \bibinfo
  {author} {\bibfnamefont {R.}~\bibnamefont {Sabatini}}, \bibinfo {author}
  {\bibfnamefont {B.}~\bibnamefont {Santra}}, \bibinfo {author} {\bibfnamefont
  {M.}~\bibnamefont {Schlipf}}, \bibinfo {author} {\bibfnamefont {A.~P.}\
  \bibnamefont {Seitsonen}}, \bibinfo {author} {\bibfnamefont {A.}~\bibnamefont
  {Smogunov}}, \bibinfo {author} {\bibfnamefont {I.}~\bibnamefont {Timrov}},
  \bibinfo {author} {\bibfnamefont {T.}~\bibnamefont {Thonhauser}}, \bibinfo
  {author} {\bibfnamefont {P.}~\bibnamefont {Umari}}, \bibinfo {author}
  {\bibfnamefont {N.}~\bibnamefont {Vast}}, \bibinfo {author} {\bibfnamefont
  {X.}~\bibnamefont {Wu}}, \ and\ \bibinfo {author} {\bibfnamefont
  {S.}~\bibnamefont {Baroni}},\ }\bibfield  {title} {\enquote {\bibinfo {title}
  {{A}dvanced capabilities for materials modelling with \textsc{QUANTUM}
  {ESPRESSO}},}\ }\href {\doibase 10.1088/1361-648X/aa8f79} {\bibfield
  {journal} {\bibinfo  {journal} {J. Phys.: Condens. Matter.}\ }\textbf
  {\bibinfo {volume} {29}},\ \bibinfo {pages} {465901} (\bibinfo {year}
  {2017})}\BibitemShut {NoStop}%
\bibitem [{\citenamefont {Perdew}\ \emph {et~al.}(2008)\citenamefont {Perdew},
  \citenamefont {Ruzsinszky}, \citenamefont {Csonka}, \citenamefont {Vydrov},
  \citenamefont {Scuseria}, \citenamefont {Constantin}, \citenamefont {Zhou},\
  and\ \citenamefont {Burke}}]{perdew2008pbesol}%
  \BibitemOpen
  \bibfield  {author} {\bibinfo {author} {\bibfnamefont {J.~P}\ \bibnamefont
  {Perdew}}, \bibinfo {author} {\bibfnamefont {A.}~\bibnamefont {Ruzsinszky}},
  \bibinfo {author} {\bibfnamefont {G.~I.}\ \bibnamefont {Csonka}}, \bibinfo
  {author} {\bibfnamefont {O.~A.}\ \bibnamefont {Vydrov}}, \bibinfo {author}
  {\bibfnamefont {G.~E.}\ \bibnamefont {Scuseria}}, \bibinfo {author}
  {\bibfnamefont {L.~A.}\ \bibnamefont {Constantin}}, \bibinfo {author}
  {\bibfnamefont {X.}~\bibnamefont {Zhou}}, \ and\ \bibinfo {author}
  {\bibfnamefont {K.}~\bibnamefont {Burke}},\ }\bibfield  {title} {\enquote
  {\bibinfo {title} {{R}estoring the density-gradient expansion for exchange in
  solids and surfaces},}\ }\href {\doibase 10.1103/PhysRevLett.100.136406}
  {\bibfield  {journal} {\bibinfo  {journal} {Phys. Rev. Lett.}\ }\textbf
  {\bibinfo {volume} {100}},\ \bibinfo {pages} {136406} (\bibinfo {year}
  {2008})}\BibitemShut {NoStop}%
\bibitem [{\citenamefont {Vanderbilt}(1990)}]{vanderbilt1990soft}%
  \BibitemOpen
  \bibfield  {author} {\bibinfo {author} {\bibfnamefont {D.}~\bibnamefont
  {Vanderbilt}},\ }\bibfield  {title} {\enquote {\bibinfo {title} {{S}oft
  self-consistent pseudopotentials in a generalized eigenvalue formalism},}\
  }\href {\doibase 10.1103/PhysRevB.41.7892} {\bibfield  {journal} {\bibinfo
  {journal} {Phys. Rev. B}\ }\textbf {\bibinfo {volume} {41}},\ \bibinfo
  {pages} {7892} (\bibinfo {year} {1990})}\BibitemShut {NoStop}%
\bibitem [{Note1()}]{Note1}%
  \BibitemOpen
  \bibinfo {note} {Ultrasoft pseudopotentials from the PSLibrary were taken
  from \protect \url {www.materialscloud.org}:
  Sr.pbesol-spn-rrkjus\textunderscore psl.1.0.0.UPF,
  Mn.pbesol-spn-rrkjus\textunderscore psl.0.3.1.UPF, and
  O.pbesol-n-rrkjus\textunderscore psl.1.0.0.UPF}\BibitemShut {NoStop}%
\bibitem [{\citenamefont {Monkhorst}\ and\ \citenamefont
  {Pack}(1976)}]{monkhorst1976special}%
  \BibitemOpen
  \bibfield  {author} {\bibinfo {author} {\bibfnamefont {H.~J.}\ \bibnamefont
  {Monkhorst}}\ and\ \bibinfo {author} {\bibfnamefont {J.~D.}\ \bibnamefont
  {Pack}},\ }\bibfield  {title} {\enquote {\bibinfo {title} {{S}pecial points
  for {B}rillouin-zone integrations},}\ }\href {\doibase
  10.1103/PhysRevB.13.5188} {\bibfield  {journal} {\bibinfo  {journal} {Phys.
  Rev. B}\ }\textbf {\bibinfo {volume} {13}},\ \bibinfo {pages} {5188}
  (\bibinfo {year} {1976})}\BibitemShut {NoStop}%
\bibitem [{\citenamefont {Rondinelli}\ and\ \citenamefont
  {Spaldin}(2011)}]{Rondinelli:2011jk}%
  \BibitemOpen
  \bibfield  {author} {\bibinfo {author} {\bibfnamefont {J.~M.}\ \bibnamefont
  {Rondinelli}}\ and\ \bibinfo {author} {\bibfnamefont {N.~A.}\ \bibnamefont
  {Spaldin}},\ }\bibfield  {title} {\enquote {\bibinfo {title} {{S}tructure and
  properties of functional oxide thin films: {I}nsights from
  electronic-structure calculations},}\ }\href {\doibase
  doi:10.1002/adma.201101152} {\bibfield  {journal} {\bibinfo  {journal} {Adv.
  Mater.}\ }\textbf {\bibinfo {volume} {23}},\ \bibinfo {pages} {3363--3381}
  (\bibinfo {year} {2011})}\BibitemShut {NoStop}%
\bibitem [{\citenamefont {Kr{\"o}ger}\ and\ \citenamefont
  {Vink}(1956)}]{KROGER1956307}%
  \BibitemOpen
  \bibfield  {author} {\bibinfo {author} {\bibfnamefont {F.~A.}\ \bibnamefont
  {Kr{\"o}ger}}\ and\ \bibinfo {author} {\bibfnamefont {H.~J.}\ \bibnamefont
  {Vink}},\ }\bibfield  {title} {\enquote {\bibinfo {title} {{R}elations
  between the concentrations of imperfections in crystalline solids},}\ }\href
  {\doibase 10.1016/S0081-1947(08)60135-6} {\bibfield  {journal} {\bibinfo
  {journal} {Solid State Phys.}\ }\textbf {\bibinfo {volume} {3}},\ \bibinfo
  {pages} {307--435} (\bibinfo {year} {1956})}\BibitemShut {NoStop}%
\bibitem [{\citenamefont {Anisimov}\ \emph {et~al.}(1991)\citenamefont
  {Anisimov}, \citenamefont {Zaanen},\ and\ \citenamefont
  {Andersen}}]{anisimov1991band}%
  \BibitemOpen
  \bibfield  {author} {\bibinfo {author} {\bibfnamefont {V.~I.}\ \bibnamefont
  {Anisimov}}, \bibinfo {author} {\bibfnamefont {J.}~\bibnamefont {Zaanen}}, \
  and\ \bibinfo {author} {\bibfnamefont {O.~K.}\ \bibnamefont {Andersen}},\
  }\bibfield  {title} {\enquote {\bibinfo {title} {{B}and theory and {M}ott
  insulators: {H}ubbard {U} instead of {S}toner {I}},}\ }\href {\doibase
  10.1103/PhysRevB.44.943} {\bibfield  {journal} {\bibinfo  {journal} {Phys.
  Rev. B}\ }\textbf {\bibinfo {volume} {44}},\ \bibinfo {pages} {943} (\bibinfo
  {year} {1991})}\BibitemShut {NoStop}%
\bibitem [{\citenamefont {Anisimov}\ \emph {et~al.}(1997)\citenamefont
  {Anisimov}, \citenamefont {Poteryaev}, \citenamefont {Korotin}, \citenamefont
  {Anokhin},\ and\ \citenamefont {Kotliar}}]{anisimov1997first}%
  \BibitemOpen
  \bibfield  {author} {\bibinfo {author} {\bibfnamefont {V.~I.}\ \bibnamefont
  {Anisimov}}, \bibinfo {author} {\bibfnamefont {A.~I.}\ \bibnamefont
  {Poteryaev}}, \bibinfo {author} {\bibfnamefont {M.~A.}\ \bibnamefont
  {Korotin}}, \bibinfo {author} {\bibfnamefont {A.~O.}\ \bibnamefont
  {Anokhin}}, \ and\ \bibinfo {author} {\bibfnamefont {G.}~\bibnamefont
  {Kotliar}},\ }\bibfield  {title} {\enquote {\bibinfo {title}
  {{F}irst-principles calculations of the electronic structure and spectra of
  strongly correlated systems: dynamical mean-field theory},}\ }\href {\doibase
  10.1088/0953-8984/9/35/010} {\bibfield  {journal} {\bibinfo  {journal} {J.
  Phys.: Condens. Matter.}\ }\textbf {\bibinfo {volume} {9}},\ \bibinfo {pages}
  {7359} (\bibinfo {year} {1997})}\BibitemShut {NoStop}%
\bibitem [{\citenamefont {Dudarev}\ \emph {et~al.}(1998)\citenamefont
  {Dudarev}, \citenamefont {Botton}, \citenamefont {Savrasov}, \citenamefont
  {Humphreys},\ and\ \citenamefont {Sutton}}]{Dudarev1998}%
  \BibitemOpen
  \bibfield  {author} {\bibinfo {author} {\bibfnamefont {S.~L.}\ \bibnamefont
  {Dudarev}}, \bibinfo {author} {\bibfnamefont {G.~A.}\ \bibnamefont {Botton}},
  \bibinfo {author} {\bibfnamefont {S.~Y.}\ \bibnamefont {Savrasov}}, \bibinfo
  {author} {\bibfnamefont {C.~J}\ \bibnamefont {Humphreys}}, \ and\ \bibinfo
  {author} {\bibfnamefont {A.~P.}\ \bibnamefont {Sutton}},\ }\bibfield  {title}
  {\enquote {\bibinfo {title} {{E}lectron-energy-loss spectra and the
  structural stability of nickel oxide: {A}n {LSDA+U} study},}\ }\href
  {\doibase 10.1103/PhysRevB.57.1505} {\bibfield  {journal} {\bibinfo
  {journal} {Phys. Rev. B}\ }\textbf {\bibinfo {volume} {57}},\ \bibinfo
  {pages} {1505} (\bibinfo {year} {1998})}\BibitemShut {NoStop}%
\bibitem [{\citenamefont {Ricca}\ \emph {et~al.}(2019)\citenamefont {Ricca},
  \citenamefont {Timrov}, \citenamefont {Cococcioni}, \citenamefont {Marzari},\
  and\ \citenamefont {Aschauer}}]{Ricca2019}%
  \BibitemOpen
  \bibfield  {author} {\bibinfo {author} {\bibfnamefont {C.}~\bibnamefont
  {Ricca}}, \bibinfo {author} {\bibfnamefont {I.}~\bibnamefont {Timrov}},
  \bibinfo {author} {\bibfnamefont {M.}~\bibnamefont {Cococcioni}}, \bibinfo
  {author} {\bibfnamefont {N.}~\bibnamefont {Marzari}}, \ and\ \bibinfo
  {author} {\bibfnamefont {U.}~\bibnamefont {Aschauer}},\ }\bibfield  {title}
  {\enquote {\bibinfo {title} {{S}elf-consistent site-dependent {DFT}+u study
  of stoichiometric and defective \ce{SrMnO3}},}\ }\href {\doibase
  10.1103/PhysRevB.99.094102} {\bibfield  {journal} {\bibinfo  {journal} {Phys.
  Rev. B}\ }\textbf {\bibinfo {volume} {99}},\ \bibinfo {pages} {094102}
  (\bibinfo {year} {2019})}\BibitemShut {NoStop}%
\bibitem [{\citenamefont {Timrov}\ \emph {et~al.}(2018)\citenamefont {Timrov},
  \citenamefont {Marzari},\ and\ \citenamefont {Cococcioni}}]{Timrov2018}%
  \BibitemOpen
  \bibfield  {author} {\bibinfo {author} {\bibfnamefont {I.}~\bibnamefont
  {Timrov}}, \bibinfo {author} {\bibfnamefont {N.}~\bibnamefont {Marzari}}, \
  and\ \bibinfo {author} {\bibfnamefont {M.}~\bibnamefont {Cococcioni}},\
  }\bibfield  {title} {\enquote {\bibinfo {title} {{H}ubbard parameters from
  density-functional perturbation theory},}\ }\href {\doibase
  10.1103/PhysRevB.98.085127} {\bibfield  {journal} {\bibinfo  {journal} {Phys.
  Rev. B}\ }\textbf {\bibinfo {volume} {98}},\ \bibinfo {pages} {085127}
  (\bibinfo {year} {2018})}\BibitemShut {NoStop}%
\bibitem [{\citenamefont {Freysoldt}\ \emph {et~al.}(2014)\citenamefont
  {Freysoldt}, \citenamefont {Grabowski}, \citenamefont {Hickel}, \citenamefont
  {Neugebauer}, \citenamefont {Kresse}, \citenamefont {Janotti},\ and\
  \citenamefont {Van~de Walle}}]{freysoldt2014first}%
  \BibitemOpen
  \bibfield  {author} {\bibinfo {author} {\bibfnamefont {C.}~\bibnamefont
  {Freysoldt}}, \bibinfo {author} {\bibfnamefont {B.}~\bibnamefont
  {Grabowski}}, \bibinfo {author} {\bibfnamefont {T.}~\bibnamefont {Hickel}},
  \bibinfo {author} {\bibfnamefont {J.}~\bibnamefont {Neugebauer}}, \bibinfo
  {author} {\bibfnamefont {G.}~\bibnamefont {Kresse}}, \bibinfo {author}
  {\bibfnamefont {A.}~\bibnamefont {Janotti}}, \ and\ \bibinfo {author}
  {\bibfnamefont {C.~G.}\ \bibnamefont {Van~de Walle}},\ }\bibfield  {title}
  {\enquote {\bibinfo {title} {{F}irst-principles calculations for point
  defects in solids},}\ }\href {\doibase 10.1103/RevModPhys.86.253} {\bibfield
  {journal} {\bibinfo  {journal} {Rev. Mod. Phys.}\ }\textbf {\bibinfo {volume}
  {86}},\ \bibinfo {pages} {253} (\bibinfo {year} {2014})}\BibitemShut
  {NoStop}%
\bibitem [{\citenamefont {Lany}\ and\ \citenamefont {Zunger}(2008)}]{lany2008}%
  \BibitemOpen
  \bibfield  {author} {\bibinfo {author} {\bibfnamefont {S.}~\bibnamefont
  {Lany}}\ and\ \bibinfo {author} {\bibfnamefont {A.}~\bibnamefont {Zunger}},\
  }\bibfield  {title} {\enquote {\bibinfo {title} {{A}ssessment of correction
  methods for the band-gap problem and for finite-size effects in supercell
  defect calculations: {C}ase studies for \ce{ZnO} and \ce{GaAs}},}\ }\href
  {\doibase 10.1103/PhysRevB.78.235104} {\bibfield  {journal} {\bibinfo
  {journal} {Phys. Rev. B}\ }\textbf {\bibinfo {volume} {78}},\ \bibinfo
  {pages} {235104} (\bibinfo {year} {2008})}\BibitemShut {NoStop}%
\bibitem [{\citenamefont {Van~de Walle}\ and\ \citenamefont
  {Neugebauer}(2004)}]{vanderwalle2004sol}%
  \BibitemOpen
  \bibfield  {author} {\bibinfo {author} {\bibfnamefont {C.~G.}\ \bibnamefont
  {Van~de Walle}}\ and\ \bibinfo {author} {\bibfnamefont {J.}~\bibnamefont
  {Neugebauer}},\ }\bibfield  {title} {\enquote {\bibinfo {title}
  {{F}irst-principles calculations for defects and impurities: {A}pplications
  to {III}-nitrides},}\ }\href {\doibase 10.1063/1.1682673} {\bibfield
  {journal} {\bibinfo  {journal} {Journal of Applied Physics}\ }\textbf
  {\bibinfo {volume} {95}},\ \bibinfo {pages} {3851--3879} (\bibinfo {year}
  {2004})}\BibitemShut {NoStop}%
\bibitem [{\citenamefont {Jain}\ \emph {et~al.}(2011)\citenamefont {Jain},
  \citenamefont {Hautier}, \citenamefont {Ong}, \citenamefont {Moore},
  \citenamefont {Fischer}, \citenamefont {Persson},\ and\ \citenamefont
  {Ceder}}]{ceder20011}%
  \BibitemOpen
  \bibfield  {author} {\bibinfo {author} {\bibfnamefont {A.}~\bibnamefont
  {Jain}}, \bibinfo {author} {\bibfnamefont {G.}~\bibnamefont {Hautier}},
  \bibinfo {author} {\bibfnamefont {S.~P.}\ \bibnamefont {Ong}}, \bibinfo
  {author} {\bibfnamefont {C.~J.}\ \bibnamefont {Moore}}, \bibinfo {author}
  {\bibfnamefont {C.~C.}\ \bibnamefont {Fischer}}, \bibinfo {author}
  {\bibfnamefont {K.~A.}\ \bibnamefont {Persson}}, \ and\ \bibinfo {author}
  {\bibfnamefont {G.}~\bibnamefont {Ceder}},\ }\bibfield  {title} {\enquote
  {\bibinfo {title} {{F}ormation enthalpies by mixing {GGA} and {GGA+U}
  calculations},}\ }\href {\doibase 10.1103/PhysRevB.84.045115} {\bibfield
  {journal} {\bibinfo  {journal} {Phys. Rev. B}\ }\textbf {\bibinfo {volume}
  {84}},\ \bibinfo {pages} {045115} (\bibinfo {year} {2011})}\BibitemShut
  {NoStop}%
\bibitem [{\citenamefont {Sit}\ \emph {et~al.}(2011)\citenamefont {Sit},
  \citenamefont {Car}, \citenamefont {Cohen},\ and\ \citenamefont
  {Selloni}}]{Sit2011}%
  \BibitemOpen
  \bibfield  {author} {\bibinfo {author} {\bibfnamefont {P.~H.~L.}\
  \bibnamefont {Sit}}, \bibinfo {author} {\bibfnamefont {R.}~\bibnamefont
  {Car}}, \bibinfo {author} {\bibfnamefont {M.~H.}\ \bibnamefont {Cohen}}, \
  and\ \bibinfo {author} {\bibfnamefont {A.}~\bibnamefont {Selloni}},\
  }\bibfield  {title} {\enquote {\bibinfo {title} {{S}imple, unambiguous
  theoretical approach to oxidation state determination via first-principles
  calculations},}\ }\href {\doibase 10.1021/ic2013107} {\bibfield  {journal}
  {\bibinfo  {journal} {Inorganic Chemistry}\ }\textbf {\bibinfo {volume}
  {50}},\ \bibinfo {pages} {10259--10267} (\bibinfo {year} {2011})}\BibitemShut
  {NoStop}%
\bibitem [{\citenamefont {King-Smith}\ and\ \citenamefont
  {Vanderbilt}(1993)}]{berryphase1}%
  \BibitemOpen
  \bibfield  {author} {\bibinfo {author} {\bibfnamefont {R.~D.}\ \bibnamefont
  {King-Smith}}\ and\ \bibinfo {author} {\bibfnamefont {D.}~\bibnamefont
  {Vanderbilt}},\ }\bibfield  {title} {\enquote {\bibinfo {title} {{T}heory of
  polarization of crystalline solids},}\ }\href {\doibase
  10.1103/PhysRevB.47.1651} {\bibfield  {journal} {\bibinfo  {journal} {Phys.
  Rev. B}\ }\textbf {\bibinfo {volume} {47}},\ \bibinfo {pages} {1651--1654}
  (\bibinfo {year} {1993})}\BibitemShut {NoStop}%
\bibitem [{\citenamefont {Vanderbilt}\ and\ \citenamefont
  {King-Smith}(1993)}]{berryphase2}%
  \BibitemOpen
  \bibfield  {author} {\bibinfo {author} {\bibfnamefont {D.}~\bibnamefont
  {Vanderbilt}}\ and\ \bibinfo {author} {\bibfnamefont {R.~D.}\ \bibnamefont
  {King-Smith}},\ }\bibfield  {title} {\enquote {\bibinfo {title} {{E}lectric
  polarization as a bulk quantity and its relation to surface charge},}\ }\href
  {\doibase 10.1103/PhysRevB.48.4442} {\bibfield  {journal} {\bibinfo
  {journal} {Phys. Rev. B}\ }\textbf {\bibinfo {volume} {48}},\ \bibinfo
  {pages} {4442--4455} (\bibinfo {year} {1993})}\BibitemShut {NoStop}%
\bibitem [{\citenamefont {Aschauer}\ \emph {et~al.}(2013)\citenamefont
  {Aschauer}, \citenamefont {Pfenninger}, \citenamefont {Selbach},
  \citenamefont {Grande},\ and\ \citenamefont {Spaldin}}]{aschauer2013strain}%
  \BibitemOpen
  \bibfield  {author} {\bibinfo {author} {\bibfnamefont {U.}~\bibnamefont
  {Aschauer}}, \bibinfo {author} {\bibfnamefont {R.}~\bibnamefont
  {Pfenninger}}, \bibinfo {author} {\bibfnamefont {S.~M.}\ \bibnamefont
  {Selbach}}, \bibinfo {author} {\bibfnamefont {T.}~\bibnamefont {Grande}}, \
  and\ \bibinfo {author} {\bibfnamefont {N.~A.}\ \bibnamefont {Spaldin}},\
  }\bibfield  {title} {\enquote {\bibinfo {title} {{S}train-controlled oxygen
  vacancy formation and ordering in \ce{CaMnO3}},}\ }\href {\doibase
  10.1103/PhysRevB.88.054111} {\bibfield  {journal} {\bibinfo  {journal} {Phys.
  Rev. B}\ }\textbf {\bibinfo {volume} {88}},\ \bibinfo {pages} {054111}
  (\bibinfo {year} {2013})}\BibitemShut {NoStop}%
\bibitem [{\citenamefont {Aschauer}\ and\ \citenamefont
  {Spaldin}(2016)}]{aschauer2016interplay}%
  \BibitemOpen
  \bibfield  {author} {\bibinfo {author} {\bibfnamefont {U.}~\bibnamefont
  {Aschauer}}\ and\ \bibinfo {author} {\bibfnamefont {N.~A.}\ \bibnamefont
  {Spaldin}},\ }\bibfield  {title} {\enquote {\bibinfo {title} {{I}nterplay
  between strain, defect charge state, and functionality in complex oxides},}\
  }\href {\doibase 10.1063/1.4958716} {\bibfield  {journal} {\bibinfo
  {journal} {Appl. Phys. lett.}\ }\textbf {\bibinfo {volume} {109}},\ \bibinfo
  {pages} {031901} (\bibinfo {year} {2016})}\BibitemShut {NoStop}%
\bibitem [{\citenamefont {Adler}(2004)}]{adler2004}%
  \BibitemOpen
  \bibfield  {author} {\bibinfo {author} {\bibfnamefont {S.~B.}\ \bibnamefont
  {Adler}},\ }\bibfield  {title} {\enquote {\bibinfo {title} {{F}actors
  governing oxygen reduction in solid oxide fuel cell cathodes},}\ }\href
  {\doibase 10.1021/cr020724o} {\bibfield  {journal} {\bibinfo  {journal}
  {Chem. Rev.}\ }\textbf {\bibinfo {volume} {104}},\ \bibinfo {pages}
  {4791--4844} (\bibinfo {year} {2004})}\BibitemShut {NoStop}%
\bibitem [{\citenamefont {Ederer}\ and\ \citenamefont
  {Spaldin}(2005)}]{Ederer2005Effect}%
  \BibitemOpen
  \bibfield  {author} {\bibinfo {author} {\bibfnamefont {C.}~\bibnamefont
  {Ederer}}\ and\ \bibinfo {author} {\bibfnamefont {N.~A.}\ \bibnamefont
  {Spaldin}},\ }\bibfield  {title} {\enquote {\bibinfo {title} {{E}ffect of
  epitaxial strain on the spontaneous polarization of thin film
  ferroelectrics},}\ }\href {\doibase 10.1103/PhysRevLett.95.257601} {\bibfield
   {journal} {\bibinfo  {journal} {Phys. Rev. Lett.}\ }\textbf {\bibinfo
  {volume} {95}},\ \bibinfo {pages} {257601} (\bibinfo {year}
  {2005})}\BibitemShut {NoStop}%
\bibitem [{\citenamefont {Varvenne}\ \emph {et~al.}(2013)\citenamefont
  {Varvenne}, \citenamefont {Bruneval}, \citenamefont {Marinica},\ and\
  \citenamefont {Clouet}}]{varvenne2013}%
  \BibitemOpen
  \bibfield  {author} {\bibinfo {author} {\bibfnamefont {C.}~\bibnamefont
  {Varvenne}}, \bibinfo {author} {\bibfnamefont {F.}~\bibnamefont {Bruneval}},
  \bibinfo {author} {\bibfnamefont {M.-C.}\ \bibnamefont {Marinica}}, \ and\
  \bibinfo {author} {\bibfnamefont {E.}~\bibnamefont {Clouet}},\ }\bibfield
  {title} {\enquote {\bibinfo {title} {{P}oint defect modeling in materials:
  {C}oupling ab initio and elasticity approaches},}\ }\href {\doibase
  10.1103/PhysRevB.88.134102} {\bibfield  {journal} {\bibinfo  {journal} {Phys.
  Rev. B}\ }\textbf {\bibinfo {volume} {88}},\ \bibinfo {pages} {134102}
  (\bibinfo {year} {2013})}\BibitemShut {NoStop}%
\end{thebibliography}%

\clearpage

\renewcommand{\thetable}{S\arabic{table}} 
\setcounter{table}{0}
\renewcommand{\thefigure}{S\arabic{figure}}
\setcounter{figure}{0}
\renewcommand{\thesection}{S\arabic{section}}
\setcounter{section}{0}
\renewcommand{\theequation}{S\arabic{equation}}
\setcounter{equation}{0}
\onecolumngrid

\begin{center}
\textbf{Supplementary information for\\\vspace{0.5 cm}
\large Interplay between polarization, strain and defect-pairs in Fe-doped \ce{SrMnO_{3-\delta}}\\\vspace{0.3 cm}}
Chiara Ricca,$^{1, 2}$ and Ulrich Aschauer$^{1, 2}$

\small
$^1$\textit{Department of Chemistry and Biochemistry, University of Bern, Freiestrasse 3, CH-3012 Bern, Switzerland}

$^2$\textit{National Centre for Computational Design and Discovery of Novel Materials (MARVEL), Switzerland}

(Dated: \today)
\end{center}

\section{Self-consistent site-dependent DFT+$U_\mathrm{SC-SD}$\label{sec:Uscsd}}

\subsection{Hubbard correction for Mn: $U_{\ce{Mn}}$}

Defect formation in transition-metal materials can result in local perturbations of the chemical environment of Hubbard sites around the defect, upon which the Hubbard parameters physically depend. For this reason, we recently proposed a self-consistent and site-dependent DFT+$U_\textrm{SC-SD}$ approach in which the $U$ values are computed for all inequivalent Hubbard sites~\citeSI{Ricca2019SI}. DFT+$U_\textrm{SC-SD}$ was found to be promising to predict properties of defective systems, in particular when excess-charge localization is restricted to atoms around the defect site and can be properly captured by site-dependent $U$ values. For example, in the case of \ce{V_O} in SMO, $U$ values were found to depend on the distance of the Hubbard site from the defect, its coordination number, its oxidation state, and on the magnetic order of the host material~\citeSI{Ricca2019SI}.

\begin{figure}[b]
	\centering
	\includegraphics[width=0.35\columnwidth]{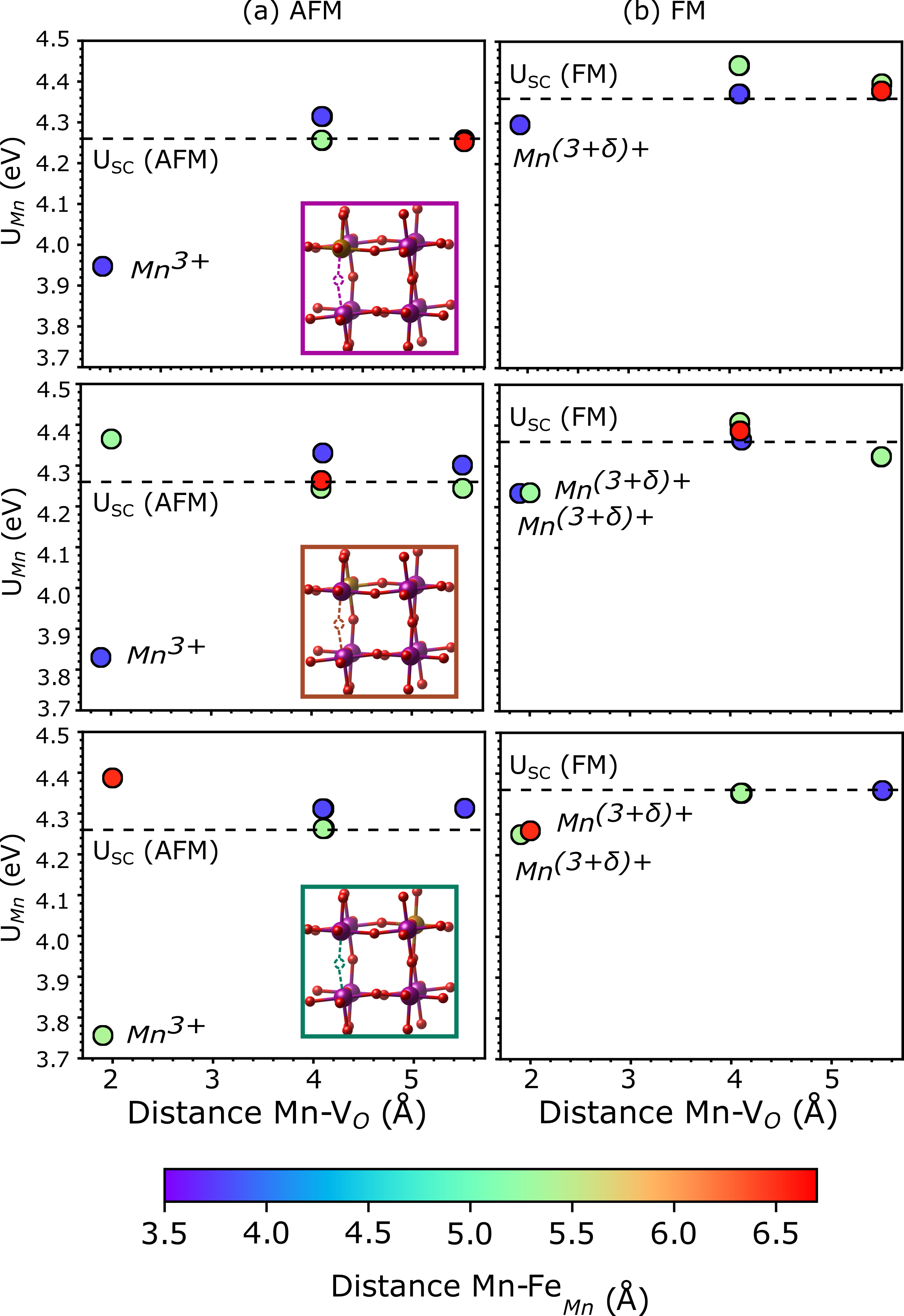}
	\caption{$U_{\textrm{SC-SD}}$ values reported as a function of the distance of each Hubbard site from the \ce{V_O} and the \ce{Fe_{Mn}} defects, computed for all Mn atoms in SMO with a NN (top), NNN (middle) and NNNN (bottom) \ce{Fe_{Mn}-V_O^{OP}} defect pair in the (a) AFM and (b) FM phases. The oxidation state is explicitly indicated for (partially) reduced Mn sites. The insets show the structure of the corresponding configuration, where the red, violet, gold, and dashed spheres represent the O, Mn, Fe atoms and the \ce{V_O}, respectively. The values of $U_{\textrm{SC}}$ for the stoichiometric phases are indicated by horizontal dashed lines for comparison.}
	\label{fig:USCSD}
\end{figure}
Fig.~\ref{fig:USCSD} shows the $U_\textrm{SC-SD}$ obtained for all inequivalent Mn Hubbard sites in SMO with \ce{Fe_{Mn}-V_O^{OP}} defect pairs, where the substitutional Fe is in NN (top), NNN (middle), or NNNN (bottom) position to \ce{V_O^{OP}}. Hubbard parameters are reported for each site as a function of the distance of the Mn site from the \ce{Fe_{Mn}} (via the color of the data points) and \ce{V_O^{OP}} defects.

\subsubsection{AFM SMO}

\begin{figure}[h]
 \centering
 \includegraphics[width=0.35\columnwidth]{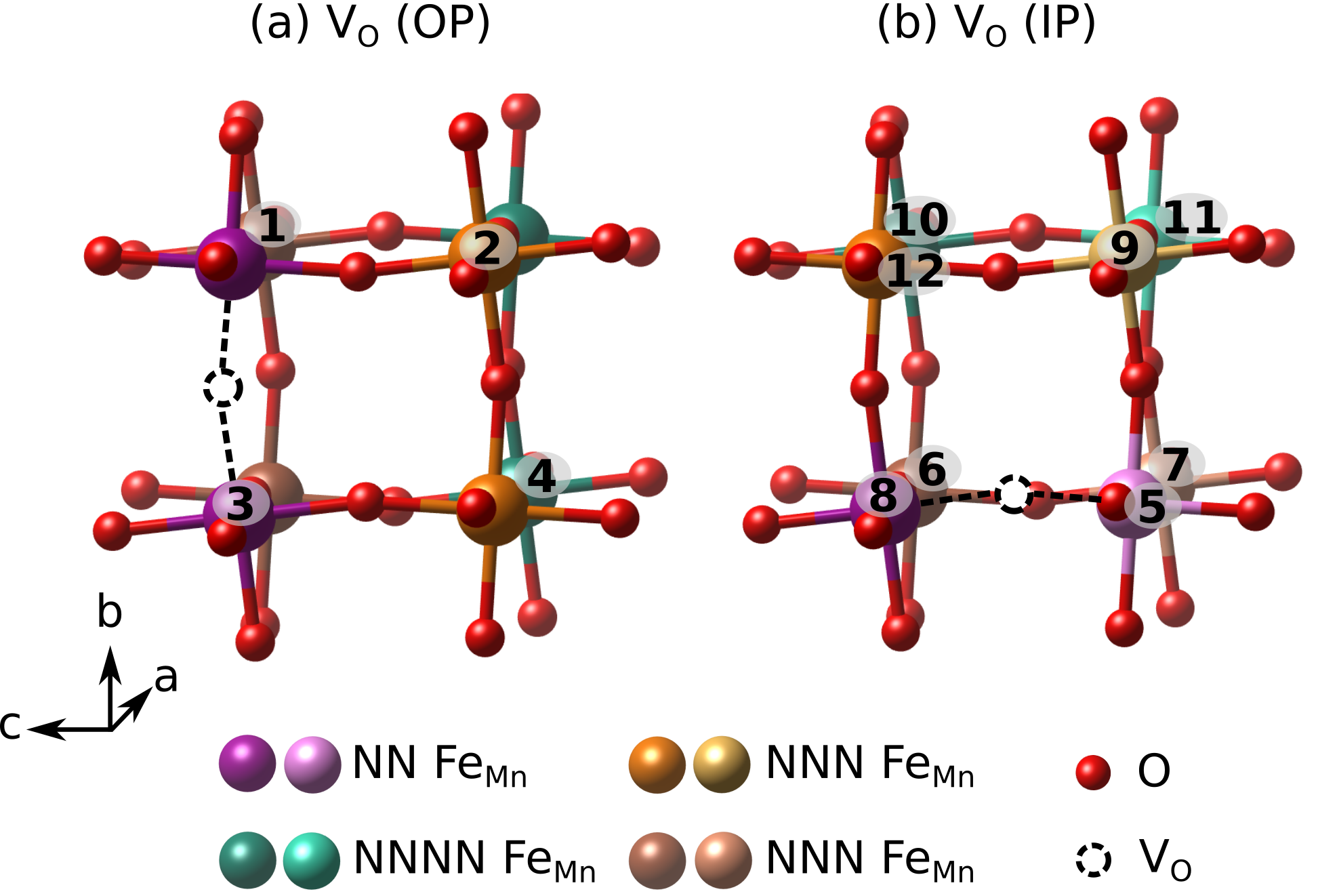}
 \caption{Defect-pair configurations as in the main text, numbered according to the location of the \ce{Fe} substitution site.}
\label{fig:SMO_structure_defect_SI}
\end{figure}

\begin{figure}[h]
 \centering
 \includegraphics[width=\columnwidth]{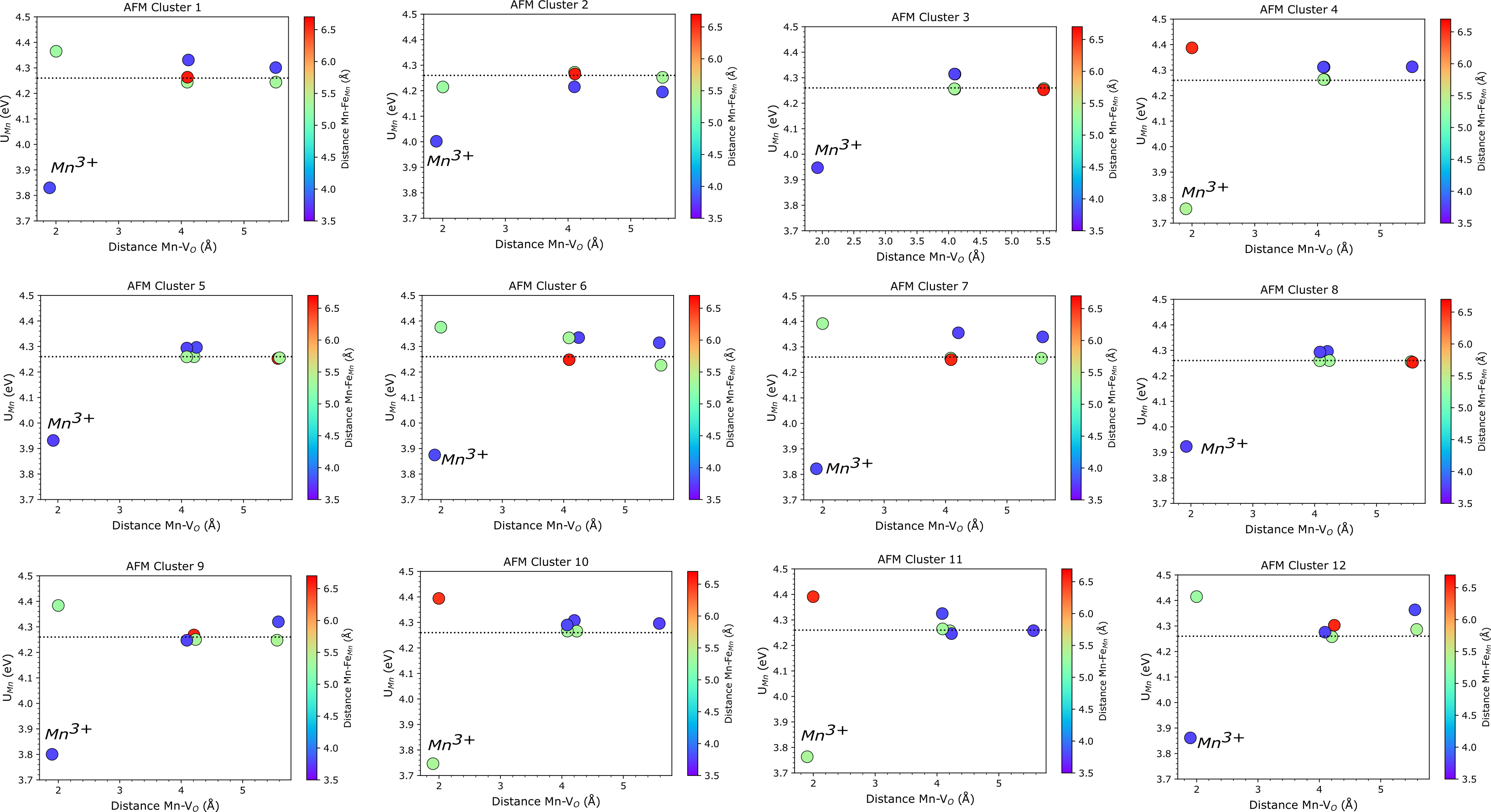}
 \caption{$U_{\textrm{SC-SD}}$ values reported as a function of the distance of each Hubbard site from \ce{Fe_{Mn}} and \ce{V_O} defects computed for the Mn atoms in AFM SMO structures in each of the 12 \ce{Fe_{Mn}-V_O} configurations shown in Fig.~\ref{fig:SMO_structure_defect_SI}. The oxidation state is indicated for the reduced Mn sites.}
\label{fig:USCSD_AFM_SI}
\end{figure}

Results for AFM SMO suggest that the distance from the \ce{V_O} has the strongest impact on the computed $U$ values (cf. Fig.~\ref{fig:USCSD}a). In fact, we observe deviations from the self-consistent $U$ computed for the stoichiometric cell ($U_\textrm{SC}$), mainly for Mn atoms adjacent to \ce{V_O} at a distance of about~1.90\AA, while Mn sites at larger distances recover $U_\textrm{SC}$ (the deviations are as small as 0.02 eV). For NNN and NNNN configurations, the two Mn sites adjacent to the vacancy show different behaviors. The Mn atom farthest from the \ce{Fe_{Mn}} defect show $U$ values slightly higher than $U_\textrm{SC}$ (+0.08 eV) with a behavior similar to the one of the Mn sites adjacent to a doubly charged \ce{V_O^x} in SMO, which can be explained in terms of a change in coordination number~\citeSI{Ricca2019SI}. Instead, a larger reduction (-0.41 eV) is observed for the Mn atom closest to the substitutional iron and for the only remaining Mn site adjacent to the \ce{V_O} in the case of a NN \ce{Fe_{Mn}-V_O^{OP}} defect couple. This latter behavior was observed for the Mn atoms adjacent to a neutral \ce{V_O^{..}} in SMO due to both the change in coordination number and oxidation state of these reduced (\ce{Mn^{3+}}) sites~\citeSI{Ricca2019SI}. Similar conclusions can be drawn for all defect pairs (schematically shown in Fig.~\ref{fig:SMO_structure_defect_SI}), including \ce{Fe_{Mn}-V_O^{IP}} defect-pairs as shown in Fig.~\ref{fig:USCSD_AFM_SI}.


\begin{figure}[h]
 \centering
 \includegraphics[width=\columnwidth]{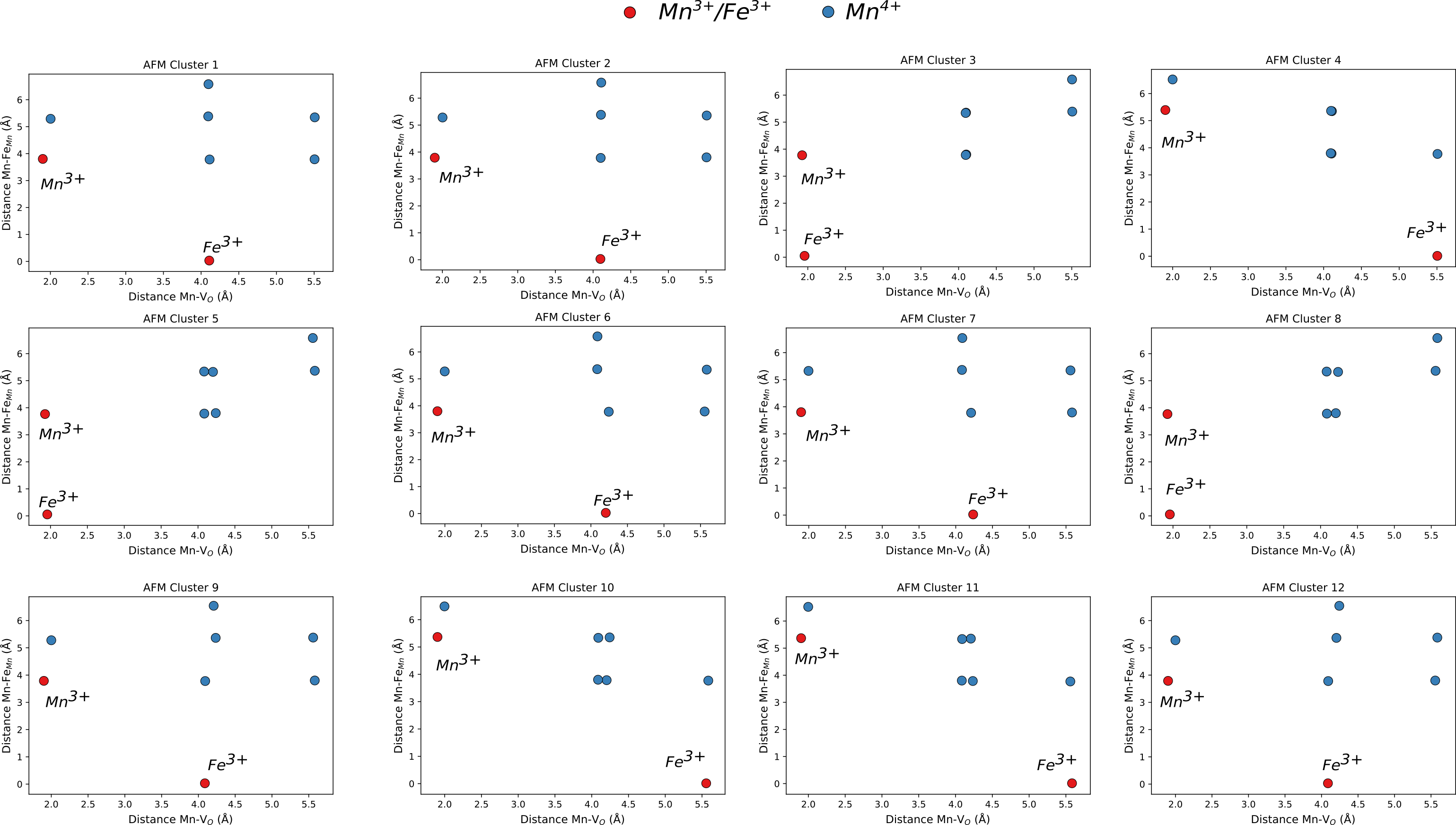}
 \caption{Oxidation state reported as a function of the distance of each site from  \ce{Fe_{Mn}} and \ce{V_O} defects computed for the Mn/Fe atoms in AFM SMO structures in each of the 12 \ce{Fe_{Mn}-V_O} configurations shown in Fig.~\ref{fig:SMO_structure_defect_SI}.}
\label{fig:OX_AFM_SI}
\end{figure}

We determine the oxidation state according to the method of Ref.~\citeSI{Sit2011SI} which is based on the occupation matrix of the $d$ orbitals of each Hubbard atom that is available when performing DFT+$U$ calculations. Each $d$ orbital is considered to be fully occupied if the corresponding matrix element is closer to unity than a given threshold. In the AFM phase, we use a threshold of 0.9 for the occupation. These oxidation states confirm the above picture (see Fig.~\ref{fig:OX_AFM_SI}) where only one of the Mn sites in nearest-neighbor position to the oxygen vacancy is reduced to \ce{Mn^{3+}}, in line with results of M{\"o}ssbauer spectroscopy suggesting a partial reduction of the Mn atoms in \ce{SrMn_{1-x}Fe_xO_{3-\delta}} samples, presumably in the vicinity of the vacant anion~\citeSI{BattleJM9960601187SI}. The Fe ion which is always in the \ce{Fe^{3+}} charge state, independently of is position relatively to the \ce{V_O}.
%

\subsubsection{FM SMO}\label{sec:SI_FM_SMO_U}

\begin{figure}[h]
 \centering
 \includegraphics[width=\columnwidth]{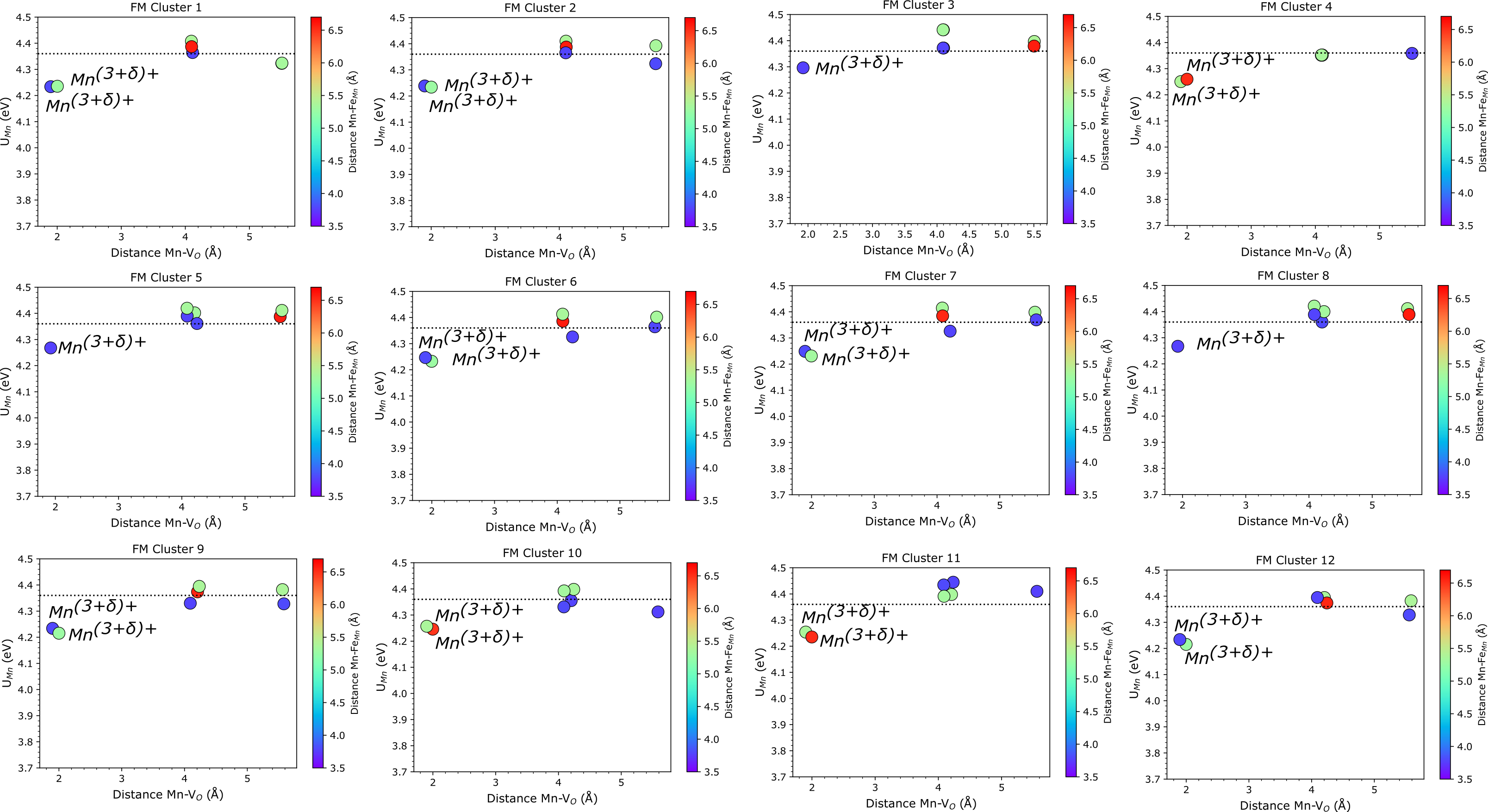}
 \caption{$U_{\textrm{SC-SD}}$ values reported as a function of the distance of each Hubbard site from  \ce{Fe_{Mn}} and \ce{V_O} defects computed for the Mn atoms in FM SMO structures in each of the 12 \ce{Fe_{Mn}-V_O} configurations shown in Fig.~\ref{fig:SMO_structure_defect_SI}. The oxidation state is reported indicated for the partially reduced Mn sites.}
\label{fig:USCSD_FM_SI}
\end{figure}

For the FM phase (cf. Figs.~\ref{fig:USCSD}b and~\ref{fig:USCSD_FM_SI}), the $U_\textrm{SC-SD}$ values deviate less from $U_\textrm{SC}$ than for AFM SMO. Interestingly, all the Mn adjacent to the \ce{V_O} show the same behavior with a small reduction of their $U_\textrm{SC-SD}$ value with respect to $U_\textrm{SC}$, suggesting a partial reduction of these sites (\ce{Mn^{(3+\delta)+}}). These observations can be explained considering the metallic nature of FM SMO, in which one expects the defect state and the corresponding changes in the local chemical environment to be more delocalized over the whole structure, compared to the semiconducting AFM phase, where a more confined impact of the defect on the local chemical environment was already observed in the case of oxygen-deficient SMO~\citeSI{Ricca2019SI}.

\begin{figure}[h]
 \centering
 \includegraphics[width=\columnwidth]{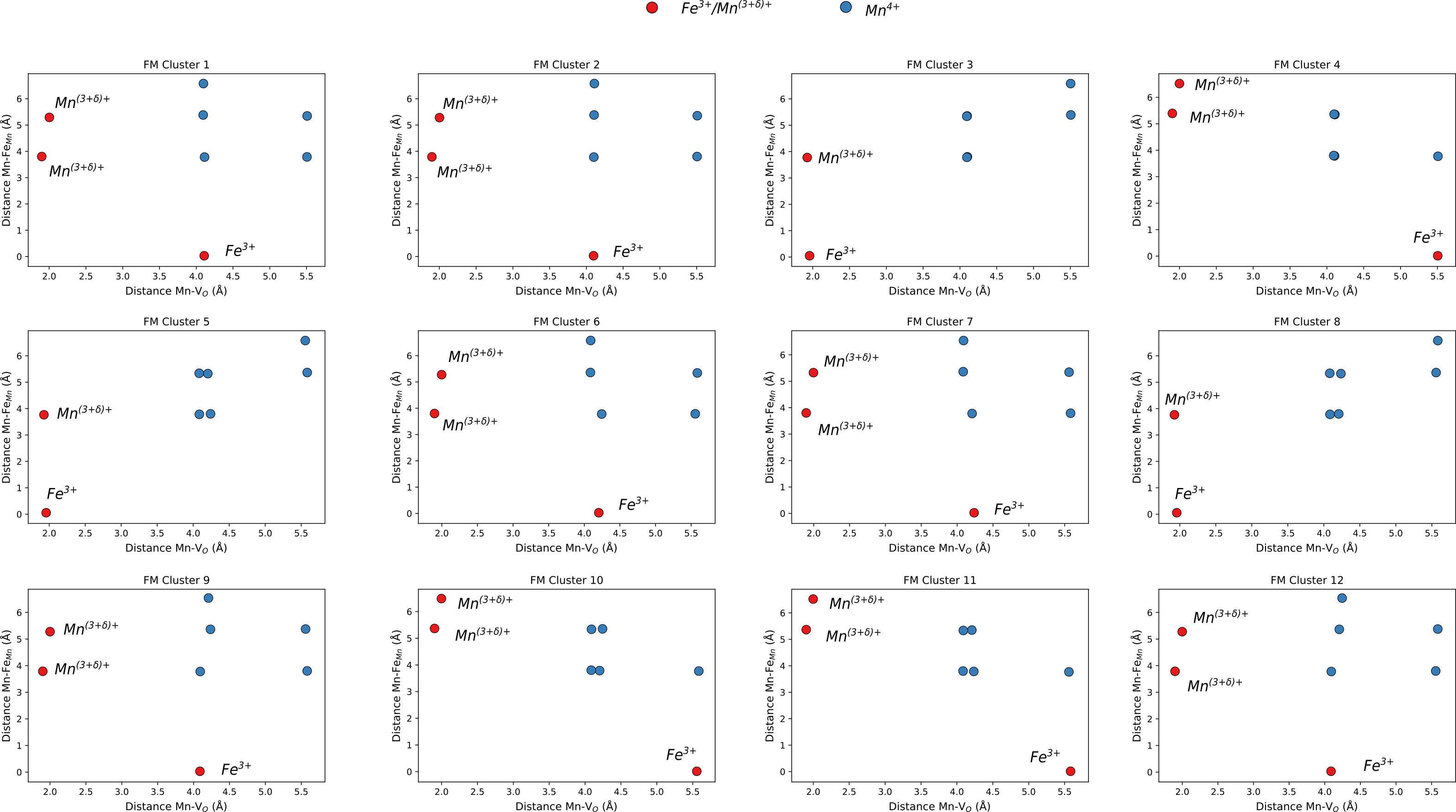}
 \caption{Oxidation state reported as a function of the distance of each site from  \ce{Fe_{Mn}} and \ce{V_O} defects computed for the Mn/Fe atoms in FM SMO structures in each of the 12 \ce{Fe_{Mn}-V_O} configurations shown in Fig.~\ref{fig:SMO_structure_defect_SI}.}
\label{fig:OX_FM_SI}
\end{figure}

These observations of a partial reduction of multiple Mn sites accompanying the \ce{Fe^{3+}} formation is again supported by the determined oxidation states according to the method of Ref.~\citeSI{Sit2011SI}. As opposed to the AFM phase, we use a lower threshold of about 0.7 for the FM order, to be able to discern one (for configurations in which the \ce{V_O} is adjacent to one Mn and to the Fe ion) or two (in all the other cases) reduced Mn atoms as shown in Fig. \ref{fig:OX_FM_SI}.

\clearpage
\subsection{Hubbard correction for Fe: $U_{\ce{Fe}}$}

\begin{figure}[h]
	\centering
	\includegraphics[width=0.35\columnwidth]{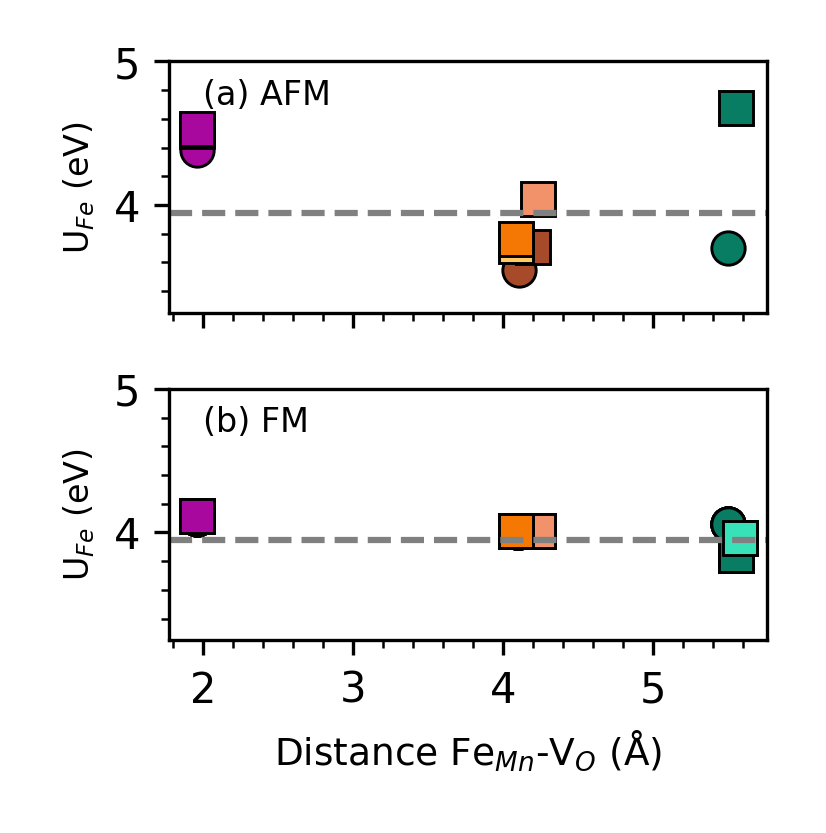}
	\caption{$U_{\textrm{SC-SD}}$ values on the \ce{Fe_{Mn}} atoms computed for all the considered \ce{Fe_{Mn}-V_O} configurations in (a) AFM and (b) FM SMO. Circle and square symbols refer to data obtained for \ce{V_O^{OP}} and \ce{V_O^{IP}}, respectively. See Fig.~\ref{fig:SMO_structure_defect} of the main text for the color code. The horizontal dashed line indicates the $U_{\textrm{SC}}$ value computed for Fe ions in the (a) G-AFM and (b) FM phases of \ce{LaFeO3} as reference.}
	\label{fig:USCSD_Fe}
\end{figure}

We also determined the $U$ value on the Fe using the SC-SD approach as shown in Fig.~\ref{fig:USCSD_Fe} for each configuration as a function of the \ce{Fe_{Mn}-V_O} distance. The horizontal dashed lines in Fig.~\ref{fig:USCSD_Fe} indicate the $U_\textrm{SC}$ value computed for iron in \ce{LaFeO3}, a perovskite material where Fe has the same octahedral coordination environment and the same oxidation state (\ce{Fe^{3+}}) as in Fe-doped SMO. Unsurprisingly, larger deviations from these $U_\textrm{SC}$ values are generally observed for configurations (in violet in Fig.~\ref{fig:USCSD_Fe}) in which the \ce{Fe_{Mn}} is nearest-neighbor position to the \ce{V_O} and in the insulating AFM phase. The larger deviation of the $U$ value for the Fe in the NNNN \ce{Fe_{Mn}-V_O^{IP}} observed only in the AFM case can be explained considering the interaction of the Fe atoms in these configurations with the \ce{V_O} in the neighboring cell, which is weaker in the more screened FM phase.

\clearpage
\section{Strain dependent defect-pair formation energy in FM SMO}

\begin{figure}[h]
 \centering
 \includegraphics[width=0.35\columnwidth]{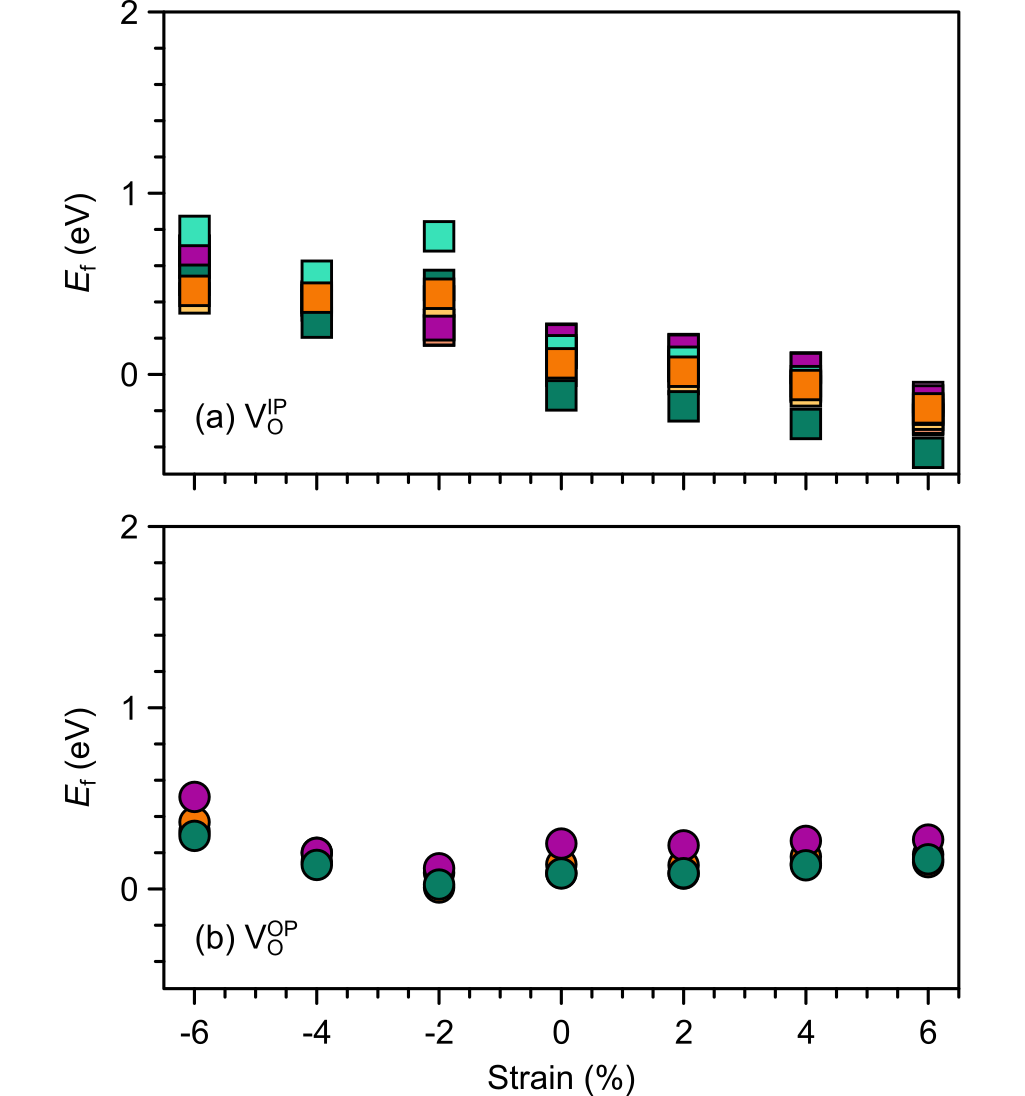}
 \caption{Strain dependent formation energy ($E_\textrm{f}$) of (a) \ce{Fe_{Mn}-V_O^{IP}} and (b) \ce{Fe_{Mn}-V_O^{OP}} defect pairs in FM SMO. See color code in Fig.~\ref{fig:SMO_structure_defect} of the main text.}
\label{fig:ef_strainFM}
\end{figure}

As shown in Fig. \ref{fig:ef_strainFM} for the FM phase, under compressive strain, the formation energy of \ce{Fe_{Mn}-V_O^{IP}} defects increases as expected from volume arguments, which can be explained in terms of a reduced sensitivity of the metallic FM phase to crystal field effects, allowing volume effects to dominate~\citeSI{aschauer2013strainSI}.
%

\section{Magnetic order in DFT+$U_\textrm{SC}$}

\begin{figure}[h]
 \centering
 \includegraphics[width=0.35\columnwidth]{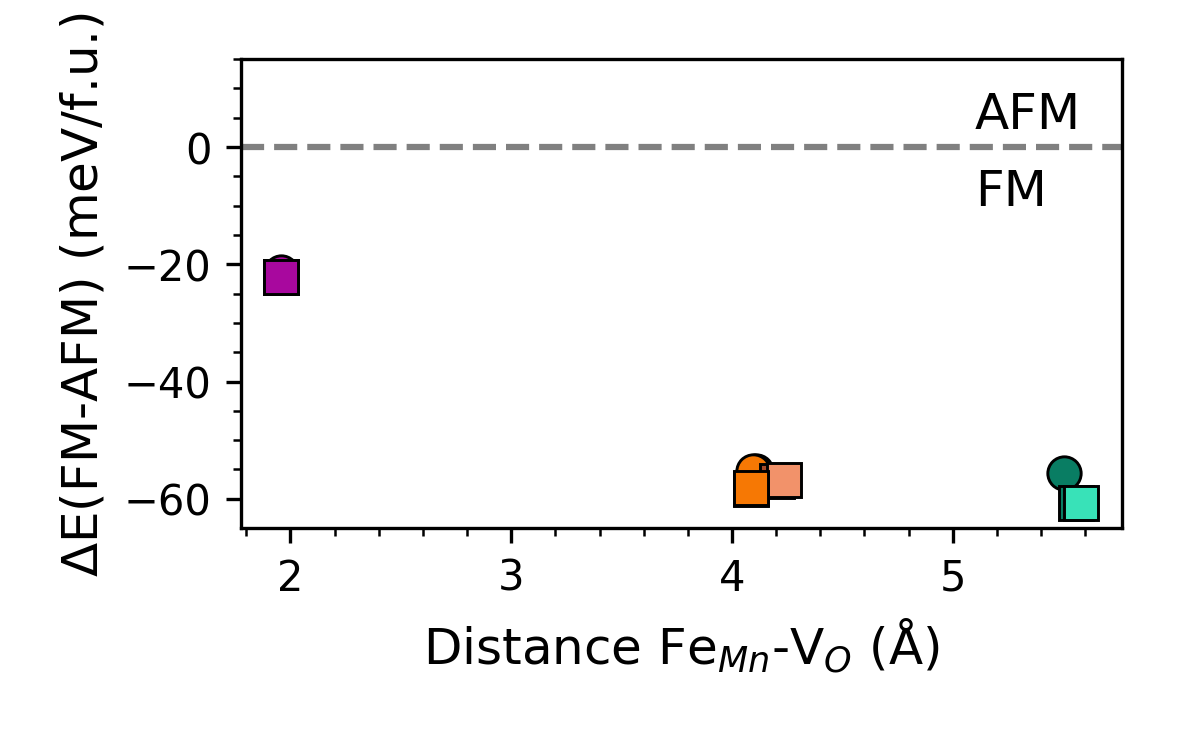}
 \caption{Total energy differences ($\Delta E(\textrm{FM-AFM})$) per formula unit between the defective cells with FM and AFM order computed with the DFT+$U_\textrm{SC}$ approach. AFM is more stable for positive and FM for negative differences. $\Delta E(\textrm{FM-AFM})$ is reported with respect to the the \ce{Fe_{Mn}-V_O} distance in each defective configuration in the unstrained SMO structure. Circle and square symbols refer to data obtained for \ce{VO_O^{OP}} and \ce{VO_O^{IP}}, respectively. See color code in Fig.~\ref{fig:SMO_structure_defect} of the main text.}
\label{fig:magnetic_order_USC}
\end{figure}

Figure~\ref{fig:magnetic_order_USC} shows that for DFT+$U_\textrm{SC}$, in which the global $U_\textrm{SC}$ computed for stoichiometric bulk SMO in the corresponding magnetic phase is applied on all the Mn sites, all the considered \ce{Fe_{Mn}-V_O} configurations are more stable in the FM phase and that the stability of the FM order increases with increasing distance of  \ce{Fe_{Mn}} from \ce{V_O}. Indeed, the increasing distance between the \ce{Fe^{3+}} and the reduced \ce{Mn^{3+}} ions promotes the ferromagnetic \ce{Mn^{3+}}/\ce{Mn^{4+}} interactions.

\clearpage
\section{Defect-induced polarization}\label{sec:SI_pol}

\subsection{Polarization in the unstrained AFM phase}

\begin{figure}[h]
 \centering
 \includegraphics[width=0.35\textwidth]{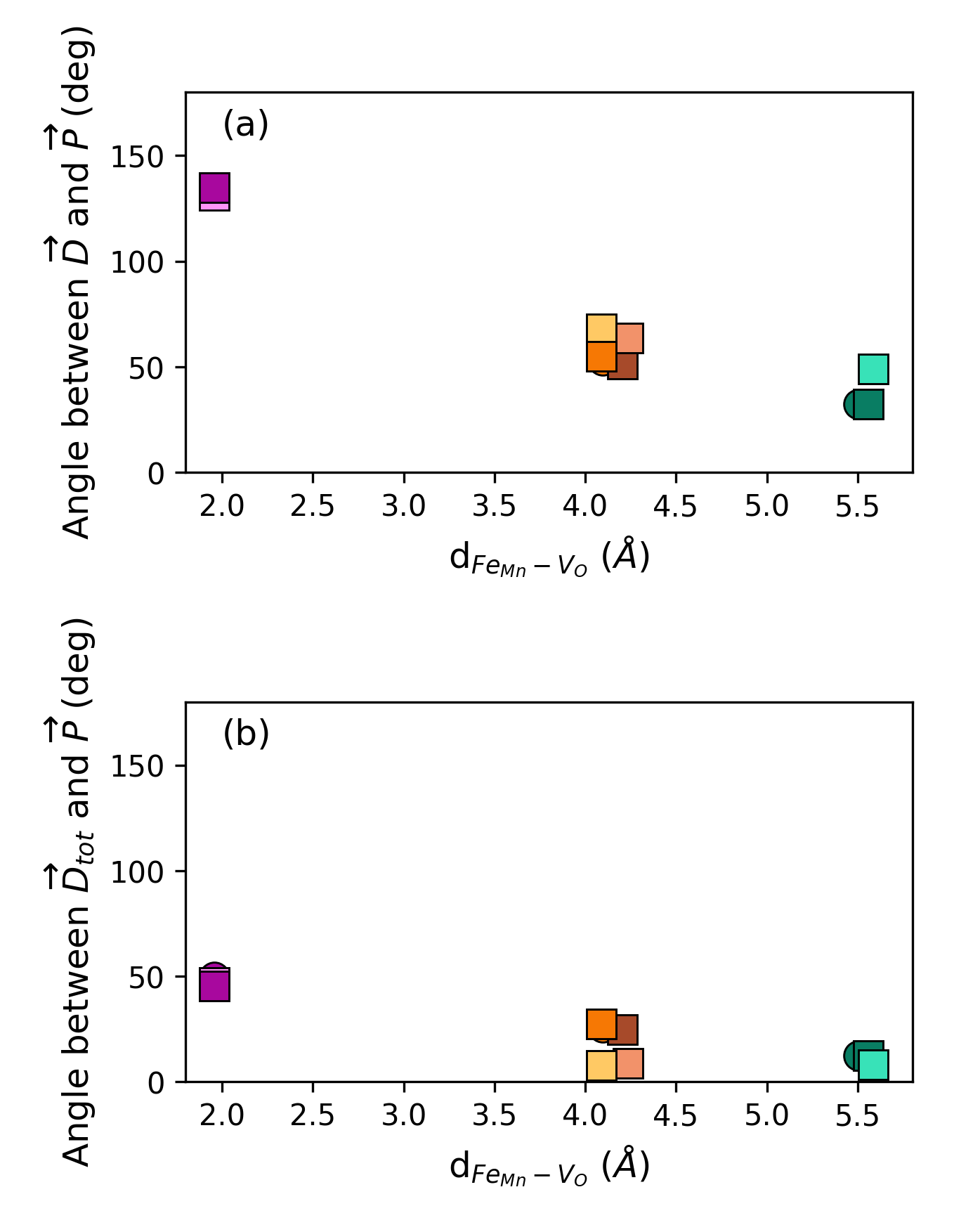}
\caption{Angle between the (a) \ce{Fe^'_{Mn}-V_O^{..}} defect dipole $\vec{D}$ or (b) the total defect dipole $\vec{D}_\mathrm{tot}$ and the polarization $\vec{P}$ as a function of the distance between \ce{Fe_{Mn}} and \ce{V_O} for the different configurations in the AFM phase. Circle and square symbols refer to data obtained for \ce{V_O^{OP}} and \ce{V_O^{IP}}, respectively. See Fig.~\ref{fig:SMO_structure_defect} in the main text for the color code.}
\label{fig:Angle_Dip_Pol_distance}
\end{figure}

Figure~\ref{fig:Angle_Dip_Pol_distance}(a) shows that the angle between the polarization $\vec{P}$ and the \ce{Fe^'_{Mn}-V_O^{..}} dipole $\vec{D}$ decreases with increasing the \ce{Fe^'_{Mn}-V_O^{..}} distance. This observation seems counterintuitive, but can be explained considering also the presence of the \ce{Mn^'_{Mn}-V_O^{..}} dipole $\vec{D'}$, with the reduced Mn always being adjacent to \ce{V_O}. The closer the \ce{Fe^'_{Mn}}, \ce{Mn^'_{Mn}}, and \ce{V_O} sites, the stronger the coupling between $\vec{D}$ and $\vec{D'}$ and consequently the larger the deviation of $\vec{P}$ from $\vec{D}$. Indeed, the polarization is mainly aligned along the direction of the vector sum of the $\vec{D}$ and $\vec{D'}$ dipoles ($\vec{D}_\mathrm{tot}$) as shown by the smaller angles between $\vec{D}_\mathrm{tot}$ and $\vec{P}$ (see Figure~\ref{fig:Angle_Dip_Pol_distance}(b)).

%

\subsection{Polarization in the unstrained FM phase\label{sec:polFM}}

The polarization for the FM phase was computed using the nominal charges of +2 for Sr, -2 for O, +3 for Fe, and +4 or +3 for stoichiometric-like or reduced Mn sites, even though results of Sec.~\ref{sec:SI_FM_SMO_U} indicated that, due to the metallic nature of this phase, the reduction of the Mn sites is only partial. While NN defect configurations, with only one reduced Mn site adjacent to the \ce{V_O}, exhibit similar or slightly higher total polarization ($\vec{P}_\mathrm{tot}$) compared to the AFM phase, the NNN and NNNN \ce{Fe_{Mn}-V_O} defect pairs show almost the same $\vec{P}_\mathrm{tot}$, regardless of the \ce{Fe_{Mn}-V_O} distance, contrarily to the increase of $\vec{P}_\mathrm{tot}$ with increasing \ce{Fe_{Mn}-V_O} distance reported for the AFM phase (cf. Fig. 7 in the main text). This behavior can be explained considering that for these configurations the polarization is computed assuming the presence of two \ce{Mn^'_{Mn}-V_O^{..}} defect pairs involving the Mn sites adjacent to \ce{V_O}, which results in the reduction/cancellation of the contribution to the polarization along the axis of the broken \ce{Mn-O-Mn} bonds. For example, for all the \ce{Fe_{Mn}-V_O^{OP}} configurations the out-of-plane $\vec{P}_{b}$ is almost null, the broken \ce{Mn-O-Mn} bond lying along the $b$-axis. The observed behavior for the FM phase can thus be explained considering the presence of only one defect dipole (the one associated with \ce{Fe^'_{Mn}-V_O^{..}}) inducing the polarization, with respect to the AFM phase where both \ce{Fe^'_{Mn}-V_O^{..}} and \ce{Mn^'_{Mn}-V_O^{..}} are responsible for the predicted properties.
\begin{figure}[h]
 \centering
 \includegraphics[width=0.35\columnwidth]{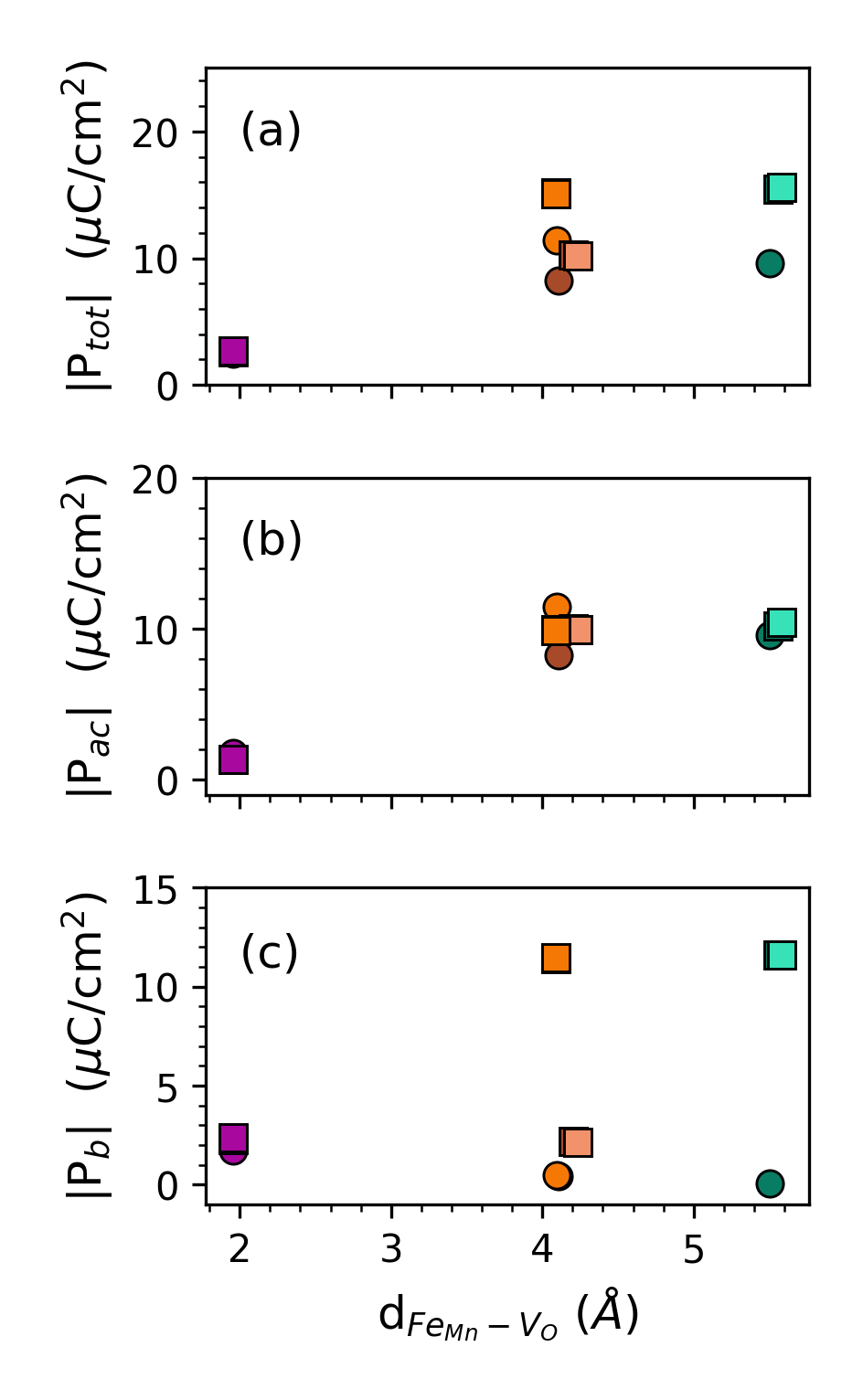}
\caption{Modulus of (a) the total polarization vector, (b) the in-plane ($|\vec{P}_{ac}|$) and (c) the out-pf-plane ($|\vec{P}_{b}|$) components for the different defect configurations in the FM phase. Circle and square symbols refer to data obtained for \ce{V_O^{OP}} and \ce{V_O^{IP}}, respectively. See Fig.~\ref{fig:SMO_structure_defect} in the main text for the color code.}
\label{fig:Pol_FM_unstrained}
\end{figure}

\clearpage
\subsection{Polarization as a function of the cell size}

\begin{figure}[h]
 \centering
 \includegraphics[width=0.35\columnwidth]{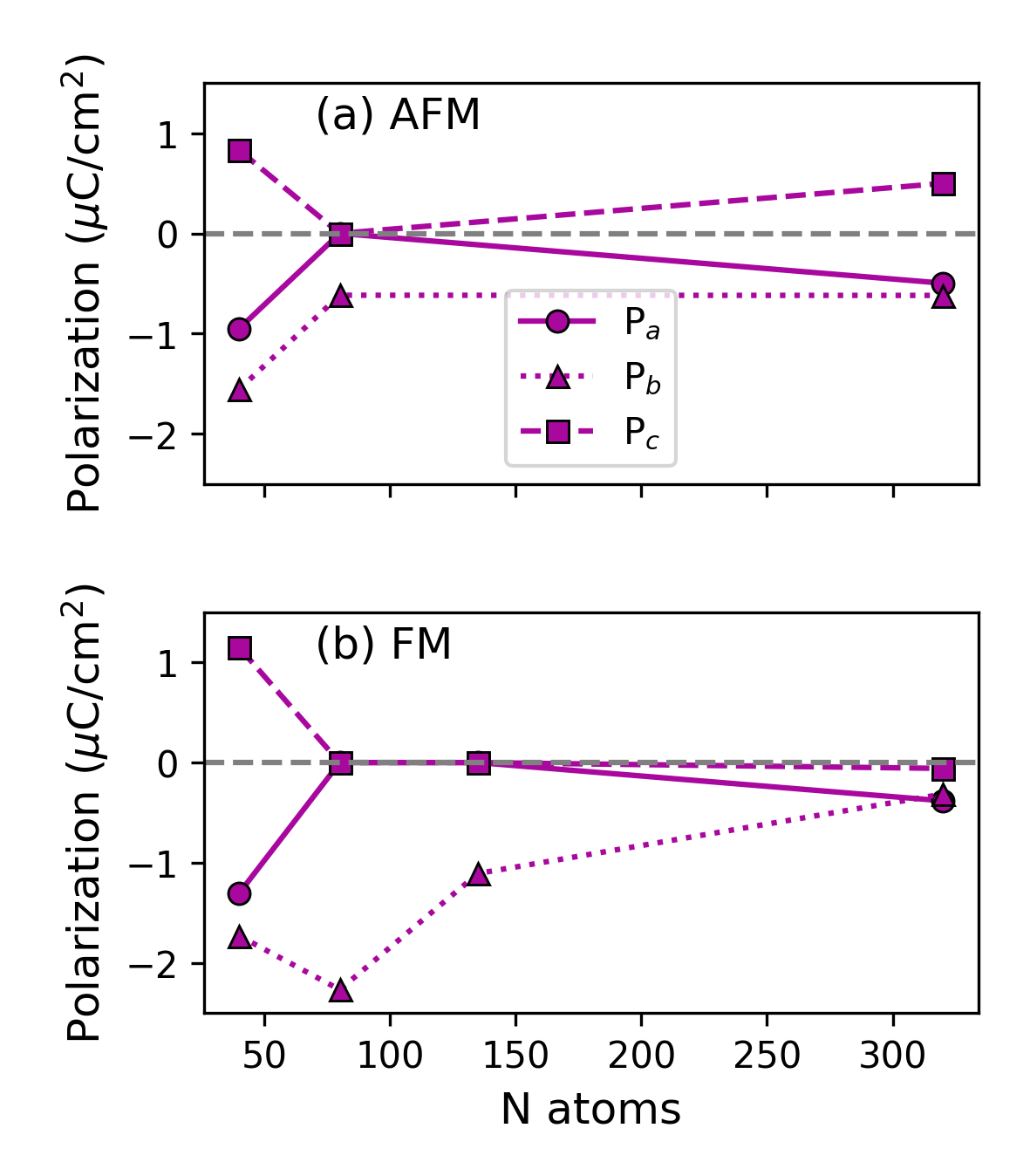}
 \caption{Evolution of the $a$, $b$, and $c$ component of the polarization for the NN \ce{Fe_{Mn}-V_O^{OP}} defective supercell as a function of the cell size in (a) the AFM and (b) the FM phase.}
\label{fig:Pol_cell_size}
\end{figure}

As shown in Fig.~\ref{fig:Pol_cell_size}, we generally observe a reduction of the polarization with increasing cell size, deviations like in the 80-atom cell likely being associated with cell-anisotropy effects. This suggests a local polarizing effect that becomes less important as the cell size increases and hence the defect-pair concentration decreases.

\clearpage
\subsection{Interplay between strain and polarization in the FM phase}

\begin{figure}[h]
 \centering
 \includegraphics[width=0.35\columnwidth]{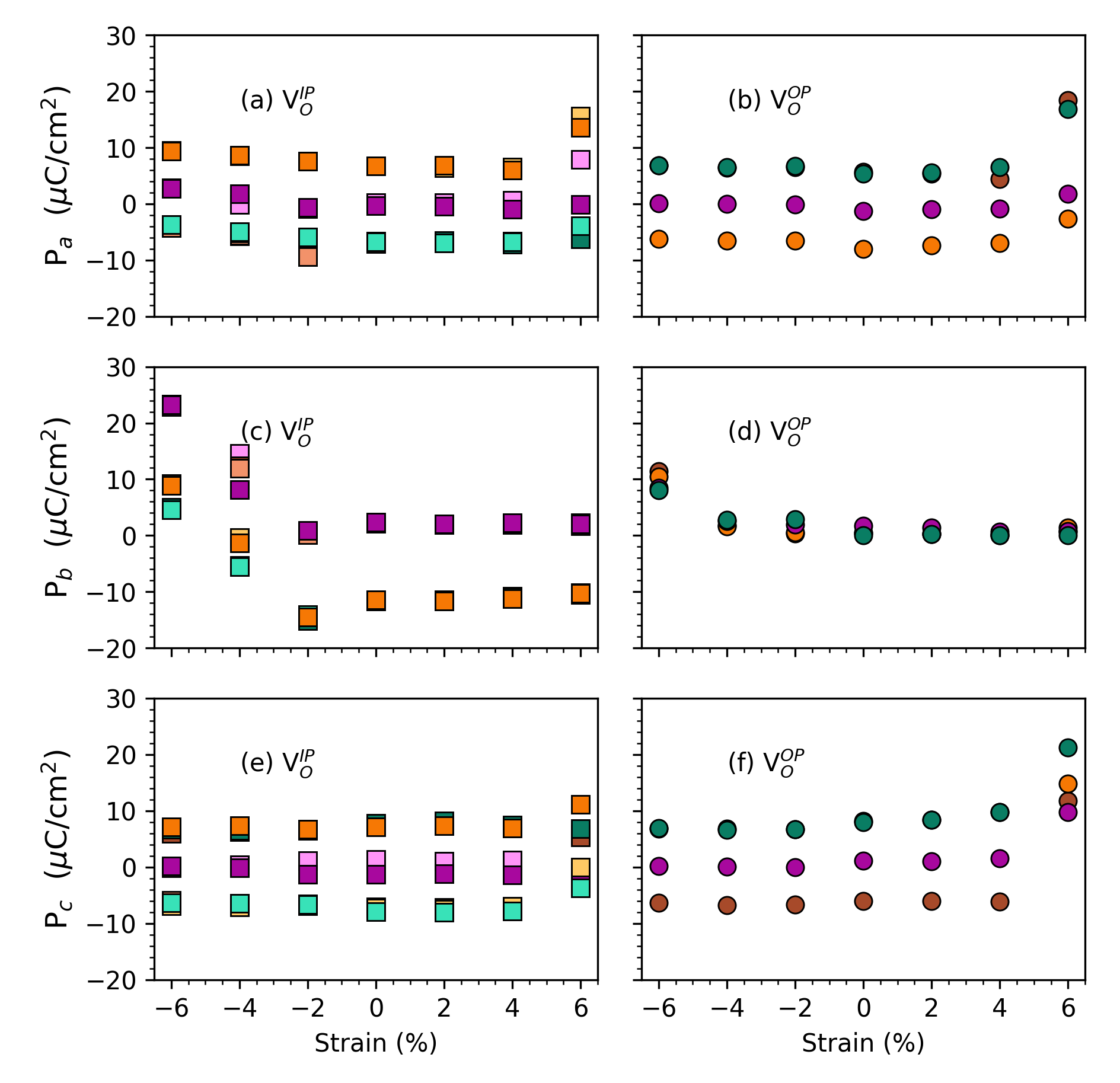}
\caption{Epitaxial-strain driven changes of the (a-b) $a$-, (c-d) $b$- and (e-f) $c$-components of the polarization for the different defect pair configurations in the FM phase of SMO. (a), (c), and (e) plots for \ce{Fe_{Mn}-V_O^{OP}} and (b), (d), and (f) plots for \ce{Fe_{Mn}-V_O^{IP}} defects. Circle and square symbols refer to data obtained for \ce{V_O^{OP}} and \ce{V_O^{OP}}, respectively. See Fig.~\ref{fig:SMO_structure_defect} in the main text for the color code.}
\label{fig:Pol_FM}
\end{figure}

\begin{figure}[h]
 \centering
 \includegraphics[width=0.35\columnwidth]{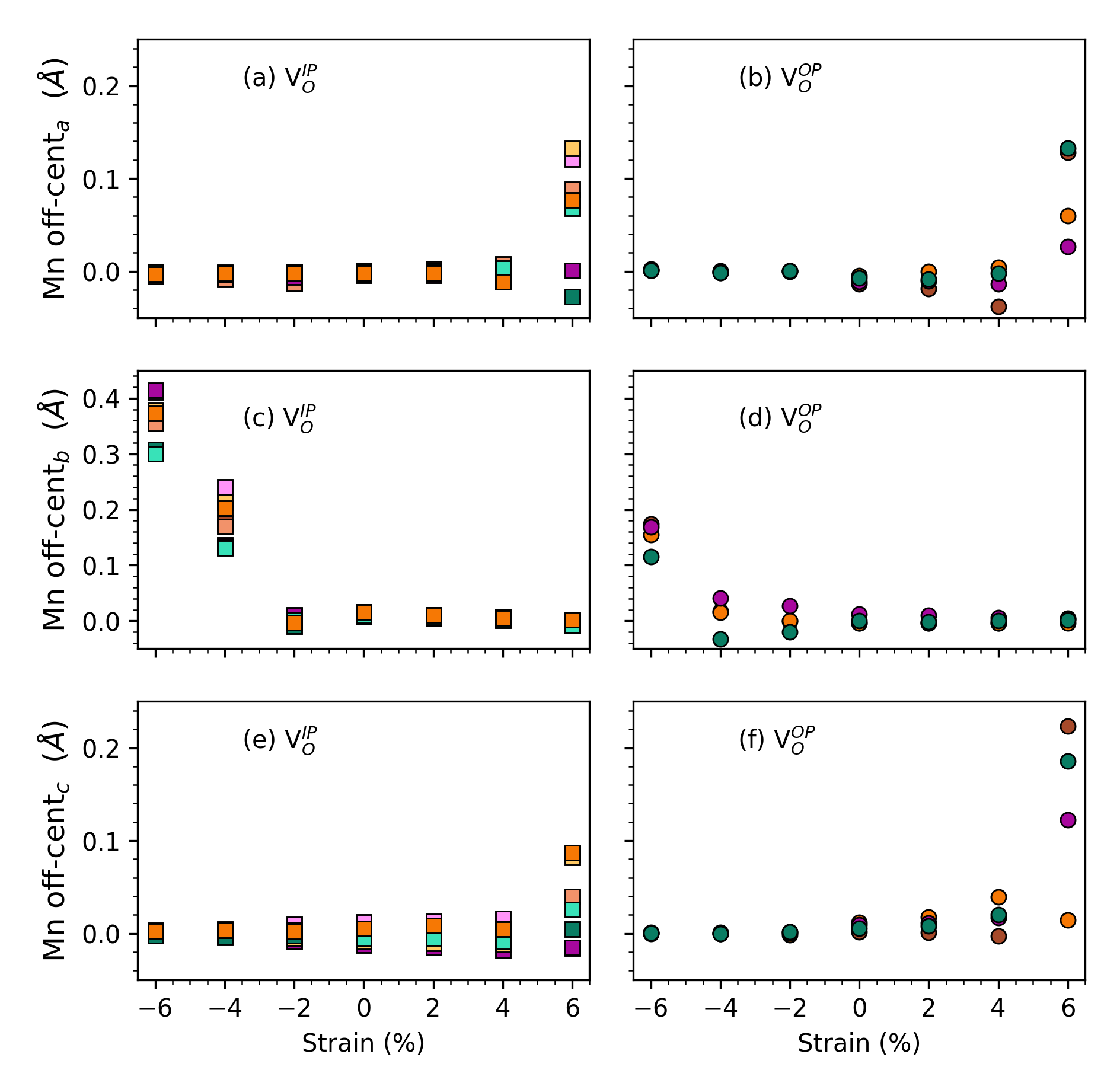}
\caption{Strain dependence of the average Mn-offcentering along the (a-b) $a$-, (c-d) $b$- and (e-f) $c$-axis for the different defect pair configurations in the FM phase of SMO. (a), (c), and (e) plots for \ce{Fe_{Mn}-V_O^{OP}} and (b), (d), and (f) plots for \ce{Fe_{Mn}-V_O^{IP}} defects. Circle and square symbols refer to data obtained for \ce{V_O^{OP}} and \ce{V_O^{IP}}, respectively. See Fig.~\ref{fig:SMO_structure_defect} in the main text for the color code.}
\label{fig:Mnoff_FM_strain}
\end{figure}

The strain dependent polarization of Fe-doped oxygen deficient SMO is the result of a complex interplay between defect chemistry and electronic and magnetic degrees of freedom. While for the insulator AFM phase, polarization can be enhanced by strain, in the FM phase the polarization is fairly constant with respect to the applied epitaxial strain (see Fig.~\ref{fig:Pol_FM} and the average Mn off-centerings remain very small (about 0.02-0.04~\AA\ in the strain range between -2 and 4\%, see Fig.~\ref{fig:Mnoff_FM_strain}) similar in magnitude to the unstrained structure. Interestingly, compressive strain values of 4\% can result in an enhancement of the $P_{b}$ component, as confirmed also by the larger Mn off-centerings observed at this strain along the $b$-axis. Also at 6\% tensile strain an increase of the components of the polarization and of the Mn off-centerings in the $ac$-plane are observed. This behavior can be understood considering the larger electronic screening of the defect dipole in the metallic FM phase, resulting in the lower sensitivity of the polarization of the defective cells to the applied strain and evolution with strain of the frequency of the polar modes in the stoichiometric FM order, where the IP modes soften only for large tensile strain of 6\% and the OP modes becomes slightly unstable at 2\% compressive strain~\citeSI{ricca2021ferroelectricitySI}.
%

\clearpage
\bibliographystyleSI{apsrev4-1}
\bibliographySI{references}

\end{document}